\documentclass[runningheads]{llncs}

\usepackage{amsmath,amssymb,amsfonts}
\usepackage{adjustbox}
\usepackage{algorithm}        
\usepackage{algpseudocode}    
\usepackage{graphicx}
\usepackage{tabularx} 
\usepackage[skins,breakable]{tcolorbox}
\usepackage{bbding}
\usepackage{booktabs}
\usepackage{multirow}
\usepackage{textcomp}
\usepackage[figuresright]{rotating}

\usepackage{capt-of}

\usepackage{latexsym}

\usepackage[T1]{fontenc}

\def\code#1{\texttt{#1}}
\usepackage{microtype}

\usepackage{graphicx}

\usepackage{xspace}
\newcommand{\posir}{\textsc{PosIR}\xspace}
\begin{document}

\title{\posir{}: Position-Aware Heterogeneous Information Retrieval Benchmark}

\author{Ziyang Zeng\inst{1} \and Dun Zhang\inst{2} \and Yu Yan\inst{1} \and Xu Sun\inst{3,4} \and Cuiqiaoshu Pan\inst{1} \and Yudong Zhou\inst{2} \and Yuqing Yang\inst{1}\thanks{Corresponding author.}}

\authorrunning{Z. Zeng et al.}

\institute{
Beijing University of Posts and Telecommunications\\
\email{ziyang1060@bupt.edu.cn, yanyu2023@bupt.edu.cn, pancuiqiaoshu@bupt.edu.cn, yangyuqing@bupt.edu.cn}
\and
Prior Shape\\
\email{dunnzhang0@gmail.com, zhouyudong@priorshape.com}
\and
Orange Research\\
\and 
Universit\'{e} Caen Normandie, ENSICAEN, CNRS, Normandie Univ, GREYC UMR6072, F-14000 Caen, France\\
\email{xu.sun@orange.com}
}

\maketitle

\begin{abstract}
In real-world documents, the information relevant to a user query may reside anywhere from the beginning to the end.
This makes position bias---a systematic tendency of retrieval models to favor or neglect content based on its location---a critical concern.
Although recent studies have identified such bias, existing analyses focus predominantly on English, fail to disentangle document length from information position, and lack a standardized framework for systematic diagnosis.
To address these limitations, we introduce \textbf{\posir} (\textbf{Pos}ition-Aware \textbf{I}nformation \textbf{R}etrieval), the first standardized benchmark designed to systematically diagnose position bias in diverse retrieval scenarios.
\posir comprises 310 datasets spanning 10 languages and 31 domains, with relevance tied to precise reference spans. 
At its methodological core, \posir employs a length-controlled bucketing strategy that groups queries by positive document length and analyzes positional effects within each bucket. This design strictly isolates position bias from length-induced performance degradation.
Extensive experiments on 10 state-of-the-art embedding-based retrieval models reveal that:
(1) retrieval performance on \posir with documents exceeding 1536 tokens correlates poorly with the MMTEB benchmark, exposing limitations of current short-text evaluations;
(2) position bias is pervasive in embedding models and even increases with document length, with most models exhibiting primacy bias while certain models show unexpected recency bias;
(3) as an exploratory investigation, gradient-based saliency analysis further uncovers two distinct internal mechanisms that correlate with these positional preferences.
We hope that \posir can serve as a valuable diagnostic framework to advance the development of position-robust retrieval systems.\footnote{\url{https://github.com/Ziyang1060/PosIR}}

\end{abstract}

\section{Introduction}
\label{sec:intro}

Information retrieval aims to identify relevant documents from a large collection in response to a user query~\cite{IIR}.
A robust retrieval system should assess a document's relevance based on its content, not on where that content happens to appear. 
However, retrieval models that systematically favor or neglect content based on its location risk missing critical information and undermining retrieval quality.
This phenomenon, known as \emph{position bias}~\cite{DBLP:conf/ecir/HofstatterLAZH21}, has attracted growing attention: recent studies~\cite{coelho-etal-2024-dwell,fayyaz-etal-2025-collapse,zeng-etal-2025-empirical} have empirically shown that embedding-based retrieval models exhibit primacy bias, disproportionately attending to the early portions of documents while neglecting evidence that appears later.

Although position bias has been recognized, existing analyses suffer from three critical limitations.
\textbf{First, the linguistic scope is narrow.}
Prior investigations focus almost exclusively on English, leaving open the question of whether position bias generalizes across typologically diverse languages and how it manifests in cross-lingual retrieval settings.
\textbf{Second, document length and information position are confounded.}
Current studies typically vary the position of relevant information without controlling for document length~\cite{zeng-etal-2025-empirical}.
As a result, when a model performs poorly on a document with late-appearing evidence, it remains ambiguous whether the failure reflects a genuine positional preference or simply a capacity limitation when processing long documents.
Without an evaluation design that isolates these two variables, position bias cannot be rigorously measured.
\textbf{Third, a standardized evaluation framework is absent.}
Different studies adopt varying document lengths, position definitions, and analytical protocols, making their conclusions difficult to compare and reproduce.
The lack of a unified benchmark prevents the community from systematically tracking progress or diagnosing model-specific weaknesses.

To address these limitations, we introduce \posir (\textbf{Pos}ition-Aware \textbf{I}nformation \textbf{R}etrieval), the first standardized benchmark for systematically diagnosing position bias in information retrieval.
\posir is characterized by three key design principles, which collectively address the limitations identified above:
1) \textbf{Heterogeneous Coverage}: \posir spans 10 languages (English, Chinese, and 8 translated languages) and 31 domains, yielding 310 datasets in total. Both automatic and human quality assessments confirm that the translated data faithfully preserves source-language semantics, ensuring that our findings generalize beyond English and are not artifacts of a specific language.
2) \textbf{Length-Controlled Analysis}: At its methodological core, \posir employs a length-controlled bucketing strategy: queries are grouped by the positive document length, and positional effects are analyzed within each bucket. This design strictly isolates position bias from length-induced performance degradation, directly addressing the \emph{length--position confound}.
3) \textbf{Position-Aware Relevance}: Unlike existing benchmarks that assign coarse document-level relevance labels~\cite{muennighoff-etal-2023-mteb,DBLP:journals/corr/abs-2104-08663,chen-etal-2025-air}, \posir associates each query with a precise reference span within the document. Relevance is verified through strict span-based contrast, enabling fine-grained quantitative analysis of how the physical location of information impacts retrieval performance.

Using \posir, we conduct extensive experiments on 10 popular embedding-based retrieval
models, yielding several key insights:
\begin{itemize}
    \item \textbf{Benchmarking Discrepancy.}
    Retrieval performance on \posir with documents exceeding 1,536 tokens correlates poorly with the MMTEB~\cite{enevoldsen2025mmteb} scores. Models that excel in short-context evaluations often degrade significantly as document length increases.

    \item \textbf{Prevalence of Bias.}
    Position bias is pervasive in embedding models and even increases with document length, across both multilingual and cross-lingual settings. While most models exhibit primacy bias, we first identify unexpected recency bias in \texttt{NV-Embed-v2}~\cite{lee2025nvembed}.

    \item \textbf{Mechanistic Origins.}
    Going beyond performance metrics, we employ an exploratory gradient-based saliency analysis to uncover the internal mechanisms driving these biases, identifying two distinct attention behaviors that correlate with embedding models' positional preferences over input tokens. 
\end{itemize}

\section{Related Work}
\subsection{Position Bias in Information Retrieval}
Prior work has widely documented the existence of position bias in neural information retrieval (IR) systems.
Previous work~\cite{Mitigating10.1007/978-3-030-72113-8_16} identified that {MS MARCO}~\cite{DBLP:conf/nips/NguyenRSGTMD16}, a foundational IR dataset, is heavily skewed toward head-located relevant spans, leading fine-tuned models to inherit and amplify this position bias~\cite{jiang-etal-2021-bert}, compromising the reliability of existing evaluations~\cite{10.1145/3640460,sun-etal-2025-posum}. 
Recent studies confirm that such bias persists even in state-of-the-art embedding models~\cite{coelho-etal-2024-dwell,fayyaz-etal-2025-collapse,zeng-etal-2025-empirical}. 
Specifically, previous work~\cite{zeng-etal-2025-empirical} utilized answer start positions in SQuAD~\cite{rajpurkar-etal-2018-know} for bias analysis; however, SQuAD is constrained to English monolingual data and brief passages (avg. 117 words), hindering the generalization of findings to multilingual or long-context scenarios. Despite these insights, a standardized benchmark to rigorously evaluate position bias, especially within multilingual and longer contexts, remains absent. Consequently, \posir is proposed to bridge this gap.

\subsection{Synthetic Data for Information Retrieval} 
Leveraging large language models (LLMs) for synthetic data generation has recently attracted growing attention in the IR community.
Existing works can be broadly categorized into two primary use cases: \emph{training} and \emph{evaluation}.
From a training perspective, LLMs are employed to address the scarcity of domain-specific or task-specific data~\cite{Zeng2025OptimizingGR}.
Methods like InPars~\cite{InPars} and Promptagator~\cite{dai2023promptagator} prompt LLMs to generate synthetic queries for specific documents, facilitating effective retriever training in zero-shot settings.
This paradigm has been further extended to support multilingual retrieval tasks~\cite{thakur-etal-2024-leveraging,wang-etal-2024-improving-text}. Regarding evaluation, recent studies have validated the reliability of synthetic test collections for benchmarking~\cite{10.1145/3626772.3657942}. 
LLM-generated pseudo-labels and queries have been leveraged for unsupervised model selection on a target corpus~\cite{10.1145/3626772.3657798}. 
Additionally, frameworks like AIR-Bench~\cite{chen-etal-2025-air} provide automated pipelines to generate evaluation data for emerging domains efficiently. 
Inspired by these advancements, we extend the application of synthetic data generation to the specific challenge of \emph{position bias}. 
Unlike prior works that focus on document-level relevance, we utilize LLMs to generate fine-grained, position-aware relevance annotations, which are crucial for analyzing position bias.
By incorporating rigorous quality control mechanisms, \posir ensures the reliability of synthetic signals, enabling efficient and effective diagnosis of position bias in retrieval models.


\begin{figure}[!t]
\centering
\includegraphics[width=1.0\textwidth]{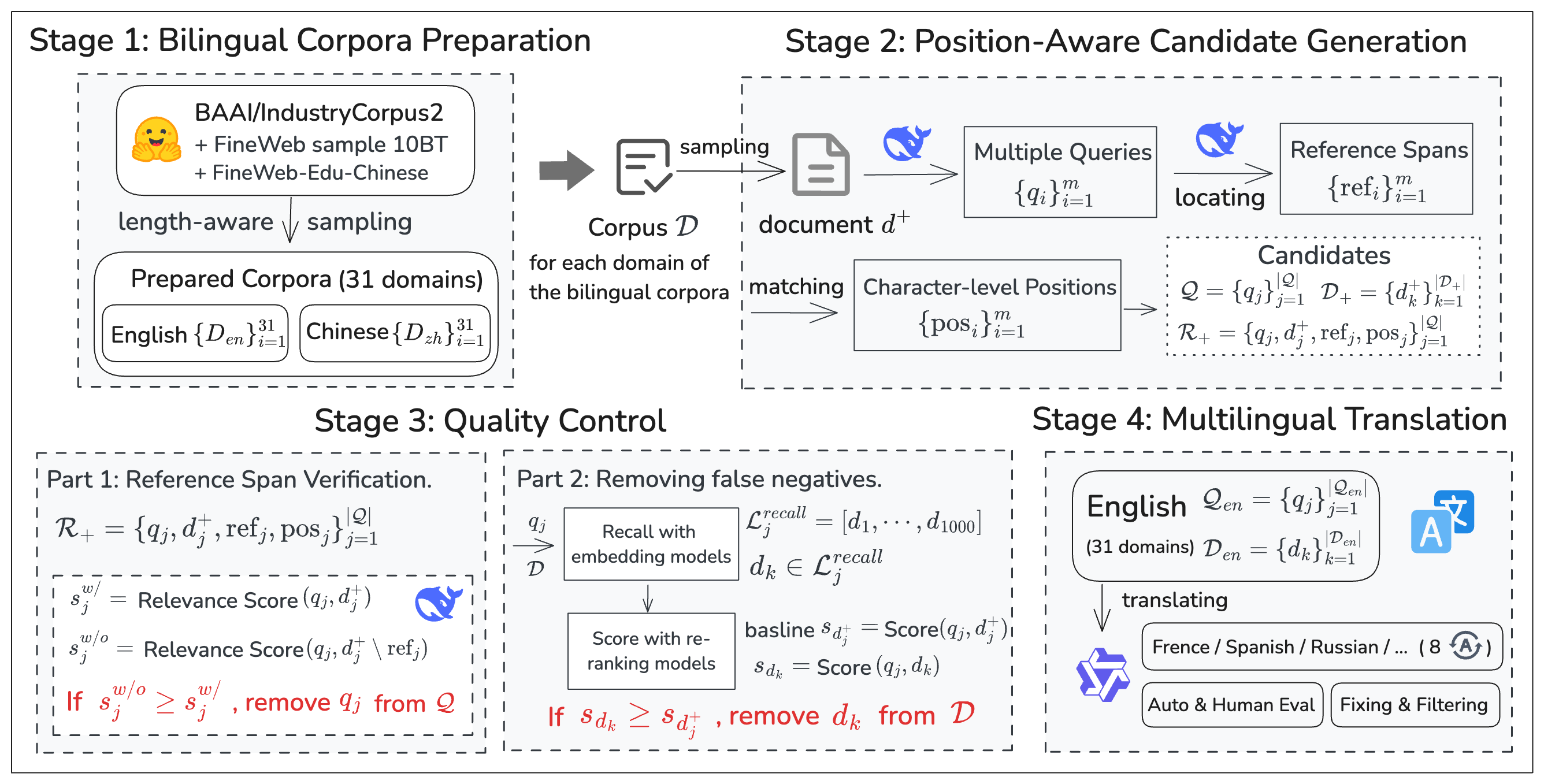}
\caption{The four-stage data generation pipeline of \posir.}
\label{fig:generation_pipeline}
\end{figure}

\section{Benchmark Construction}

The entire data generation pipeline of \posir consists of four stages: 1) Bilingual corpora preparation, 2) Position-aware candidate generation, 3) Quality control, and 4) Multilingual translation.
\subsection{Bilingual Corpora Preparation}
\label{sec:corpus_preparation}
As illustrated in Figure~\ref{fig:generation_pipeline}, the first stage involves preparing high-quality bilingual corpora across diverse domains.
Specifically, we leverage \texttt{IndustryCorpus2}\footnote{\url{https://huggingface.co/datasets/BAAI/IndustryCorpus2}}, an open-source and carefully curated English-Chinese dataset that covers 31 industry categories, broadly representing mainstream real-world application domains.
We exclude the original \textit{Others} category to maintain domain specificity and introduce \textit{FineWeb} to represent the general domain: for English, we use the \texttt{fineweb} sampled 10BT subset\footnote{\url{https://huggingface.co/datasets/HuggingFaceFW/fineweb/tree/main/sample/10BT}}, while for Chinese we adopt \texttt{Fineweb-Edu-Chinese-V2.1}\footnote{\url{https://huggingface.co/datasets/opencsg/Fineweb-Edu-Chinese-V2.1}}.
After this adjustment, the resulting corpus still consists of 31 domains.
Then, we adopt a length-aware sampling strategy that moderately over-samples shorter documents, which are more prevalent in real-world corpora. Specifically, using the Qwen3 tokenizer~\cite{yang2025qwen3technicalreport}, we partition documents into eight length buckets with an interval of 256 tokens (covering 0 to 2048 tokens) and apply a sampling ratio of 3:3:3:3:2:2:2:2, such that documents under 1024 tokens are sampled at 1.5 $\times$ the rate of longer ones.
Following this strategy, we sample approximately 70k documents for each domain from both the English and Chinese corpora.
Formally, for a given domain, we denote the English and Chinese corpora as
$\mathcal{D}_{en} = \{d_i\}_{i=1}^{n_{en}}$ and
$\mathcal{D}_{zh} = \{d_i\}_{i=1}^{n_{zh}}$,
containing $n_{en}$ and $n_{zh}$ documents, respectively.
More generally, we use
$\mathcal{D} = \{ d_i \}_{i=1}^{n}$
to denote a corpus in an arbitrary language, where $n$ denotes the number of documents in that corpus.

\subsection{Position-Aware Candidate Generation}
\label{sec:candidate_generation}
A retrieval dataset typically consists of three components: a document corpus, a set of queries, and relevance judgments (i.e., qrels).\footnote{In TREC parlance, \emph{qrels} refer to relevance judgments that indicate whether a document is relevant to a given query.}
After preparing the corpus in the initial stage, the candidate generation stage constructs the remaining two components of the dataset—queries and qrels—while explicitly accounting for the positional nature of relevance.
As shown in Figure~\ref{fig:generation_pipeline}, the candidate generation process for a given domain proceeds as follows:
1) Sample one document from the corpus $\mathcal{D}$ as the positive document $d^{+}$. 
2) Prompt LLMs to generate multiple queries $\{q_i\}_{i=1}^{m}$ for $d^{+}$ under a randomly sampled positional constraint (see Appendix~\ref{appendix_sec:prompt_qa}). 
3) Prompt LLMs to locate the reference span $\mathrm{ref}_{i}$ corresponding to each generated question $q_i$ (see Appendix~\ref{appendix_sec:prompt_ref}). 
4) Discard queries for which reference span localization fails (e.g., no unambiguous span can be identified, or multiple candidate spans exist), and apply regular-expression matching to determine the character-level position span $\mathrm{pos}_{i}(\mathrm{start,end})$ of $\mathrm{ref}_{i}$ within $d^{+}$.
In this stage, we use \texttt{DeepSeek-V3.1 (Thinking Mode)}~\cite{deepseekai2025deepseekv3technicalreport} for Steps 2 and 3, as these steps require multi-step reasoning over document content.
After repeating the above steps thousands of times, we obtain the query set $\mathcal{Q}=\{q_j\}_{j=1}^{|\mathcal{Q}|}$, the positive document set $\mathcal{D}_{+}=\{d_{k}^{+}\}_{k=1}^{|\mathcal{D}_{+}|}$, the position-aware fine-grained qrels $\mathcal{R}_{+}=\{q_j,d_{j}^{+},\mathrm{ref}_{j}, \mathrm{pos}_{j}(\mathrm{start}, \mathrm{end})\}_{j=1}^{|\mathcal{Q}|}$.

\subsection{Quality Control}
In this stage, we design a set of comprehensive quality control strategies to improve the reliability and precision of the generated dataset.
As illustrated in Figure~\ref{fig:generation_pipeline}, the quality control process consists of two complementary components.

\paragraph{Reference Span Verification.}
Since all reference spans in the qrels $\mathcal{R}_{+}$ are generated by LLMs, they may suffer from localization errors or hallucinations.
To identify and remove such cases, we evaluate the necessity of each reference span using relevance contrast.
Specifically, we employ two re-ranking models (e.g., \texttt{bge-reranker-v2-minicpm-layerwise} and \texttt{Qwen3-Reranker-4B}~\cite{zhang2025qwen3embeddingadvancingtext}) to compute two relevance scores for each query $q_j$:
$s^{w/}_{j}$, the relevance score between $q_j$ and the full positive document $d_j^{+}$, and
$s^{w/o}_{j}$, the relevance score between $q_j$ and a modified document obtained by removing the reference span $\mathrm{ref}_j$ from $d_j^{+}$.
We retain only those queries for which $s^{w/}_{j} > s^{w/o}_{j}$ under both re-ranking models, indicating that the reference span is indispensable for establishing query-document relevance.
For the queries that pass this initial filter, we perform a second verification step using an LLM-based relevance assessor (see Appendix~\ref{appendix_sec:prompt_relevance}).
Following the same contrastive setup, the LLM produces two 5-level relevance labels, $s'^{w/}_{j}$ and $s'^{w/o}_{j}$, ranging from 0 (completely irrelevant) to 4 (perfectly relevant).
We retain a query only if $s'^{w/}_{j} > s'^{w/o}_{j}$ and $s'^{w/}_{j} \geq 3$, corresponding to a \emph{highly relevant} judgment.
We acknowledge that removing the reference span may disrupt local coherence, potentially causing relevance scores ($s^{w/o}$) to decrease even when the span is not truly indispensable. 
To mitigate this, the LLM-based verification assesses relevance at a semantic level, providing a complementary check that is less susceptible to surface-level fluency disruption.
Through this dual verification process, we ensure that each retained query-document pair is relevant specifically due to a well-defined reference span at a particular position---a property essential for accurately studying position bias.

\paragraph{Removing False Negatives.} False negatives refer to potential relevant documents  in the corpus $\mathcal{D}$ that are overlooked in the current qrels $\mathcal{R}_{+}$. Given a query $q_j$, we design a two-step pipeline to remove false negatives, thereby reducing noise in retrieval evaluation.
1) \textit{Recall with embedding model}. Use the embedding model \texttt{bge-m3}~\cite{chen-etal-2024-m3} to search top-1000 relevant documents $\mathcal{L}^{recall}_{j} = \left[d_1, \cdots, d_{1000} \right]$ from the corpus $\mathcal{D}$ for $q_j$. 
2) \textit{Score with re-ranking models}. Use two relatively small re-ranking models (\texttt{bge-reranker-v2-m3} and \texttt{Qwen3-Reranker-0.6B}) to score $\mathcal{L}^{recall}_{j}$ for better efficiency.
Each document $d_k \in \mathcal{L}^{recall}_{j}$ is then assigned a relevance score $s_{d_k}(\mathcal{M})$ by the re-ranking model $\mathcal{M}$.
Specifically, if any $s_{d_k}(\mathcal{M})$ is higher than the baseline score $s_{d^{+}_j}(\mathcal{M})$, indicating a potential false negative, we execute a strict filtering protocol: permanently remove $d_k$ from the corpus $\mathcal{D}$.
Consequently, to ensure dataset integrity, we also discard any query $q_{*}$ (along with its associated entries in $\mathcal{R}_{+}$) for which $d_k$ served as the ground-truth positive document.
Rather than relabeling such documents as additional positives, we adopt this conservative strategy to avoid introducing ambiguous relevance signals and to preserve a single-positive-document setting for position-aware evaluation.

After applying the above quality control process to each query, we obtain the refined query set $\mathcal{Q}'$, corpus $\mathcal{D}'$, and qrels $\mathcal{R}_+'$, which together constitute the final bilingual dataset for each domain.
Overall, this intentionally conservative quality control process prioritizes annotation precision over coverage, which is essential for isolating and analyzing fine-grained positional effects.
To further validate data quality, we reviewed 500 generated English query–document–reference span triples (details in Appendix~\ref{appendix_sec:hm_review}), confirming that our automated pipeline introduces minimal risk of accumulated hallucinations or semantic drift.

\subsection{Multilingual Translation}

After completing the earlier stages, we obtain 62 high-quality English and Chinese datasets covering 31 domains.
To extend linguistic coverage, we employ \texttt{Qwen3-30B-A3B-Instruct-2507}~\cite{yang2025qwen3technicalreport} to translate the English datasets across 31 domains into 8 additional languages (French, Spanish, Russian, etc.), yielding 248 datasets.
Specifically, for each domain, we translate the English datasets $\mathcal{Q}_{en}'$ and $\mathcal{D}_{en}'$ into the target language datasets $\mathcal{Q}_{*}'$ and $\mathcal{D}_{*}'$, while the qrels $\mathcal{R}_+'$ are shared across source and translated languages to ensure consistent position-aware relevance labels.
While character-level span indices inevitably shift across languages due to differences in morphology and syntax, our position bias analysis operates on coarse-grained relative positions, making it robust to such minor deviations.
To evaluate translation quality, we employ both LLM-based automatic evaluation~\cite{kocmi-federmann-2023-large} and human evaluation on a sampled subset, both confirming overall high quality.
However, we observe that token lengths grow disproportionately across languages, sometimes causing the translated corpus to exceed the predefined 2048-token limit.
We also encounter issues such as invalid prefixes, refusal to translate, and verbose outputs. 
To address these issues and further improve translation quality, we implement a set of fixing and filtering strategies for translated languages.
For more details on the multilingual machine translation stage, please refer to Appendix~\ref{appendix_sec:mmt}.

\section{The \posir Benchmark}
\begin{figure}[t] 
    \centering
    
    \begin{minipage}[t]{0.48\textwidth}
        \vspace{0pt} 
        \centering
        \captionof{table}{Summary of \posir.}
        \vspace{2mm}
        \label{tab:dataset_overview}
        \resizebox{\textwidth}{!}{%
        \begin{tabular}{lr}
        \toprule
        \textbf{Metric} & \textbf{Value} \\
        \midrule
        Languages & 10 \\
        Domains & 31 \\
        Language-Domain Pairs & 310 \\
        -\,\,\; Avg \#Queries & 1,360 \\
        -\,\,\; Avg \#Documents & 55,902 \\
        Total Queries Tokens & 9,238,156 \\
        -\,\,\; Avg Tokens per Query & 21.91 \\
        Total Corpus Tokens & 22,784,263,943 \\
        -\,\,\; Avg Tokens per Document & 1,314.75 \\
        \bottomrule
        \end{tabular}}
    \end{minipage}\hfill
    \begin{minipage}[t]{0.49\textwidth}
        \vspace{0pt} 
        \centering
        \includegraphics[width=\linewidth]{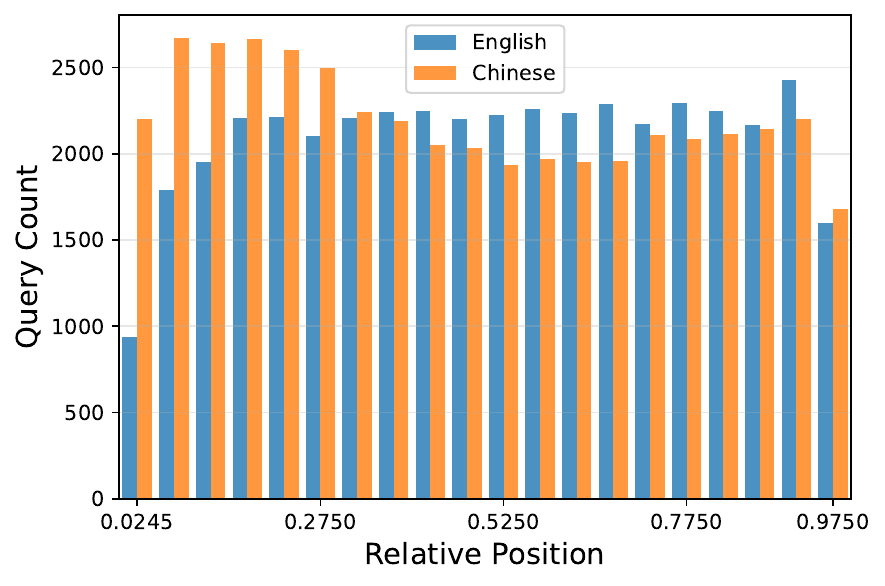}
        \caption{Distribution of normalized reference span positions in the English and Chinese datasets.}
        \label{fig:relative_position_eng_zh}
    \end{minipage}
    
\end{figure}

Table~\ref{tab:dataset_overview} provides a statistical overview of \posir, which comprises 310 retrieval datasets spanning 10 languages\footnote{Arabic, Chinese, German, English, French, Italian, Korean, Portuguese, Russian, Spanish.} and 31 domains\footnote{The 31 domains are grouped into the following high-level categories: Technology and Science, Industry and Economics, Health and Medicine, Society and Law, Entertainment and Culture, Finance, and Miscellaneous.}.
More details on \posir are available in Appendix~\ref{appendix_sec:posir_datasets}.

\subsection{Positional Distribution Analysis}
\label{sec:positional_analysis}
In the position-aware fine-grained qrels
$\mathcal{R}_{+}'=\{q_j, d_{j}^{+}, \mathrm{ref}_{j}, \mathrm{pos}_{j}(\mathrm{start}, \mathrm{end})\}_{j=1}^{|\mathcal{Q}'|}$,
we explicitly incorporate positional information through $\mathrm{pos}_j$.
To analyze the distributional characteristics of this positional information, we aggregate the qrels across all domains separately for the English and Chinese datasets.
For each query-document pair, we compute the mean position of $\mathrm{pos}_{j}(\mathrm{start}, \mathrm{end})$ and normalize it by the character length of the corresponding positive document $d_{j}^{+}$, yielding a relative position in the range $[0,1]$.
We then discretize the relative positions into 20 equal-width bins and count the number of queries falling into each bin.
The resulting positional distributions are illustrated in Figure~\ref{fig:relative_position_eng_zh}, confirming that reference spans are broadly distributed across document positions, without a pronounced bias toward early segments as observed in {MS-MARCO}~\cite{Mitigating10.1007/978-3-030-72113-8_16,coelho-etal-2024-dwell}.
Such a broad distribution is crucial for evaluating retrieval models’ sensitivity to position bias, as it ensures that relevance signals occur throughout the document rather than being dominated by leading content.
This also serves as evidence that our automated generation pipeline does not introduce systematic positional artifacts: if the pipeline itself were biased toward particular positions~\cite{liu-etal-2024-lost}, the distribution in Figure~\ref{fig:relative_position_eng_zh} would exhibit a pronounced skew, which is not observed.

\subsection{Positioning of \posir}
Existing retrieval benchmarks serve complementary purposes: {MMTEB}~\cite{muennighoff-etal-2023-mteb,enevoldsen2025mmteb} and {BEIR}~\cite{DBLP:journals/corr/abs-2104-08663} evaluate general-purpose embedding quality, {MIRACL}~\cite{zhang-etal-2023-miracl} targets multilingual retrieval, and {AIR-Bench}~\cite{chen-etal-2025-air} meets the demand for efficient evaluation in emerging domains. However, all of them assign position-agnostic relevance labels, leaving position bias undiagnosed. \posir fills this gap as a multilingual, position-aware retrieval benchmark. By associating each query with a localized reference span within length-diverse documents, \posir enables controlled analysis of how retrieval performance varies with evidence position---a dimension largely unexplored in existing benchmarks.

\begin{table}[!ht]
\centering
\caption{
nDCG@10~$\uparrow$ performance of 10 multilingual retrieval models on MMTEB (Retrieval Task) and \posir.
MMTEB scores are sourced from the official leaderboard.
For \posir, the results are first weighted-averaged across 31 domains and then macro-averaged across 10 languages.
Q1--Q4 represent query buckets partitioned by the token length of positive documents (512-token intervals). 
Spearman rank correlation coefficients between MMTEB and \posir (overall and per length bucket) are reported at the bottom. The detailed results for each language in \posir are available in Table~\ref{tab:detailed_experiment_results}.
}
\label{tab:comparison_results}
\begin{adjustbox}{max width=\textwidth}
\begin{tabular}{lcccccc}
\toprule
\textbf{Model} & \textbf{MMTEB} & \textbf{\posir} & \textbf{Q1(512)} & \textbf{Q2(1024)} & \textbf{Q3(1536)} & \textbf{Q4(2048)} \\
\midrule
gte-multilingual-base~\cite{zhang-etal-2024-mgte}  & 57.16 & 47.37& 61.28& 48.79 & 39.21 & 32.01 \\
bge-m3~\cite{chen-etal-2024-m3} & 54.60 & 43.22 & 57.16 & 43.48 & 35.13 & 29.72 \\
Qwen3-Embedding-0.6B~\cite{zhang2025qwen3embeddingadvancingtext} & 64.65 & 53.63 & 62.93 & 54.41 & 48.11 & 43.32 \\
inf-retriever-v1-1.5b & 62.96 & 58.81 & 67.82 & 58.40 & 53.90 & 51.20 \\
Qwen3-Embedding-4B~\cite{zhang2025qwen3embeddingadvancingtext} & 69.60 & 62.26 & 71.63 & 62.91 & 56.74 & 50.96 \\
inf-retriever-v1 & 66.48 & 65.01 & 74.71 & 65.03 & 59.82 & 54.68 \\
NV-Embed-v2~\cite{lee2025nvembed}  & 56.72 & 45.02 & 70.48 & 47.12 & 26.33 & 16.27 \\
llama-embed-nemotron-8b~\cite{babakhin2025llamaembednemotron8buniversaltextembedding}  & 68.69 & 64.09 & 75.76 & 64.16 & 57.47 & 52.28 \\
Qwen3-Embedding-8B~\cite{zhang2025qwen3embeddingadvancingtext}  & 70.88 & 64.08 & 72.68 & 64.33 & 59.00 & 53.87 \\
KaLM-Embedding-12B~\cite{zhao2025kalmembeddingv2superiortrainingtechniques}  & 75.66 & 51.87 & 74.01 & 54.64 & 35.11 & 27.65 \\
\midrule
\textbf{Spearman Corr.} & -- & 0.62 & 0.73 & 0.71 & 0.44 & 0.39 \\
(\textit{p}-value)      & -- & (0.05) & (0.01) & (0.02) & (0.20) & (0.20) \\
\bottomrule
\end{tabular}
\end{adjustbox}
\end{table}

\section{Experiments}

In this section, we aim to address the following research questions:

\noindent \textbf{RQ1:} To what extent do model rankings on \posir align with or diverge from those on established retrieval benchmarks, and in which scenarios does \posir reveal discrepancies that are invisible under standard benchmarks?

\noindent \textbf{RQ2:} What systematic patterns of position bias (e.g., early vs. late relevance, document length effects, and cross-lingual variation) are exposed by the \posir benchmark?

\noindent \textbf{RQ3:} What internal model behaviors and representational dynamics contribute to the emergence of position bias in dense retrieval models?


\subsection{Comparison with MMTEB (RQ1)}
\label{sec:rq1}

\paragraph{Experimental Setup.} We adopt MMTEB (Retrieval Task) as a representative multilingual retrieval benchmark. We evaluate 10 popular embedding models on both MMTEB and \posir using nDCG@10, and compute the Spearman rank correlation coefficient between model rankings on the two benchmarks as a measure of consistency. To further analyze the effect of document length, we partition queries in \posir into four length buckets (Q1--Q4) with 512-token intervals, based on the token length of their positive documents $d_j^{+}$ in the qrels  $\mathcal{R}_{+}'$. For translated datasets, token lengths are measured based on their original English counterparts to ensure consistent bucket assignment across languages. We then compute Spearman correlations between MMTEB and the average performance of models within each length bucket.

\paragraph{Main Results.}
As shown in Table~\ref{tab:comparison_results}, the Spearman rank correlation between MMTEB and \posir is 0.62 ($p=0.05$), indicating a moderate correlation.
This suggests that while \posir partially aligns with existing retrieval evaluations, it captures distinct performance characteristics not fully reflected by MMTEB.
When controlling for document length, a clear trend emerges: the correlation is strongest for Q1 (documents up to 512 tokens), reaching 0.73 ($p=0.01$), and progressively decreases as document length increases.
For Q4 (documents up to 2048 tokens), the correlation drops to 0.39 ($p=0.2$), indicating no statistically significant correlation.
This length-dependent divergence implies MMTEB may primarily reflect model performance on short-document retrieval. In contrast, \posir, through its length-aware sampling, exposes substantial discrepancies in how models handle long-context retrieval.
Notably, most models exhibit pronounced performance degradation in Q4, despite nominally supporting long input contexts.

A pronounced pattern is observed for \texttt{NV-Embed-v2} and \texttt{KaLM-Embedding-12B}, which achieve competitive performance on MMTEB and short-document queries (Q1), but suffer substantial degradation in Q3 and Q4.
This behavior is consistent with their reported training configurations~\cite{lee2025nvembed,zhao2025kalmembeddingv2superiortrainingtechniques}, as both models are primarily trained with short-context inputs (e.g., up to 512 tokens), limiting their ability to effectively encode long documents. Furthermore, \texttt{NV-Embed-v2}, while strong in Q1, degrades severely in longer-document buckets in multilingual settings. 
This aligns with the fact that it is trained predominantly on English data, suggesting that long-context cross-lingual generalization poses a significant challenge when both document length and language shift are combined.
\begin{table}[!t]
\centering
\caption{
Position Sensitivity Index (PSI)~$\downarrow$ of 10 multilingual retrieval models on \posir.
In both multilingual retrieval and cross-lingual retrieval (translated queries retrieving English documents) settings, the results are first weighted-averaged across 31 domains and then macro-averaged across 10 languages.
NV-Embed-v2 improves the mean* pooling method with a latent attention layer.
The detailed results for each language in \posir are available in Table~\ref{tab:detailed_experiment_results}.
}
\label{tab:psi_res}
\begin{adjustbox}{max width=\textwidth}
\begin{tabular}{lcccccccc}
\toprule
\multirow{2}{*}{\textbf{Model}} & \multirow{2}{*}{\textbf{Dimension}} & \multirow{2}{*}{\textbf{Attention}} & \multirow{2}{*}{\textbf{Pooling}} & \textbf{\posir} & \textbf{Q1(512)} & \textbf{Q2(1024)} & \textbf{Q3(1536)} & \textbf{Q4(2048)} \\
&  &  &  & \textbf{nDCG@10}~$\uparrow$ & \textbf{PSI}~$\downarrow$& \textbf{PSI}~$\downarrow$& \textbf{PSI}~$\downarrow$& \textbf{PSI}~$\downarrow$ \\
\midrule
\multicolumn{9}{l}{\textit{Multilingual Retrieval}} \\
\midrule
gte-multilingual-base & 768 & bidirectional & CLS & 47.37 & 0.21 & 0.44 & 0.56 & 0.62 \\
bge-m3 & 1024 & bidirectional & CLS & 43.22  & 0.30 & 0.43 & 0.49 & 0.44 \\
Qwen3-Embedding-0.6B & 1024 & causal & last & 53.63 & 0.21 & 0.38 & 0.47& 0.49 \\
inf-retriever-v1-1.5b & 1536 & bidirectional & last & 58.81 & 0.20 & 0.28 & 0.33 & 0.27 \\
Qwen3-Embedding-4B & 2560 & causal & last & 62.26  &  0.13 & 0.30 & 0.39 & 0.44 \\
inf-retriever-v1 & 3584 & bidirectional & last & 65.01 & 0.14 & 0.21 & 0.17 & 0.18 \\
NV-Embed-v2 & 4096 & bidirectional & mean* & 45.02 & 0.19 & 0.53 & 0.73 & 0.81 \\
llama-embed-nemotron-8b & 4096 & bidirectional & mean & 64.09 & 0.14 & 0.18 & 0.15 & 0.22 \\
Qwen3-Embedding-8B & 4096 & causal & last & 64.08  & 0.12 & 0.23 & 0.31 & 0.34 \\
KaLM-Embedding-12B & 3840 & bidirectional & last & 51.87 & 0.11 & 0.26 & 0.28 & 0.32 \\
\midrule
\multicolumn{4}{r}{\textbf{Average}} & 55.54 & 0.18 & 0.32 & 0.39 & 0.41 \\
\midrule
\multicolumn{9}{l}{\textit{Cross-lingual Retrieval}} \\
\midrule
gte-multilingual-base & 768 & bidirectional & CLS & 46.49 & 0.15 & 0.32 & 0.36 & 0.43 \\
bge-m3 & 1024 & bidirectional & CLS & 36.30 & 0.34 & 0.47 & 0.52 & 0.47 \\
Qwen3-Embedding-0.6B & 1024 & causal & last & 49.66 & 0.14 & 0.37 & 0.48 & 0.51 \\
inf-retriever-v1-1.5b & 1536 & bidirectional & last & 53.11 & 0.18 & 0.27 & 0.24 & 0.19 \\
Qwen3-Embedding-4B & 2560 & causal & last & 60.76 &  0.13 & 0.30 & 0.43 & 0.47 \\
inf-retriever-v1& 3584 & bidirectional & last & 62.87  & 0.13 & 0.22 & 0.19 & 0.16 \\
NV-Embed-v2 & 4096 & bidirectional & mean* & 45.97  & 0.16 & 0.14 & 0.52 & 0.62 \\
llama-embed-nemotron-8b & 4096 & bidirectional & mean & 62.68  & 0.10 & 0.20 & 0.16 & 0.18 \\
Qwen3-Embedding-8B & 4096 & causal & last & 61.93 & 0.12 & 0.26 & 0.41 & 0.45 \\
KaLM-Embedding-12B  & 3840 & bidirectional & last  & 55.48 & 0.10 & 0.22 & 0.29 & 0.24 \\
\midrule
\multicolumn{4}{r}{\textbf{Average}} & 53.53 & 0.16 & 0.27 & 0.36 & 0.37 \\
\bottomrule
\end{tabular}
\end{adjustbox}
\end{table}

\begin{figure*}[t] 
    \centering
    \includegraphics[width=1\textwidth]{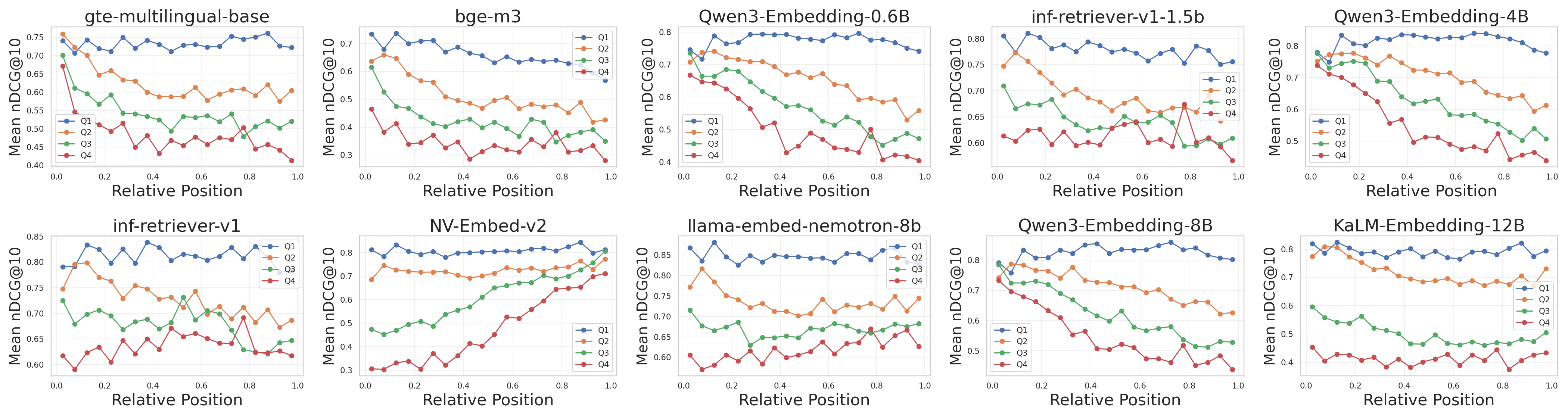}
    \caption{Mean nDCG@10 scores of 10 IR models across 20 relative position bins on the English subset of \posir. 
    }
    \label{fig:multilingual-eng}
\end{figure*}

\begin{figure*}[t] 
    \centering
    \includegraphics[width=1\textwidth]{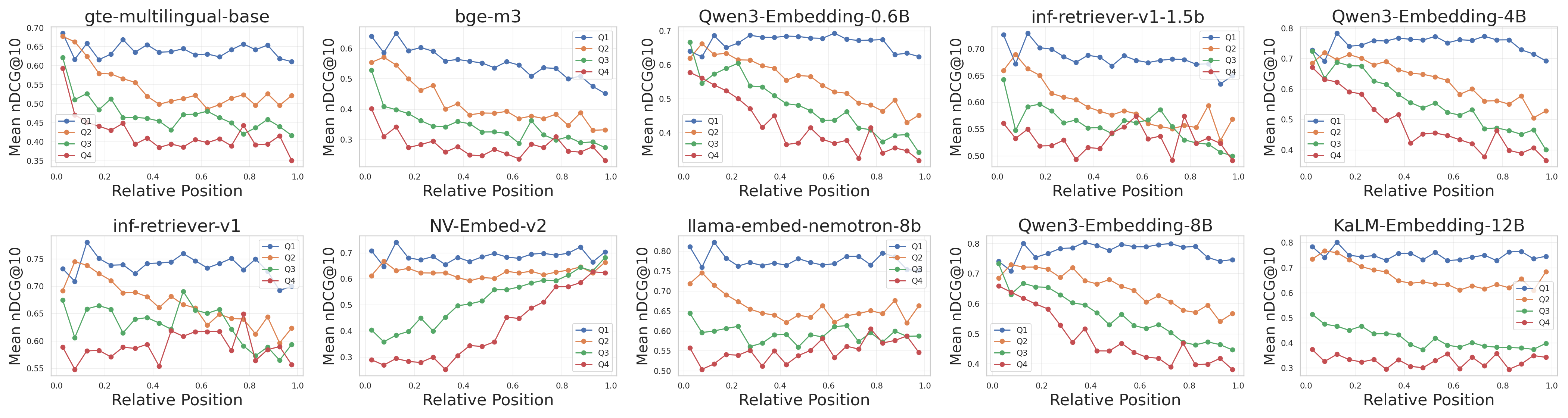}
    \caption{Mean nDCG@10 scores of 10 IR models across 20 relative position bins in the French-to-English cross-lingual setting of \posir. 
    }
    \label{fig:crosslingual-fra}
\end{figure*}

\subsection{Patterns of Position Bias (RQ2)}
\paragraph{Experimental Setup.}
We adopt the Position Sensitivity Index (PSI) introduced in~\cite{zeng-etal-2025-empirical} as an intuitive diagnostic metric for quantifying position bias from a worst-case perspective. We divide the queries into four length buckets (Q1--Q4) according to the strategy in Section~\ref{sec:rq1}.
Within each bucket, we calculate the average nDCG@10 scores across the 20 equal-width relative position bins (denoted as $\mathbf{s} = \{s_1,\ldots,s_{20}\}$), following the discretization described in Section~\ref{sec:positional_analysis}. 
Formally, PSI is defined as: $\mathrm{PSI}=1 - \frac{\min(\mathbf{s})}{\max(\mathbf{s})}$.
A lower PSI indicates that the model's performance is more consistent across document positions, reflecting reduced sensitivity to the location of relevant content.
For a more detailed discussion of the PSI metric, please refer to the original work.

\paragraph{Main Results.}
As shown in Table~\ref{tab:psi_res}, we observe a clear trend: position bias tends to increase with document length across both multilingual and cross-lingual retrieval tasks.\footnote{We consider a cross-lingual retrieval setting where non-English queries are used to retrieve an English corpus, which is a common scenario.}
For short documents (Q1), most models exhibit relatively low PSI values, suggesting that position bias is minimal when processing short inputs. 
However, as document length increases (Q3 and Q4), PSI rises markedly for several models, indicating heightened sensitivity to position in longer-context retrieval scenarios.
Moreover, cross-lingual retrieval exhibits a lower overall PSI compared to the monolingual setting. 
We hypothesize this is partly because the generally lower performance in cross-lingual tasks compresses the score range, thereby reducing the discriminative power of the PSI metric across position bins. 
We also observe that for certain models, the PSI in Q3 or Q4 is counter-intuitively lower than in Q2. 
We attribute this to the models' failure to effectively encode longer document representations, leading to uniformly degraded performance where position bias is no longer the dominant factor in score fluctuations.
Moreover, we observe no clear correlation between positional sensitivity and architectural factors such as model size, vector dimension, attention mechanism (bidirectional / causal), or pooling strategy (CLS / last / mean).\footnote{However, the latent attention layer preceding the mean pooling strategy may make \texttt{NV-Embed-v2} more unique.}
This suggests that these architectural choices alone may not be the primary determinants of position bias, or their effects are overshadowed by training data distributions. 

We visualize the fine-grained model performance across 20 relative position bins in Figure~\ref{fig:multilingual-eng} and~\ref{fig:crosslingual-fra}.
Additional figures are available in the official GitHub repository.\footnote{\url{https://github.com/Ziyang1060/PosIR/tree/main/figs}}
As shown in these figures, most models exhibit a common pattern of primacy bias, where retrieval performance degrades as the relevant information moves towards the end of the document.
An unexpected case is \texttt{NV-Embed-v2}, which exhibits a recency bias skewed towards the end of the document, diverging from the typical patterns observed in other models.

\subsection{Exploring Mechanisms of Position Bias (RQ3)}

We present a preliminary mechanistic analysis to generate hypotheses about the internal behaviors underlying the observed retrieval biases, rather than to provide definitive causal explanations.

\paragraph{Experimental Setup.}
We conduct exploratory analyses on the internal behaviors of two representative models: \texttt{Qwen3-Embedding-8B} (exhibiting primacy bias) and \texttt{NV-Embed-v2} (exhibiting recency bias).
To quantitatively assess the contribution of tokens at different positions, we employ gradient-based saliency analysis~\cite{DBLP:journals/corr/SimonyanVZ13}. 
Given a query $q$ and a relevant long document $d^+$ (with length $\ge$ 1024 tokens), where the reference span is located near the middle of the document (i.e., relative position within $[0.4, 0.6]$), we compute the gradient of the cosine similarity score $s(q, d^+)$ with respect to the input document embeddings.
The magnitude of the gradient serves as a proxy for token importance, indicating how strongly each token influences the final document representation used for relevance matching.
Specifically, for each document token $w_i$ at absolute position $i$, we compute the L2 norm of the gradient vector $||\nabla_{w_i} s(q, d^+)||_2$, which measures the sensitivity of the relevance score to that token.\footnote{Special tokens at the sequence boundaries are excluded.}
To facilitate comparison across documents of varying lengths, we apply max normalization over all document tokens and rescale token positions to the range $[0,1]$ using linear interpolation, discretized into 100 relative position bins.
We aggregate the saliency over 3,131 English query-document pairs that satisfy the length and position constraints to ensure statistical robustness.

\begin{figure}[t] 
    \centering
    \includegraphics[width=0.6\linewidth]{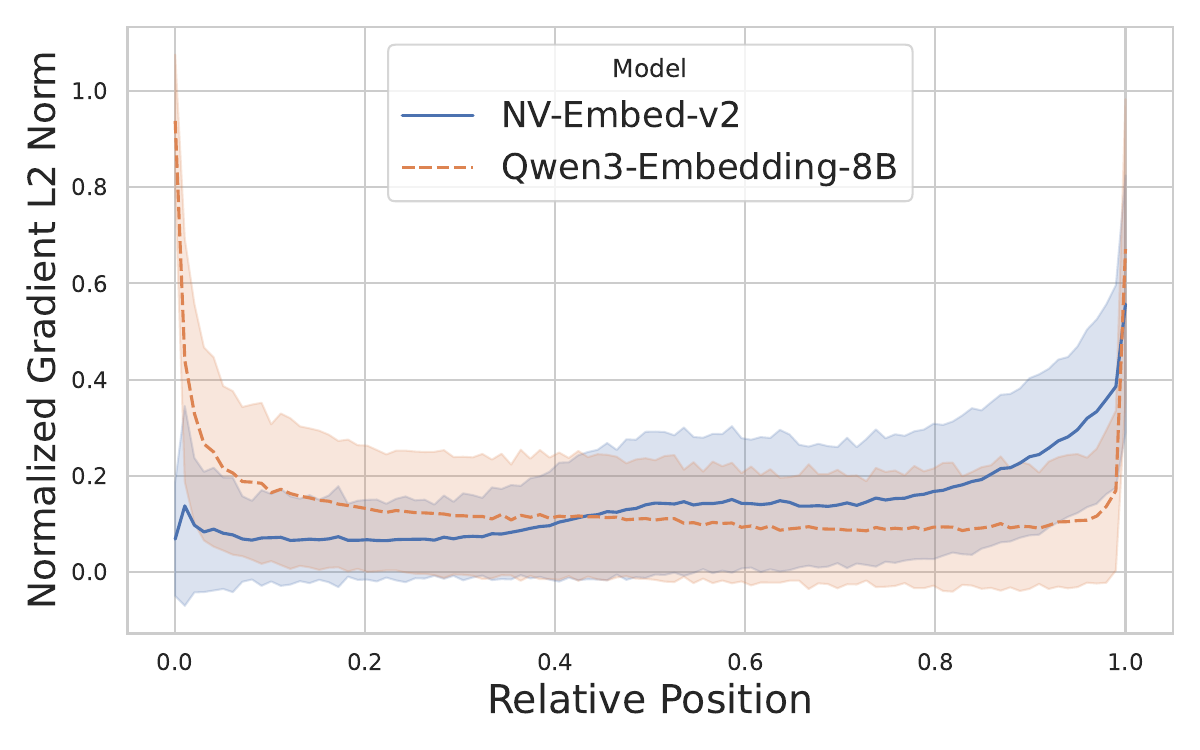}
    \caption{Gradient-based saliency maps comparing the token-level sensitivity of Qwen3-Embedding-8B and NV-Embed-v2. The x-axis represents the normalized relative position within a document, and the y-axis shows the normalized L2 norm of the gradients. Shaded regions indicate one standard deviation.}
    \label{fig:saliency_comparison}
\end{figure}

\paragraph{Main Results.}
We visualize the gradient-based saliency maps in Figure~\ref{fig:saliency_comparison}. The results reveal starkly contrasting internal behaviors between the two models. The saliency curve for \texttt{Qwen3-Embedding-8B} (Orange) is characterized by an extreme peak at the very beginning. The subsequent sharp decay suggests the model may struggle to effectively propagate gradient signal from later sections, offering a candidate mechanistic explanation for its tendency to under-weight mid-document information. We further conjecture that the spike at the final position arises as a structural artifact of last-token pooling, though causal verification remains to be established.
In stark contrast, \texttt{NV-Embed-v2} (Blue) exhibits suppressed sensitivity to the beginning of the document, with gradients remaining relatively low for the first 20\% of tokens. Instead, it demonstrates a continuous rising trend starting from the mid-point (0.5) and peaking at the end. This J-shaped profile is consistent with the hypothesis that \texttt{NV-Embed-v2}'s encoding mechanism progressively prioritizes recent information, potentially overwriting or diluting early context. We conjecture this pattern underlies its atypical recency bias and its failure to retrieve information located at the start of long documents. 
We emphasize that the above analyses are correlational. Intervention experiments such as token ablation and positional perturbation would be needed to establish causality, which we leave to future work.

\section{Conclusion}
In this paper, we introduce \posir, the first standardized benchmark for diagnosing position bias in diverse retrieval scenarios, covering 10 languages and 31 domains.
Our experiments reveal prevalent primacy and unexpected recency biases across both multilingual and cross-lingual settings, while also exposing significant discrepancies with existing short-text benchmarks. 
These findings contribute new empirical evidence on position bias in dense retrieval, paving the way for more position-robust retrieval systems.


\section*{Limitations}
Despite our efforts, this study has several limitations: 
1) Although we employ LLMs to extract reference spans and implement a strict contrastive verification pipeline, the localization of relevant information is inherently approximate. LLM-extracted spans may not perfectly capture the minimal necessary evidence, and our contrastive filtering validates span necessity only indirectly---by confirming that removing the span degrades relevance scores. However, span removal may also disrupt the document's overall semantic coherence, potentially conflating span necessity with document fluency degradation. As a result, a marginal risk remains that certain spans are imprecise or that alternative relevant spans are overlooked.
2) While LLM-based translation significantly extends the linguistic diversity of \posir, it is intrinsically difficult to completely eliminate subtle semantic drift or translation artifacts across all datasets. Moreover, positional annotations derived from the source language may shift to varying degrees due to differences in linguistic structure (e.g., word order, morphological expansion), potentially introducing noise into position bias evaluation for translated languages. Although our sampling-based automatic and human evaluations indicate high fidelity, translation quality in certain low-resource languages might still be influenced by the underlying model bias.
3) Our empirical analysis focuses on embedding-based dense retrieval models. Although prior research~\cite{zeng-etal-2025-empirical} indicates that cross-encoders may exhibit superior position robustness, a systematic large-scale investigation of alternative architectures, including generative retrieval, remains for future work. Nevertheless, \posir provides the necessary model-agnostic framework to facilitate such benchmarking.



\bibliographystyle{splncs04}
\bibliography{references}


\newpage
\appendix
\appendix 
\section*{Appendix for \posir}

For detailed information of the models appearing in this paper, please refer to Table~\ref{tab:model_information}.

\section{\posir Datasets}
\label{appendix_sec:posir_datasets}
\subsection{Overview}
\posir comprises 310 datasets (17,329,673 documents and 421,708 queries) spanning 10 languages and 31 domains, forming a large-scale multilingual benchmark for information retrieval.
For each dataset, we use the same format as BEIR, i.e. corpus, queries and qrels, which will be all available in the Hugging Face Hub.
For each dataset, we use the same format as BEIR, i.e. corpus, queries and qrels, which are all available in the Hugging Face Hub.\footnote{\url{https://huggingface.co/datasets/infgrad/PosIR-Benchmark-v1}}
Table~\ref{tab:corpus_statistics} summarizes language-level statistics, showing well-balanced corpus sizes across languages, with each language contributing approximately 1.7M--1.9M documents. 
Token length statistics reveal substantial cross-lingual variation that reflects differences in writing systems and morphological complexity. Logographic languages exhibit more compact representations (e.g., Chinese and English), whereas morphologically rich alphabetic languages require longer sequences (e.g., Arabic and Russian). 
Table~\ref{tab:corpus_statistics_by_domain} reports domain-level aggregates and highlights pronounced variation in domain scale and document characteristics. Domain sizes range from several hundred thousand documents to over three quarters of a million, while token length distributions reflect domain-specific content properties. Technical domains such as Information Security tend to contain longer documents, whereas more formal or symbolic domains, such as Mathematics and Statistics, exhibit comparatively shorter texts. This diversity enables evaluation of IR systems under heterogeneous conditions.
For comprehensive coverage, Tables~\ref{tab:dataset_stats_accommodationcateringhotel}--\ref{tab:dataset_stats_waterresourcesocean} provide detailed statistics for all 31 domains, presented individually for each domain. 
These tables report language-specific document counts, query counts, average token lengths, and token length ratios relative to English. Such fine-grained reporting supports reproducibility and facilitates the analysis of domain--language interaction effects in multilingual IR.

\begin{table}[!ht]
\centering
\caption{Summary of corpus statistics aggregated by language.}
\label{tab:corpus_statistics}
\resizebox{\textwidth}{!}{%
\begin{tabular}{l|rrr|rrrrrr}
\toprule
 & \multicolumn{3}{c|}{\textbf{Document Count}} & \multicolumn{6}{c}{\textbf{Token Length}} \\
\textbf{Language} & \textbf{Corpus} & \textbf{Queries} & \textbf{Ratio} & \textbf{Min} & \textbf{Q1} & \textbf{Median} & \textbf{Q3} & \textbf{Max} & \textbf{Mean} \\
\midrule
Arabic & 1,712,847 & 41,894 & 9.89\% & 5 & 717 & 1401 & 2250 & 4719 & 1509.5 \\
Chinese & 1,870,228 & 43,931 & 10.78\% & 13 & 473 & 869 & 1410 & 2050 & 950.8 \\
German & 1,718,556 & 41,983 & 9.92\% & 3 & 655 & 1278 & 2067 & 4296 & 1387.3 \\
English & 1,719,701 & 42,015 & 9.92\% & 14 & 416 & 842 & 1392 & 2051 & 922.4 \\
French & 1,718,832 & 42,001 & 9.92\% & 2 & 645 & 1269 & 2054 & 4352 & 1375.1 \\
Italian & 1,718,831 & 42,005 & 9.92\% & 3 & 660 & 1295 & 2090 & 4298 & 1401.3 \\
Korean & 1,715,299 & 41,913 & 9.90\% & 5 & 704 & 1403 & 2258 & 4685 & 1509.5 \\
Portuguese & 1,718,848 & 42,002 & 9.92\% & 3 & 607 & 1195 & 1941 & 4020 & 1298.1 \\
Russian & 1,718,056 & 41,969 & 9.91\% & 2 & 721 & 1424 & 2300 & 4828 & 1539.2 \\
Spanish & 1,718,475 & 41,995 & 9.92\% & 4 & 604 & 1187 & 1925 & 3956 & 1287.9 \\
\midrule
Total/Avg & 17,329,673 & 421,708 & 100.00\% & 5 & 620 & 1216 & 1968 & 3925 & 1318.1 \\
\bottomrule
\end{tabular}%
}

\end{table}

\begin{table}[!ht]
\centering
\caption{Summary of corpus statistics aggregated by domain.}
\label{tab:corpus_statistics_by_domain}
\resizebox{\textwidth}{!}{%
\begin{tabular}{l|rrr|rrrrrr}
\toprule
 & \multicolumn{3}{c|}{\textbf{Document Count}} & \multicolumn{6}{c}{\textbf{Token Length}} \\
\textbf{Domain} & \textbf{Corpus} & \textbf{Queries} & \textbf{Ratio} & \textbf{Min} & \textbf{Q1} & \textbf{Median} & \textbf{Q3} & \textbf{Max} & \textbf{Mean} \\
\midrule
Accommodation Catering Hotel & 439,993 & 14,612 & 2.56\% & 10 & 607 & 1173 & 1886 & 4432 & 1294.6 \\
Aerospace & 581,933 & 13,546 & 3.36\% & 9 & 610 & 1207 & 1955 & 4585 & 1326.1 \\
Agriculture Forestry Animal Husbandry Fishery & 581,329 & 16,175 & 3.37\% & 9 & 589 & 1231 & 2013 & 4719 & 1346.3 \\
Artificial Intelligence Machine Learning & 551,318 & 11,154 & 3.17\% & 5 & 538 & 1113 & 1884 & 4540 & 1276.7 \\
Automobile & 565,317 & 13,047 & 3.26\% & 5 & 584 & 1120 & 1870 & 4346 & 1271.3 \\
Biomedicine & 616,910 & 13,948 & 3.55\% & 7 & 580 & 1195 & 1972 & 4709 & 1327.3 \\
Computer Communication & 620,254 & 14,144 & 3.57\% & 6 & 607 & 1172 & 1908 & 4544 & 1300.5 \\
Computer Programming Code & 519,815 & 6,522 & 2.97\% & 3 & 701 & 1201 & 1853 & 4127 & 1307.0 \\
Current Affairs Government Administration & 606,998 & 12,832 & 3.49\% & 2 & 619 & 1225 & 1985 & 4785 & 1341.8 \\
Electric Power Energy & 599,389 & 13,119 & 3.45\% & 6 & 613 & 1220 & 2001 & 4676 & 1352.1 \\
Film Entertainment & 596,575 & 16,911 & 3.46\% & 10 & 592 & 1195 & 1906 & 4605 & 1295.0 \\
Finance Economics & 607,364 & 12,948 & 3.49\% & 5 & 626 & 1227 & 1982 & 4571 & 1343.5 \\
Fineweb & 698,633 & 16,748 & 4.03\% & 5 & 575 & 1175 & 1919 & 4645 & 1301.7 \\
Fire Safety Food Safety & 313,846 & 12,852 & 1.84\% & 13 & 672 & 1197 & 1953 & 4612 & 1361.1 \\
Game & 442,997 & 13,014 & 2.57\% & 6 & 646 & 1234 & 1970 & 4634 & 1339.5 \\
Law Judiciary & 603,544 & 12,947 & 3.47\% & 10 & 626 & 1221 & 1971 & 4676 & 1338.7 \\
Literature Emotion & 598,538 & 15,717 & 3.46\% & 3 & 613 & 1181 & 1915 & 4653 & 1299.6 \\
Mathematics Statistics & 618,864 & 10,324 & 3.54\% & 2 & 588 & 1140 & 1827 & 4180 & 1244.3 \\
Medicine Health Psychology Traditional Chinese Medicine & 633,903 & 13,925 & 3.65\% & 14 & 609 & 1211 & 1987 & 4828 & 1342.9 \\
Mining & 485,528 & 13,345 & 2.81\% & 9 & 510 & 1114 & 1890 & 4771 & 1266.1 \\
News Media & 539,996 & 12,879 & 3.12\% & 5 & 588 & 1230 & 1964 & 4738 & 1326.3 \\
Other Information Services Information Security & 321,187 & 8,590 & 1.86\% & 13 & 669 & 1246 & 2026 & 4402 & 1384.2 \\
Other Manufacturing & 596,244 & 16,025 & 3.45\% & 6 & 594 & 1131 & 1884 & 4444 & 1283.5 \\
Petrochemical & 577,348 & 12,803 & 3.32\% & 6 & 570 & 1180 & 1962 & 4583 & 1316.6 \\
Real Estate Construction & 567,575 & 16,259 & 3.29\% & 13 & 594 & 1211 & 1963 & 4693 & 1328.5 \\
Sports & 566,912 & 13,787 & 3.27\% & 13 & 565 & 1163 & 1892 & 4531 & 1281.9 \\
Subject Education Education & 618,409 & 16,796 & 3.58\% & 9 & 625 & 1221 & 1975 & 4569 & 1339.0 \\
Technology Scientific Research & 614,697 & 14,198 & 3.54\% & 3 & 595 & 1195 & 1960 & 4592 & 1318.4 \\
Tourism Geography & 549,553 & 14,974 & 3.18\% & 5 & 560 & 1176 & 1927 & 4633 & 1295.9 \\
Transportation & 547,312 & 14,129 & 3.16\% & 10 & 573 & 1190 & 1965 & 4640 & 1323.3 \\
Water Resources Ocean & 547,392 & 13,438 & 3.16\% & 7 & 571 & 1194 & 1960 & 4563 & 1325.8 \\
\midrule
Total/Avg & 17,329,673 & 421,708 & 100.00\% & 7 & 600 & 1189 & 1939 & 4581 & 1316.1 \\
\bottomrule
\end{tabular}%
}

\end{table}

\subsection{Diversity Analysis}
\subsubsection{Query Diversity}
To analyze the diversity of query types in \posir, we adopt a keyword-based heuristic that categorizes queries according to their leading interrogative terms (e.g., \textsc{what}), with non-interrogative queries grouped as \textsc{claim}.
Table~\ref{tab:query_type_diversity} summarizes the distribution of query types in the English and Chinese retrieval datasets.\footnote{The \textsc{OTHERS} category mainly includes yes/no questions—typically those starting with auxiliary verbs (e.g., do/does/did, can, will) or be verbs (e.g., is/are)—as well as other infrequent or implicit question forms that do not match predefined interrogative patterns.}
Several observations can be made from the results.
First, \textsc{what} queries constitute the largest proportion of the datasets, followed by \textsc{how} and \textsc{why} queries, indicating that the generated queries predominantly seek factual explanations and procedural information.
Notably, Chinese queries contain a higher proportion of \textsc{why} and \textsc{who}, whereas English queries exhibit a larger share of queries categorized as \textsc{others}, suggesting differences in query formulation across languages.
In addition, a small fraction of queries are categorized as \textsc{claim}, representing declarative statements that express information needs without explicit interrogative forms.

\begin{figure*}[!ht]
\centering
\includegraphics[width=1.0\textwidth]{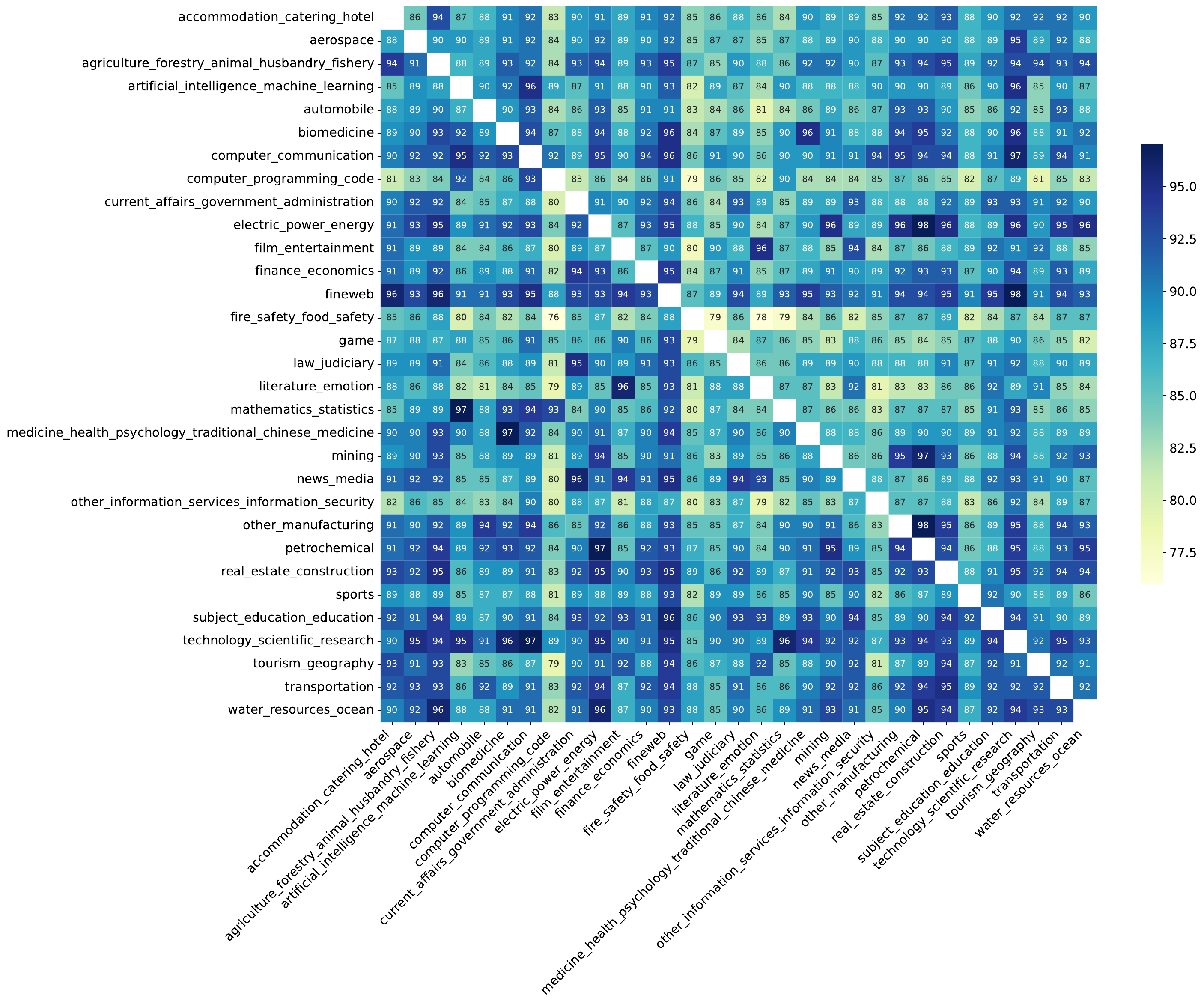}
\caption{Inter-domain similarity heatmap.
The lower triangular part of the matrix shows similarity scores computed from the English corpus, while the upper triangular part corresponds to those from the Chinese corpus. The diagonal is omitted for clarity.}
\label{fig:domain_similarity_heatmap}
\end{figure*}

\subsubsection{Corpus Diversity}
Following the approach in MTEB~\cite{muennighoff-etal-2023-mteb}, we compute inter-domain similarities separately for English and Chinese corpora, each consisting of 31 domains, in \posir.
Specifically, we adopt the highest-scoring model on the MTEB (eng, v2) STS task,
QZhou-Embedding (91.65)~\cite{yu2025qzhouembeddingtechnicalreport},
to generate embeddings for all documents within each domain.
We then average document embeddings at the domain level and compute pairwise cosine similarities between domains.
The resulting similarity matrix is visualized as a heatmap in Figure~\ref{fig:domain_similarity_heatmap}.

\begin{table}[!ht]
\centering
\caption{Distribution of query types in the English and Chinese retrieval datasets.}
\label{tab:query_type_diversity}
\begin{tabular}{lcc}
\toprule
\textbf{Query Type} & \textbf{English (\%)} & \textbf{Chinese (\%)} \\
\midrule
\textsc{WHAT} & 34.46 & 41.44\\
\textsc{HOW} & 26.60 & 17.93\\
\textsc{WHY} & 12.33 & 23.18 \\
\textsc{WHICH} & 5.53 & 1.75\\
\textsc{WHEN} & 3.14 & 5.68 \\
\textsc{WHERE} & 2.96 & 1.86 \\
\textsc{WHO} & 2.67 & 5.25 \\
\textsc{CLAIM} & 0.42 & 0.03 \\
\textsc{OTHERS} & 11.88 & 2.90 \\
\midrule
Total & 100.00 & 100.00 \\
\bottomrule
\end{tabular}

\end{table}

\section{Human Evaluation of Generated Queries and Reference Spans}
\label{appendix_sec:hm_review}
\subsection{Evaluation Setup}
To validate the quality of the generated queries and reference spans in \posir, we conducted targeted human evaluation on a representative subset. Given the large scale of \posir (310 datasets spanning 10 languages and 31 domains), we focused our human evaluation on the English subset to enable a focused and thorough assessment.
Specifically, we randomly sampled 500 English query-document-reference span triples across 10 diverse domains (50 samples per domain). Each sample consists of: (1) a query generated by LLMs, (2) the corresponding positive document, and (3) the reference span annotated in the qrels, which we evaluated to verify whether it accurately localizes the answer within the document.

\subsection{Evaluation Protocol}
One annotator evaluated each sample across three dimensions using a structured evaluation form:

\begin{itemize}
  \item \textbf{Query Quality (1-5 Scale):} This dimension assesses the overall quality of the query for retrieval. Scores range from 5 (a perfect query that is clear, specific, fully relevant to the document) to 1 (query with significant issues in relevance, clarity, or answerability).
  \item \textbf{Span Necessity (Binary 0/1):} This dimension evaluates whether the reference span contains only the necessary content. A score 1 indicates the span captures the core answer without irrelevant context, and a score 0 indicates over-extraction with unnecessary surrounding content.
  \item \textbf{Span Sufficiency (Binary 0/1)}: This dimension evaluates whether the reference span contains complete information to answer the query. A score 1 indicates the span is sufficient with the core answer completely contained, and a score 0 indicates under-extraction where some supporting details are missing.
\end{itemize}

We also evaluated whether the generated spans fabricate information not grounded in the document.

\subsection{Evaluation Results}
The evaluation results demonstrate the high quality of \posir's position-aware annotations. Query quality achieved a mean score of 4.91, with 496 of 500 samples (99.2\%) rated $\ge$ 4, indicating Good or Excellent quality. For span precision, 490 of 500 spans (98.0\%) satisfied both necessity and sufficiency criteria. Among the 10 remaining cases, 3 exhibited over-extraction and 7 exhibited under-extraction. Notably, no hallucinations were observed.
These results confirm that \posir's fine-grained position-aware relevance annotations are reliable and suitable for rigorous evaluation of position bias in retrieval models\footnote{We acknowledge that a single annotator limits the assessment of inter-rater reliability; however, the near-ceiling scores across all dimensions suggest that annotation quality is unambiguous in the vast majority of cases.}.

\section{Multilingual Machine Translation}
\label{appendix_sec:mmt}
In this section, we detail the selection process of the translation model and the rigorous quality control protocols applied to construct the multilingual subset of \posir.

\paragraph{Data Construction.}
To construct a representative evaluation set for translation quality, we perform stratified sampling from the 31 domain-specific English corpora utilized in \posir. 
Within each domain, we select 12 documents stratified by token length into four bins (0--512, 512--1024, 1024--1536, and 1536--2048 tokens, measured by the Qwen3 tokenizer), yielding a total of 372 source texts. 
This design ensures balanced domain coverage while capturing a broad spectrum of document lengths and linguistic complexities.

\paragraph{Translation Models.}
We evaluate four representative translation systems: 
(1) \textit{Google Translate}\footnote{\url{https://translate.google.com}}, a widely used commercial neural machine translation service; 
(2) \textit{Hunyuan-MT-7B}~\cite{zheng2025hunyuanmttechnicalreport}, a specialized translation model achieved first place in the WMT25 competition; 
(3) \textit{Qwen3-30B-A3B-Instruct-2507}~\cite{yang2025qwen3technicalreport}, a 31B-parameter open-source general MoE model; 
and 
(4) \textit{GPT-4o}~\cite{openai2024gpt4technicalreport}, a frontier general model serving as a strong baseline. 
Each system translates the sampled texts into 9 target languages: Arabic (ara-Arab), French (fra-Latn), German (deu-Latn), Italian (ita-Latn), Korean (kor-Kore), Portuguese (por-Latn), Russian (rus-Cyrl), and Spanish (spa-Latn).
The prompt for the translation task is provided in Appendix~\ref{appendix_sec:prompt_mmt}.

\subsection{Automatic Evaluation}
\subsubsection{Evaluation Protocol}
\label{sec:Automatic_Evaluation}
To efficiently evaluate the translation quality across 8 languages and 31 domains, we employ an LLM-based reference-free evaluation approach~\cite{kocmi-federmann-2023-large} (see Appendix~\ref{appendix_sec:prompt_tqv}). 
Specifically, we utilize GPT-4o as an automatic evaluator to assess translation quality along four dimensions: \textit{Fluency} (naturalness and readability of the target text), \textit{Accuracy} (semantic fidelity to the source), \textit{Completeness} (preservation of all source information), and \textit{Style Consistency} (appropriateness of tone and register). Each dimension is rated on a 5-point Likert scale (1 = Poor, 5 = Excellent), and an overall score is computed as the mean of the four dimensions. In total, this results in 2,976 evaluated translation outputs (372 samples $\times$ 8 languages).

\begin{table}[!ht]
\centering
\caption{Automatic evaluation of multilingual translation quality by languages and dimensions (1-5 scale). Evaluation dimensions include Fluency (Flu), Accuracy (Acc), Completeness (Com), Style Consistency (Sty), and Overall Average Score (Ove). ``Qwen3-30B-A3B'' denotes the ``Qwen3-30B-A3B-Instruct-2507'' model.}
\label{tab:translation_quality}
\small
\setlength{\tabcolsep}{2pt}
\resizebox{\textwidth}{!}{%
\begin{tabular}{l|ccccc|ccccc|ccccc|ccccc}
\toprule
\multirow{2}{*}{\textbf{Language}} & \multicolumn{5}{c}{\textbf{Qwen3-30B-A3B}} & \multicolumn{5}{c}{\textbf{Google Translate}} & \multicolumn{5}{c}{\textbf{Hunyuan-MT-7B}} & \multicolumn{5}{c}{\textbf{GPT-4o}} \\
 & \textbf{Flu} & \textbf{Acc} & \textbf{Com} & \textbf{Sty} & \textbf{Ove} & \textbf{Flu} & \textbf{Acc} & \textbf{Com} & \textbf{Sty} & \textbf{Ove} & \textbf{Flu} & \textbf{Acc} & \textbf{Com} & \textbf{Sty} & \textbf{Ove} & \textbf{Flu} & \textbf{Acc} & \textbf{Com} & \textbf{Sty} & \textbf{Ove} \\
\midrule
{Arabic} & 4.38 & 4.18 & 4.19 & 4.33 & 4.27 & 4.33 & 4.24 & 4.26 & 4.33 & 4.29 & 4.30 & 3.70 & 3.28 & 4.14 & 3.85 & 4.25 & 4.16 & 4.14 & 4.22 & 4.19 \\
{French} & 4.85 & 4.63 & 4.54 & 4.84 & 4.71 & 4.64 & 4.47 & 4.43 & 4.64 & 4.54 & 4.40 & 3.74 & 3.27 & 4.24 & 3.91 & 4.57 & 4.48 & 4.38 & 4.56 & 4.50 \\
{German} & 4.61 & 4.45 & 4.51 & 4.60 & 4.54 & 4.58 & 4.35 & 4.38 & 4.55 & 4.47 & 4.40 & 3.74 & 3.33 & 4.21 & 3.92 & 4.55 & 4.50 & 4.45 & 4.55 & 4.51 \\
{Italian} & 4.67 & 4.48 & 4.44 & 4.65 & 4.56 & 4.39 & 4.37 & 4.37 & 4.39 & 4.38 & 4.37 & 3.67 & 3.15 & 4.14 & 3.83 & 4.44 & 4.34 & 4.32 & 4.40 & 4.37 \\
{Korean} & 4.39 & 4.12 & 4.13 & 4.35 & 4.25 & 4.24 & 4.17 & 4.25 & 4.24 & 4.23 & 4.40 & 3.91 & 3.54 & 4.28 & 4.03 & 3.88 & 3.74 & 3.78 & 3.87 & 3.82 \\
{Portuguese} & 4.81 & 4.61 & 4.59 & 4.81 & 4.71 & 4.59 & 4.45 & 4.47 & 4.56 & 4.52 & 4.42 & 3.75 & 3.29 & 4.24 & 3.93 & 4.71 & 4.60 & 4.53 & 4.70 & 4.63 \\
{Russian} & 4.69 & 4.44 & 4.47 & 4.68 & 4.57 & 4.51 & 4.28 & 4.30 & 4.45 & 4.39 & 4.30 & 3.57 & 3.14 & 4.11 & 3.78 & 4.35 & 4.23 & 4.19 & 4.33 & 4.28 \\
{Spanish} & 4.77 & 4.60 & 4.61 & 4.77 & 4.69 & 4.52 & 4.39 & 4.40 & 4.52 & 4.46 & 4.37 & 3.71 & 3.22 & 4.22 & 3.88 & 4.63 & 4.52 & 4.49 & 4.62 & 4.57 \\
\midrule
{Average} & {4.65} & {4.44} & {4.43} & {4.63} & {4.54} & {4.47} & {4.34} & {4.36} & {4.46} & {4.41} & {4.37} & {3.72} & {3.28} & {4.20} & {3.89} & {4.42} & {4.32} & {4.29} & {4.41} & {4.36} \\
\bottomrule
\end{tabular} }

\end{table}

\begin{figure*}[!ht]
\centering
\includegraphics[width=1.0\textwidth]{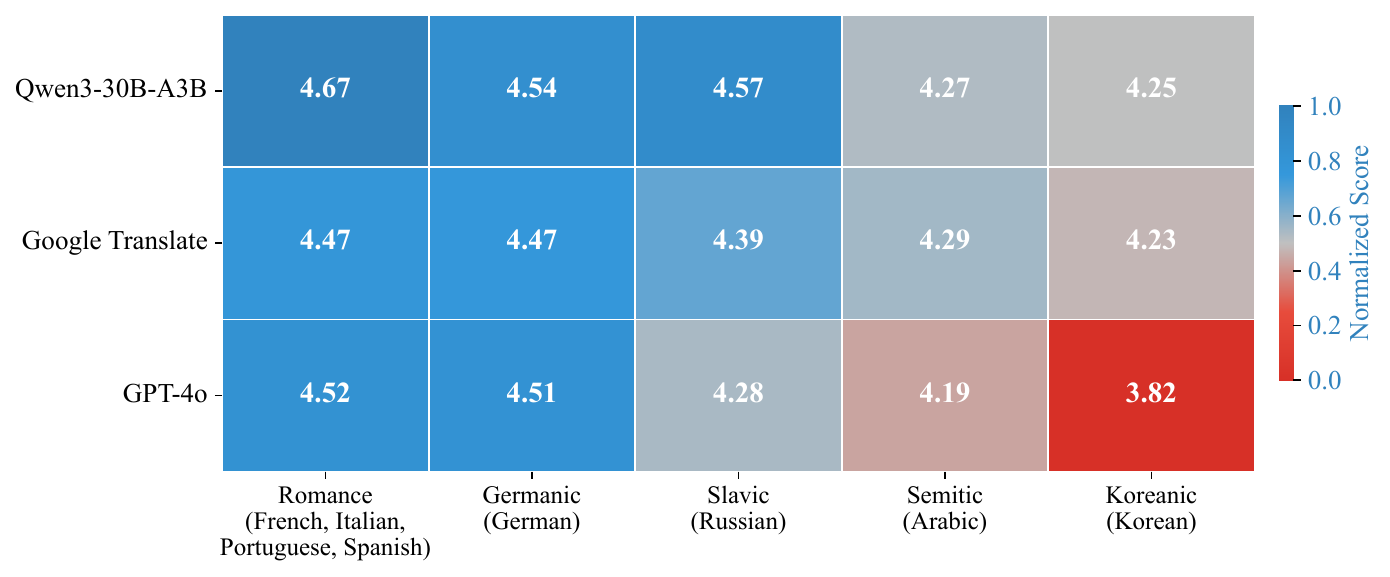}
\caption{Cross-lingual translation quality heatmap (using Overall Average Score) across 3 translation models and 5 language families. 
``Qwen3-30B-A3B'' denotes the ``Qwen3-30B-A3B-Instruct-2507'' model.}
\label{fig:heatmap_by_family}
\end{figure*}

\subsubsection{Evaluation Results}
Table~\ref{tab:translation_quality} present aggregated evaluation results across four translation models and five quality dimensions. 
Qwen3-30B-A3B-Instruct-2507 achieves the highest overall score (4.54), followed by Google Translate (4.41), GPT-4o (4.36), and Hunyuan-MT-7B (3.89). 
Dimension-wise analysis reveals that Qwen3-30B-A3B-Instruct-2507 ranks first across all metrics, demonstrating particular strengths in \textit{fluency} (4.65) and \textit{style consistency} (4.63), while maintaining competitive performance in \textit{accuracy} (4.44) and \textit{completeness} (4.43). 
Google Translate exhibits a balanced profile with relatively uniform scores across dimensions, indicating stable but less exceptional translation quality. In stark contrast, Hunyuan-MT-7B shows severe degradation in \textit{completeness} (3.28) and \textit{accuracy} (3.72), with performance gaps exceeding 1.0 point compared to top-performing models. 
Despite achieving comparable \textit{fluency} score (4.37), qualitative analysis reveals that Hunyuan-MT-7B frequently sacrifices semantic fidelity to preserve surface-level linguistic smoothness, resulting in content omissions and meaning distortions, particularly when handling complex or information-dense source texts.
Moreover, as shown in Figure~\ref{fig:heatmap_by_family}, a cross-linguistic analysis reveals that Romance languages such as French achieve relatively higher translation quality, whereas Arabic and Korean exhibit comparatively lower performance.

\subsection{Human Evaluation}
\label{appendix:human_eval}
Initial inspection using automatic evaluation suggests that Qwen3-30B-A3B-Instruct-2507 produces high-quality translations across \posir, achieving competitive performance in comparison to other methods. However, given that translation quality appears strong, rigorous human validation is essential. While automatic evaluation can detect catastrophic failures, it may be insensitive to subtle but critical errors in high-quality translations, such as nuanced mistranslations, contextual inappropriateness, or domain-specific inaccuracies. 
Due to limited resources, we conduct focused human evaluation on three typologically diverse languages (Arabic, French, and Russian) to provide expert assessment of translation quality, facilitating a direct comparison between human evaluation and LLM-based assessments.

\begin{table}[!ht]
\centering
\caption{Human evaluation scores by languages and dimensions (1-5 scale).}
\label{tab:human_eval}
\setlength{\tabcolsep}{0pt} 
\begin{tabular*}{\columnwidth}{@{\extracolsep{\fill}}lcccc}
\toprule
\textbf{Dimension} & \textbf{French} & \textbf{Russian} & \textbf{Arabic} & \textbf{Avg.} \\
\midrule
Fluency & 4.71 & 4.65 & 4.58 & 4.65 \\
Accuracy & 4.52 & 4.48 & 4.35 & 4.45 \\
Completeness & 4.89 & 4.82 & 4.76 & 4.82 \\
Style Consistency & 4.68 & 4.61 & 4.52 & 4.60 \\
\midrule
Overall Avg. Score & 4.70 & 4.64 & 4.55 & 4.63 \\
\bottomrule
\end{tabular*}

\end{table}

\subsubsection{Annotator Recruitment}
Five annotators were recruited from graduate programs in linguistics, natural language processing, and applied foreign languages via established academic networks. 
All annotators were advanced graduate students or recent PhD graduates with expertise in computational linguistics and/or translation studies.
For Arabic and Russian, native speakers of the respective languages conducted the annotations. 
For French, two annotators holding DALF C1/C2 certificates (Diplôme Approfondi de Langue Française, the highest level of French language certification) completed the evaluation.
All annotators were compensated at a fair hourly rate, in compliance with applicable labor regulations.

\subsubsection{Evaluation Protocol}

\begin{figure*}[!ht] 
    \centering
    \includegraphics[width=1.0\textwidth]{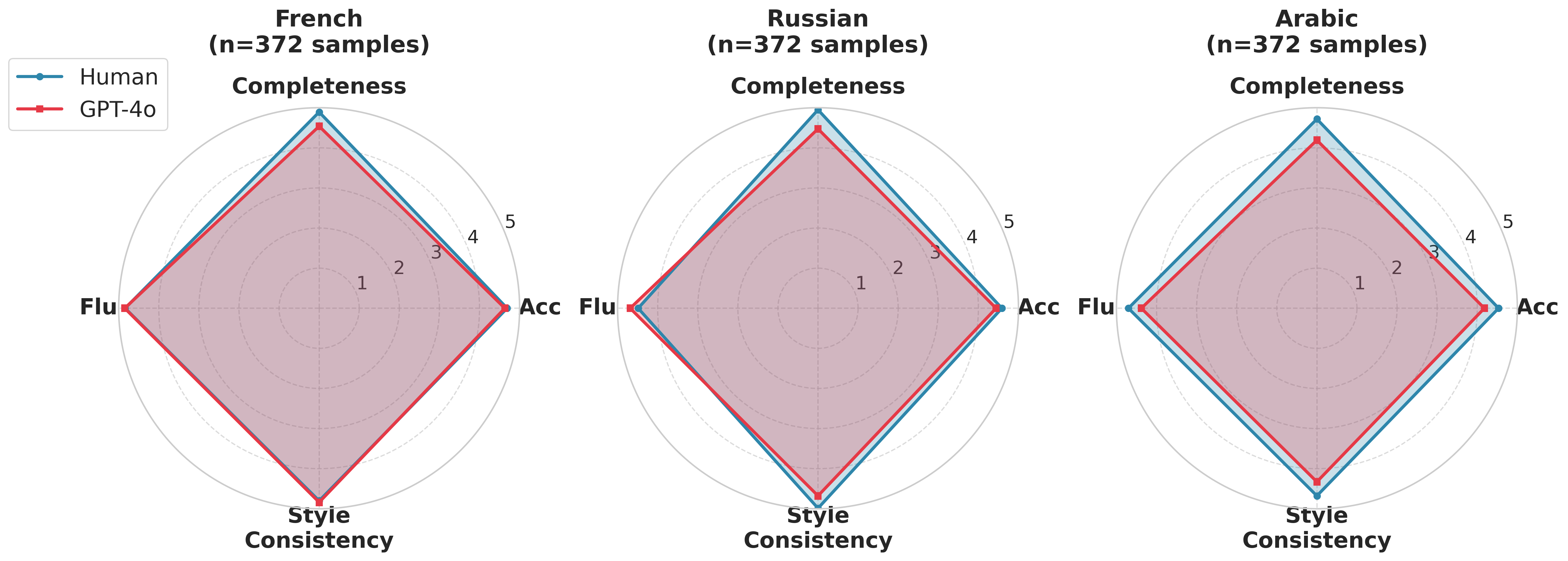}
    \caption{Human vs. GPT-4o score comparison by language. Evaluation dimensions include Fluency (Flu), Accuracy (Acc), Completeness, and Style Consistency. The overlapping radar shapes indicate strong alignment in quality assessment patterns.}
    \label{fig:radar}
\end{figure*}

\begin{figure*}[!ht]
    \centering
    \includegraphics[width=0.9\textwidth]{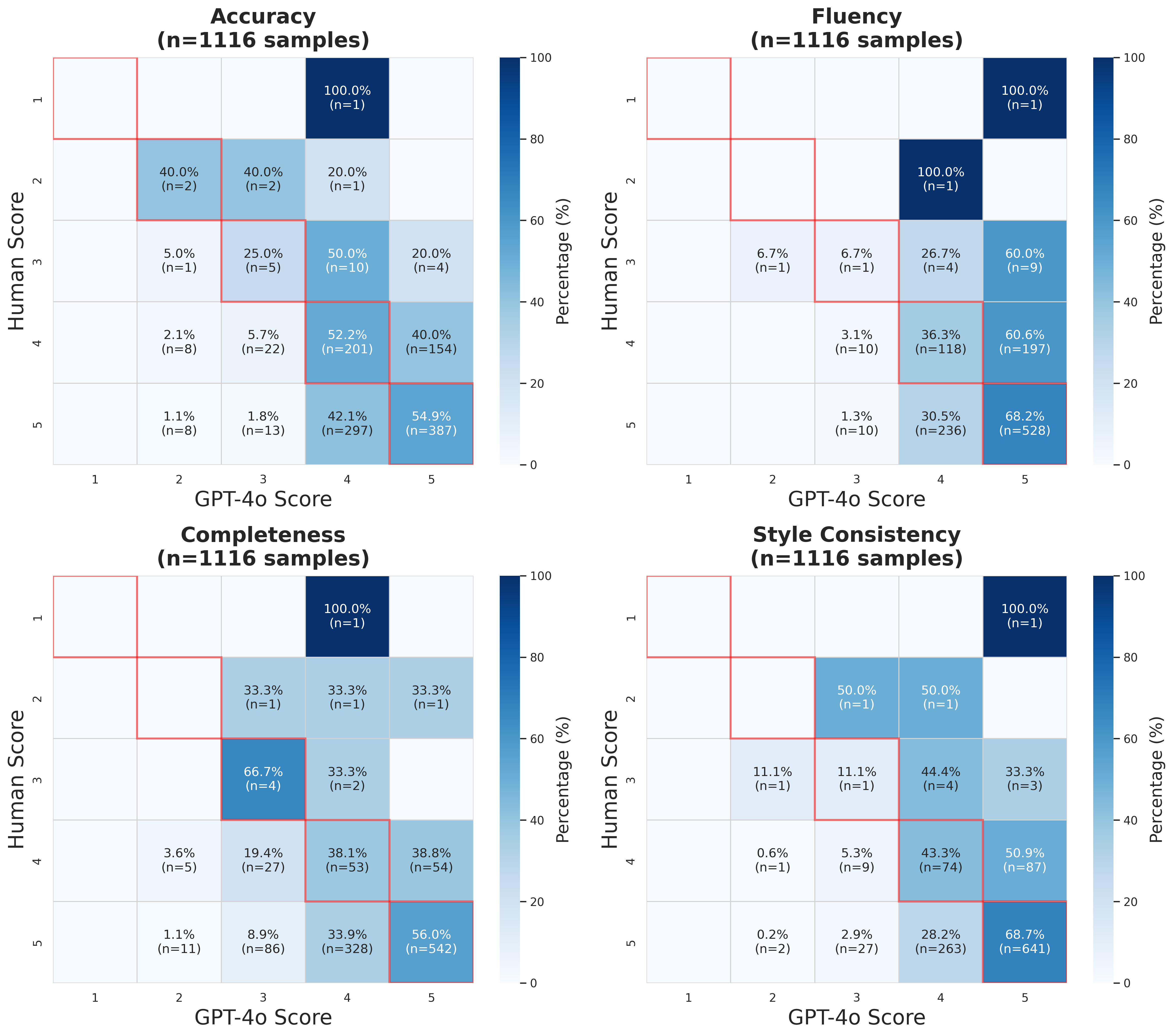}
    \caption{Confusion matrix of human vs. GPT-4o scores. Each row shows distribution of GPT-4o scores for a given human score. Data concentrates in the high-score diagonal (4--5), with GPT-4o showing systematic strictness for top-rated samples.}
    \label{fig:confusion}
\end{figure*}

Annotators evaluated 372 translation pairs across 3 languages (1,116 samples in total) generated by Qwen3-30B-A3B-Instruct-2507. The evaluations were conducted in accordance with the consistency standards outlined in the automatic evaluation, as described in Appendix~\ref{sec:Automatic_Evaluation}. 
Annotators were given both the English source text and the target language translation, with no access to reference translations to prevent anchoring bias.

\subsubsection{Evaluation Results}

Table~\ref{tab:human_eval} presents the mean scores across languages and dimensions. All language-dimension combinations achieved scores between 4.35 and 4.89, indicating consistently high translation quality. Over 97\% of translations received scores $\geq$4 across all dimensions.
Several patterns emerge from the results. First, \textit{completeness} scores consistently achieve the highest ratings (4.76--4.89), indicating that Qwen3-30B-A3B-Instruct-2507 rarely omits source information and preserves semantic integrity well. Second, \textit{fluency} scores consistently exceed \textit{accuracy} scores across all languages (4.58--4.71 vs. 4.35--4.52), suggesting the model generates natural-sounding translations in target languages while maintaining acceptable fidelity. 
Third, language-specific differences reflect the typological distance from English: French achieves the highest average scores (4.70), likely due to abundant training data and linguistic proximity; Arabic shows relatively lower \textit{accuracy} (4.35), potentially due to its complex morphological system; Russian performs in between (4.64), despite its challenging case system.
The score distribution exhibits a pronounced ceiling effect: 67.1\% of samples received 5 (Excellent), 30.4\% received 4 (Good), and only 2.5\% scored $\leq$3. This confirms that the Qwen model produces high-quality translations suitable for \posir, with $>$97\% of samples rated Good or Excellent. 

\begin{table}[!t]
\centering
\caption{Human-GPT-4o agreement within ±1 point tolerance.}
\label{tab:agreement}
\begin{tabular}{lcc}
\toprule
\textbf{Dimension} & \textbf{Agreement (±1)} & \textbf{MAE} \\
\midrule
Fluency & 98.1\% & 0.44 \\
Accuracy & 96.9\% & 0.51 \\
Completeness & 90.6\% & 0.57 \\
Style Consistency & 96.9\% & 0.39 \\
\bottomrule
\end{tabular}

\end{table}

\subsection{GPT-4o Human Evaluation Alignment}
We validate GPT-4o as a translation evaluator by comparing its ratings against human scores on the same 1,116 samples.
Figure~\ref{fig:radar} shows the score distribution comparison between human annotators and GPT-4o across languages. The radar charts demonstrate high shape alignment: both raters identify \textit{completeness} as the strongest dimension and \textit{accuracy} as relatively lower, with mean differences within 0.3 points across all dimensions.
Under a ±1 tolerance threshold (appropriate for ordinal Likert scales), GPT-4o achieves 96.15\% agreement with human ratings. 
Table~\ref{tab:agreement} shows per-dimension results. Figure~\ref{fig:confusion} presents the detailed discrepancy analysis, revealing two notable patterns. In the high-score region, data concentrates along the 4--5 diagonal, explaining the strong agreement rates. However, for samples rated 5 by human annotators, GPT-4o averages 4.5--4.7, exhibiting a systematic offset of $-0.3$ to $-0.5$ points. This suggests GPT-4o acts as a stricter judge, more inclined to identify potential flaws than to endorse perfection, effectively serving as a conservative lower-bound estimator. Conversely, in the sparsely populated low-score region, GPT-4o tends toward leniency: for the rare samples rated 1--3 by humans (only 2.5\% of data), GPT-4o often assigns 4--5 scores, suggesting potential blind spots in detecting severe semantic errors. Fortunately, such low-quality samples are rare in our dataset, so GPT-4o's conservative behavior on high-quality translations, where the vast majority of data resides, provides reliable quality assurance for \posir.
In summary, the strong human-GPT-4o alignment validates automated quality verification for the full dataset, establishing confidence in the translation reliability of the Qwen model.

\subsection{Model Selection Rationale}
Based on comprehensive evaluation results and practical considerations, we select \emph{Qwen3-30B-A3B-Instruct-2507} as the primary translation model for constructing the \posir multilingual corpus. 
This decision is justified by three key factors: (1) \textit{superior translation quality}, consistently achieving the highest scores across all evaluation dimensions and target languages, with an overall advantage of 0.13 points over Google Translate and 0.21 points over GPT-4o; (2) \textit{strong cross-lingual robustness}, maintaining stable performance across diverse linguistic typologies and writing systems without significant quality degradation; (3) \textit{favorable resource efficiency}, delivering high translation quality with moderate computational requirements, which enables large-scale multilingual data construction under practical inference-time constraints. 
Additionally, as an open-source and freely accessible model, it offers further flexibility and cost-effectiveness for large-scale deployment and application.

\subsection{Fixing and Filtering Strategies}
Through manual inspection of the translated outputs, we observed recurring artifacts that could degrade data quality, such as invalid prefixes, refusal-style responses, and repetitive or verbose translations.
To systematically mitigate these issues while preserving corpus-query-qrels consistency, we design and apply a three-stage cleaning pipeline.

\subsubsection{Prefix Removal}
The Qwen3-30B-A3B-Instruct-2507 model sometimes prepends explanatory text (e.g., ``Here is the translation:'') to its outputs. To identify and remove such artifacts, we adopt a dual-strategy prefix detection heuristic. For corpus texts, we employ a co-occurrence pattern matching strategy that identifies the presence of translation-related keywords (``translate'' or ``translation'') and newline characters within a proximity window of 500 characters, starting from the first 500 characters and extending up to 2,000 characters. For queries, which are typically shorter, we apply full-text matching. Candidate prefixes are validated as English using langdetect\footnote{\url{https://github.com/fedelopez77/langdetect}} and manually reviewed before compilation into a blacklist (118 patterns). Overall, this procedure affected 54 instances across 40 files, accounting for 0.0003\% of the 17.7M records and spanning 8 non-English languages.

\subsubsection{Refusal Detection}
We employ a two-stage filtering approach to identify translation refusals. In the first stage, we apply lightweight blacklist matching combined with language detection to flag candidate refusals, prioritizing high recall with minimal computational overhead. In the second stage, flagged instances are validated using Qwen3-Max\footnote{\url{https://qwen.ai/blog?id=qwen3-max}}, which inspects the first 1,500 characters to distinguish refusal-related meta-commentary (e.g., ``I cannot translate'') from legitimate translations. This process identified 2,249 refusals, predominantly in Arabic (1,378 instances) and Korean (323). We remove 2,223 corpus entries and 26 queries, with cascading deletion of 26 corresponding qrels to maintain data consistency.

\subsubsection{Statistical Outlier Filtering}
We perform token-length-based outlier filtering at the language–domain level using the Qwen3 tokenizer. For each stratum, we compute the interquartile range (IQR) and define asymmetric thresholds, with a lower bound of $Q1 - 3.0\times IQR$ and an upper bound of $Q3 + 1.5\times IQR$. 
This asymmetric design focuses on filtering abnormally long texts, which are more likely to result from translation errors, duplication, or unintended document concatenation. It enforces a stricter upper bound to remove such cases, while allowing reasonable length variation among shorter documents. We process only corpus outliers, cascading deletions to associated queries and qrels. 

In total, 15,641 corpus documents (0.11\%) and 332 queries (0.10\%) are removed. The largest reductions occur in Arabic (5,490 corpus entries; 0.25\%) and Korean (4,082; 0.24\%), followed by Russian (1,557; 0.11\%). Overall, the removal ratios remain consistently low across all languages, indicating that the filtering procedure effectively eliminates anomalous long texts while preserving the structural integrity and cross-lingual balance of the benchmark.

\section{Prompts}
\subsection{Prompt for Position-Aware QA Generation}
\label{appendix_sec:prompt_qa}

We use the following prompt to instruct an LLM to generate diverse and realistic question--answer (QA) pairs under specific positional constraints.
Unlike explicitly segmenting the document into segments, we always feed the \emph{entire} document to the LLM. Position control is achieved solely by injecting a sampled positional constraint into a dedicated slot in the prompt.

\paragraph{Sampling and Constraint Injection.}
For each generation instance, we randomly sample one positional constraint from a predefined set and insert it into the prompt via \textbf{Positional Requirement Slot}. This constraint guides the LLM to produce questions whose answers primarily come from a particular region of the document, while the model still has access to the complete document context.

\noindent The Positional Requirement Slot is filled with exactly one of the following options:
\begin{itemize}
    \item \emph{Focus on first third: The question should primarily reference the first third of the document}.
    \item \emph{Focus on middle third: The question should primarily reference the middle third of the document}.
    \item \emph{Focus on final third: The question should primarily reference the final third of the document}.
\end{itemize}

\paragraph{Prompt Usage.}
The prompt below enforces key requirements including answerability, naturalness, and the avoidance of meta references to the source document. The positional constraint is an \emph{internal control signal} and must not be mentioned explicitly in the generated questions.
The question type list is intentionally over-complete to support diverse generation; not all types are expected to be instantiated for every document.

\begin{tcolorbox}[breakable]
\# Task Description \\
Generate user questions in realistic search engine scenarios based on a given document. \\

\# Core Requirements \\
1. Answerability: Answers must be explicitly stated or directly inferable from the document (no outside knowledge). \\
2. Authenticity: Questions must reflect natural human phrasing. \\
3. No Referencing: Avoid terms like ``this text'', ``the author'', or ``this document''. \\
4. \textbf{\code{\{\{Positional Requirement Slot\}\}}} \\

\# Question Configuration \\
Configure two elements: Question Type and Expression Style. Select configurations contextually. \\

\#\# Question Type \\
1. **Fact Query**: Retrieve specific information (e.g., dates, metrics, definitions). \\
2. **Operation Guide**: Request steps/methods to accomplish tasks (focuses on actionable steps). \\
3. **Cause Analysis**: Investigate reasons behind occurrences or phenomena. \\
4. **Comparison Choice**: Contrast differences/advantages to facilitate decisions. \\ 
5. **Concept Explanation**: Clarify meanings, categories, or structures of terms/concepts. \\
6. **Viewpoint Verification**: Validate accuracy/reliability of claims/data (requires credibility evaluation). \\
7. **Prediction/Speculation**: Inquire about future outcomes or plausible inferences strictly grounded in the document content. \\
8. **True/False Judgment**: Questions answerable with "yes" or "no". \\
9. **Background Exploration**: Understand context, history, current status, or related environments. \\
10. **Seeking Advice**: Request subjective opinions, experiences, or recommendations (emphasizes actionable suggestions). \\
11. **Open Discussion**: Pose broad/debatable topics to stimulate dialogue. \\
12. **Emotional Resonance**: Express feelings (frustration/seeking help/excitement) to obtain empathy/support. \\
13. **Feasibility Assessment**: Query viability, costs, or resource requirements for plans/ideas. \\
14. **Mathematical Reasoning**: Problems needing numerical calculations or logical deductions (require step-by-step solutions). \\
15. **Content Creation**: Request original text, narratives, or creative output. \\
16. **Relationship Counseling**: Focuses on conflict resolution in interpersonal interactions, communication strategies, or emotional advice. \\
17. **Diagnostic Troubleshooting**: Analyzes the root causes of issues based on anomalies and provides solutions. \\
18. **Hypothetical Scenarios**: Explores possibilities, ethical implications, or logical inferences under fictional or extreme conditions. \\
19. **Decision Advisory**: Assists in analyzing pros and cons, weighing factors for significant or personalized choices. \\
20. **Personal Growth**: Requests specific strategies, methods, or learning path planning for skill development, habit formation, efficiency enhancement, goal setting, or mindset adjustment. \\

\#\# Expression Style \\
- Concise | Casual | Informal \\
- Formal | Professional \\
- Technical | Academic \\

\# Steps \\
Follow the steps below and output results for each step in order. \\
1. Scenario Setup: Envision potential users and contexts where this document would be relevant. \\
2. Type Selection: Identify appropriate question types based on the content emphasized by the sampled positional constraint. \\
3. Style Matching: Select compatible expression styles. \\
4. Final Generation: Generate the final QA pairs and present them as a JSON array enclosed in a Markdown code block.
\begin{verbatim}
[
  {
    "question": "Generated question",
    "answer": "Answer text"
  }
]
\end{verbatim}

\# Important Tips \\
1. Users do not know the document content. Questions must (a) strictly avoid document references, and (b) use natural phrasing. \\
2. The positional constraint is for generation control only; do not reveal or mention it in the questions. \\
3. For subjective question types (e.g., advice), answers must be grounded in the document’s stated information; do not introduce external opinions.
4. If the document quality prevents question generation, output: ``Document quality insufficient for generation.'' \\

Now, let's start this task: \\
<document>\textbf{\code{\{\{DOCUMENT\}\}}}</document>
\end{tcolorbox}

\subsection{Prompt for Reference Locating}
\label{appendix_sec:prompt_ref}
Although explicit positional constraints are imposed during QA generation, the resulting questions may still violate these constraints in practice.
To address this issue, we design the following prompt to instruct an LLM to locate supporting reference text within the original document, thereby enhancing the factual grounding and answerability of the generated questions.

\begin{tcolorbox}[breakable]
\# Task Description \\
Given a list of questions and a document, for each question, **precisely identify and extract continuous text segments** from the document that **directly answer or support answering** this question. \\

\#\# Requirements\\
1. **Text Integrity**: Copy document content verbatim. Do not paraphrase, summarize, or splice across paragraphs. Preserve original punctuation, whitespace, and line breaks exactly. \\
2. **Extraction Principles**: Prioritize complete clauses/semantic units as minimal extraction units; If multiple relevant segments exist, output all segments completely without omission. \\
3. **Relevance Criteria**: Extracted segments must satisfy **at least one** of: (a) Containing direct answers to the question; (b) Providing essential contextual support for answering the question \\
4. **Question Coverage**: Process **every question** in the provided question list.\\
5. **Accurately Assess Question Answerability**: If the document can answer the question (directly or indirectly), extract the reference text as supporting evidence; If the document cannot provide useful information, return an empty list.\\

\#\# Output Specification: Use strictly this JSON format.
\begin{verbatim}
[
  {
    "question": "Original Question",
    "reference_texts": [
      "Document Segment 1",
      "Document Segment 2",
      ...
    ]
  }
]
\end{verbatim}

Now, let's start this task: \\
<document>\textbf{\code{\{\{DOCUMENT\}\}}}</document> \\
<questions>\textbf{\code{\{\{QUESTIONS\}\}}}</questions>
\end{tcolorbox}

\subsection{Prompt for Relevance Assessment}
\label{appendix_sec:prompt_relevance}
Previous work \cite{llm_accessors} demonstrate that large language models such as GPT-4 can achieve labeling quality comparable to human annotators when producing gold-standard relevance labels for search systems.
Moreover, \cite{zeng2025explainabledoctorrecommendationlarge} shows that incorporating fine-grained relevance labels into prompts for LLM-based rerankers significantly improves zero-shot reranking performance by better distinguishing borderline and ambiguous relevance cases.
Motivated by these findings, we employ DeepSeek-V3.1 (Non-thinking Mode)\footnote{\url{https://api-docs.deepseek.com/news/news250821}} as a relevance labeler,
adopting a 5-level relevance labeling scheme to capture fine-grained distinctions in relevance.
This labeling process is used solely for auxiliary verification rather than primary relevance annotation.

\begin{tcolorbox}[breakable]
\# Task

Given a list of questions and a document, evaluate the relevance between each question and the document on a 0-4 scale:\\
0 = Completely irrelevant  \\
1 = Slightly relevant (minimal connection)  \\
2 = Moderately relevant (The document addresses parts of the question but lacks completeness, depth, or direct alignment)  \\
3 = Highly relevant (covers most aspects of the question)  \\
4 = Perfectly relevant (can correctly address the question)\\

\# Output Format\\
Output all the relevance scores in the **same order** as the input questions in the format of a JSON array in Markdown. An example, assuming there are three questions, then your output is:
\begin{verbatim}
[
  1,
  3,
  0
]
\end{verbatim}

Now, let's start this task: \\
<document>\textbf{\code{\{\{DOCUMENT\}\}}}</document> \\
<questions>\textbf{\code{\{\{QUESTIONS\}\}}}</questions>
\end{tcolorbox}

\subsection{Prompt for Multilingual Machine Translation}
\label{appendix_sec:prompt_mmt}
\begin{tcolorbox}[breakable]
You are a world-class professional translator, fluent in both English and \textbf{\code{\{\{LANGUAGE\}\}}}.\\

Your task is to translate the following text into \textbf{\code{\{\{LANGUAGE\}\}}}.\\
Please adhere to the highest standards of translation, ensuring the output is:\\
1. **Faithful:** Accurately convey the original meaning, content, and intent without omission or distortion.\\
2. **Expressive:** Make the translation smooth, natural, and easy to understand for a native speaker of the \textbf{\code{\{\{LANGUAGE\}\}}}.\\
3. **Elegant:** Preserve the style, tone, and literary grace of the original text, making it aesthetically pleasing to read.\\

Here is the text:\\
\textbf{\code{\{\{TEXT\}\}}}
\end{tcolorbox}

\subsection{Prompt for Translation Quality Evaluation}
\label{appendix_sec:prompt_tqv}
\begin{tcolorbox}[breakable]
\# Task \\
You are a professional translation quality evaluator. Evaluate the English to \textbf{\code{\{\{LANGUAGE\}\}}} translation using a 5-point integer scale (1-5, whole numbers only). \\

\#\# Evaluation Criteria \\
1. **Accuracy**: Semantic fidelity to original meaning - no mistranslations, omissions, or additions. \\
2. **Fluency**: Natural language quality, grammar, and readability in target language. \\
3. **Completeness**: All source content translated without missing information or concepts. \\
4. **Style Consistency**: Appropriate tone, register, and style matching the source text context. \\

\#\# Scoring Scale \\
- **5**: Excellent (professional publication quality, no issues). \\
- **4**: Good (minor issues that don't affect understanding).\\
- **3**: Acceptable (noticeable issues but meaning is clear).\\
- **2**: Poor (significant problems affecting comprehension).\\
- **1**: Unacceptable (major errors, meaning unclear or wrong).\\

\#\# Output Format \\
Provide ONLY this JSON format (no additional text):
\begin{verbatim}
{
    "accuracy": <1-5>,
    "fluency": <1-5>,
    "completeness": <1-5>,
    "style_consistency": <1-5>,
    "comments": "Brief justification"
}
\end{verbatim} 

Now, let's start this task: \\
\#\# Input \\
\#\#\# Original Text (English)\\
\textbf{\code{\{\{ORIGINAL\_TEXT\}\}}}\\

\#\#\# Translation (\textbf{\code{\{\{LANGUAGE\}\}}})\\
\textbf{\code{\{\{TRANSLATED\_TEXT\}\}}}
\end{tcolorbox}

\begin{table}[t]
    \centering

        \caption{Detailed evaluation results of 10 multilingual retrieval models on \posir (Part 1 of 2). In both multilingual retrieval and cross-lingual retrieval (translated queries retrieving English documents) settings, the results for each language are weighted-averaged across 31 domains. Q1--Q4 represent query buckets partitioned by the token length of positive documents (512-token intervals). ``KaLM-Embedding-12B'' denotes the ``KaLM-Embedding-Gemma3-12B-2511'' model. For Part 2, refer to Table~\ref{tab:detailed_experiment_results_part2}.
    }
    \label{tab:detailed_experiment_results}
    \setlength{\extrarowheight}{2pt}
    \resizebox{\linewidth}{!}{
        \begin{tabular}{c l | c c |c c|c c|c c|c c|c c|c c|c c|c c|c c}
        \toprule
        & \multirow{2}{*}{\textbf{Model}} & \multicolumn{2}{c}{\textbf{Arabic}}  & \multicolumn{2}{c}{\textbf{Chinese}}  & \multicolumn{2}{c}{\textbf{German}}  & \multicolumn{2}{c}{\textbf{English}} & \multicolumn{2}{c}{\textbf{French}} & \multicolumn{2}{c}{\textbf{Italian}} & \multicolumn{2}{c}{\textbf{Korean}} & \multicolumn{2}{c}{\textbf{Portuguese}} & \multicolumn{2}{c}{\textbf{Russian}} & \multicolumn{2}{c}{\textbf{Spanish}} \\
        & & nDCG@10 & PSI & nDCG@10 & PSI & nDCG@10 & PSI & nDCG@10 & PSI & nDCG@10 & PSI & nDCG@10 & PSI & nDCG@10 & PSI & nDCG@10 & PSI & nDCG@10 & PSI & nDCG@10 & PSI \\
        \midrule
        \multirow{22}{*}{\rotatebox{90}{\posir}} & \multicolumn{9}{l}{\textit{Multilingual Retrieval}} \\
        \cmidrule(lr){2-22}
        & gte-multilingual-base & 34.08 & 0.52 & 51.52 & 0.26 & 47.96 & 0.44 & 61.59 & 0.19 & 49.59 & 0.41 & 47.63 & 0.45 & 37.54 & 0.48 & 48.94 & 0.42 & 44.89 & 0.46 & 49.95 & 0.42 \\
        & bge-m3 & 34.08 & 0.41 & 47.93 & 0.34 & 46.59 & 0.29 & 50.79 & 0.30 & 44.07 & 0.33 & 42.50 & 0.34 & 38.19 & 0.35 & 44.27 & 0.36 & 40.60 & 0.36 & 43.16 & 0.35 \\
        & Qwen3-Embedding-0.6B & 42.11 & 0.45 & 57.49 & 0.30 & 53.79 & 0.37 & 65.10 & 0.22 & 55.33 & 0.34 & 54.49 & 0.36 & 45.68 & 0.38 & 55.68 & 0.35 & 50.20 & 0.40 & 56.43 & 0.33 \\
        & inf-retriever-v1-1.5b & 46.48 & 0.32 & 64.12 & 0.17 & 58.83 & 0.23 & 69.26 & 0.09 & 60.85 & 0.21 & 59.39 & 0.25 & 51.52 & 0.25 & 60.94 & 0.23 & 55.49 & 0.23 & 61.21 & 0.23 \\
        & Qwen3-Embedding-4B & 54.10 & 0.32 & 60.32 & 0.31 & 63.67 & 0.25 & 69.96 & 0.22 & 62.47 & 0.27 & 63.84 & 0.26 & 57.26 & 0.27 & 64.78 & 0.26 & 60.51 & 0.30 & 65.67 & 0.25 \\
        & inf-retriever-v1 & 55.03 & 0.13 & 68.58 & 0.15 & 65.59 & 0.10 & 72.78 & 0.08 & 66.47 & 0.11 & 65.76 & 0.10 & 58.65 & 0.13 & 67.01 & 0.11 & 63.02 & 0.08 & 67.18 & 0.10 \\
        & NV-Embed-v2 & 18.32 & 0.75 & 37.46 & 0.56 & 51.28 & 0.49 & 68.27 & 0.32 & 52.40 & 0.48 & 50.79 & 0.50 & 28.17 & 0.72 & 52.41 & 0.51 & 37.42 & 0.62 & 53.67 & 0.49 \\
        & llama-embed-nemotron-8b & 55.41 & 0.07 & 60.01 & 0.11 & 66.34 & 0.06 & 73.48 & 0.06 & 65.73 & 0.05 & 65.79 & 0.06 & 58.78 & 0.07 & 66.67 & 0.07 & 61.94 & 0.06 & 66.71 & 0.06 \\
        & Qwen3-Embedding-8B & 57.43 & 0.20 & 60.36 & 0.32 & 65.30 & 0.18 & 70.27 & 0.20 & 65.32 & 0.15 & 65.70 & 0.18 & 59.57 & 0.17 & 66.57 & 0.20 & 62.95 & 0.19 & 67.32 & 0.19 \\
        & KaLM-Embedding-12B & 43.74 & 0.11 & 50.82 & 0.11 & 53.33 & 0.08 & 64.22 & 0.06 & 52.40 & 0.06 & 52.15 & 0.07 & 43.80 & 0.10 & 51.71 & 0.07 & 51.73 & 0.08 & 54.79 & 0.06 \\
        \cmidrule(lr){2-22}
        & \multicolumn{9}{l}{\textit{Cross-lingual Retrieval}} \\
        \cmidrule(lr){2-22}
        & gte-multilingual-base & 31.28 & 0.28 & - & - & 50.75 & 0.26 & - & - & 53.47 & 0.24 & 51.62 & 0.27 & 32.88 & 0.23 & 53.38 & 0.24 & 44.45 & 0.28 & 54.06 & 0.24 \\
        & bge-m3 & 24.04 & 0.39 & - & - & 43.16 & 0.35 & - & - & 41.70 & 0.35 & 40.37 & 0.35 & 28.06 & 0.43 & 41.26 & 0.36 & 31.83 & 0.39 & 39.97 & 0.36 \\
        & Qwen3-Embedding-0.6B & 34.91 & 0.38 & - & - & 53.97 & 0.29 & - & - & 55.14 & 0.28 & 54.28 & 0.29 & 39.50 & 0.32 & 55.52 & 0.28 & 47.68 & 0.29 & 56.30 & 0.27 \\
        & inf-retriever-v1-1.5b & 35.68 & 0.23 & - & - & 57.33 & 0.14 & - & - & 60.08 & 0.12 & 58.19 & 0.13 & 42.74 & 0.18 & 60.14 & 0.13 & 50.34 & 0.17 & 60.39 & 0.13 \\
        & Qwen3-Embedding-4B & 52.20 & 0.32 & - & - & 63.38 & 0.25 & - & - & 62.88 & 0.27 & 63.49 & 0.26 & 55.74 & 0.28 & 64.09 & 0.27 & 59.50 & 0.28 & 64.80 & 0.26 \\
        & inf-retriever-v1 & 52.28 & 0.17 & - & - & 65.34 & 0.10 & - & - & 66.77 & 0.10 & 65.89 & 0.10 & 56.62 & 0.14 & 66.94 & 0.10 & 61.82 & 0.12 & 67.28 & 0.10 \\
        & NV-Embed-v2 & 20.60 & 0.34 & - & - & 56.47 & 0.30 & - & - & 58.42 & 0.30 & 57.20 & 0.30 & 20.29 & 0.39 & 58.75 & 0.30 & 37.48 & 0.34 & 58.52 & 0.30 \\
        & llama-embed-nemotron-8b & 52.73 & 0.08 & - & - & 65.80 & 0.04 & - & - & 66.18 & 0.06 & 66.21 & 0.05 & 56.35 & 0.08 & 66.74 & 0.06 & 60.47 & 0.05 & 66.94 & 0.05 \\
        & Qwen3-Embedding-8B & 53.97 & 0.26 & - & - & 64.33 & 0.22 & - & - & 64.47 & 0.22 & 64.20 & 0.23 & 57.28 & 0.25 & 64.92 & 0.23 & 60.81 & 0.24 & 65.47 & 0.22 \\
        & KaLM-Embedding-12B & 47.61 & 0.11 & - & - & 57.65 & 0.07 & - & - & 58.18 & 0.07 & 57.85 & 0.07 & 50.90 & 0.10 & 58.23 & 0.07 & 54.92 & 0.08 & 58.54 & 0.07 \\
        \midrule
        \multirow{22}{*}{\rotatebox{90}{Q1(512)}} & \multicolumn{9}{l}{\textit{Multilingual Retrieval}} \\
        \cmidrule(lr){2-22}
        & gte-multilingual-base & 49.55 & 0.33 & 61.52 & 0.22 & 61.47 & 0.21 & 73.24 & 0.07 & 64.12 & 0.20 & 61.56 & 0.22 & 54.06 & 0.28 & 62.82 & 0.19 & 59.86 & 0.24 & 64.62 & 0.18 \\
         & bge-m3 & 46.29 & 0.40 & 62.37 & 0.34 & 59.89 & 0.21 & 66.00 & 0.23 & 58.53 & 0.28 & 56.63 & 0.29 & 52.32 & 0.32 & 58.36 & 0.28 & 53.41 & 0.31 & 57.82 & 0.29 \\
         & Qwen3-Embedding-0.6B & 50.25 & 0.35 & 69.10 & 0.19 & 62.12 & 0.21 & 77.34 & 0.10 & 63.80 & 0.21 & 62.89 & 0.18 & 54.66 & 0.23 & 64.43 & 0.21 & 59.04 & 0.29 & 65.70 & 0.17 \\
         & inf-retriever-v1-1.5b & 55.72 & 0.32 & 71.77 & 0.16 & 67.87 & 0.22 & 77.89 & 0.07 & 69.96 & 0.18 & 68.72 & 0.22 & 60.70 & 0.24 & 70.10 & 0.21 & 64.80 & 0.23 & 70.66 & 0.18 \\
         & Qwen3-Embedding-4B & 63.67 & 0.18 & 73.96 & 0.17 & 71.79 & 0.13 & 81.54 & 0.11 & 70.39 & 0.12 & 71.76 & 0.12 & 66.46 & 0.14 & 72.93 & 0.12 & 69.65 & 0.13 & 74.16 & 0.09 \\
         & inf-retriever-v1 & 65.89 & 0.19 & 75.98 & 0.14 & 75.03 & 0.14 & 81.09 & 0.07 & 76.27 & 0.13 & 76.08 & 0.16 & 70.56 & 0.20 & 76.73 & 0.11 & 72.39 & 0.13 & 77.08 & 0.12 \\
         & NV-Embed-v2 & 43.77 & 0.54 & 62.36 & 0.20 & 77.81 & 0.12 & 80.81 & 0.08 & 77.95 & 0.11 & 78.12 & 0.10 & 59.23 & 0.37 & 79.26 & 0.09 & 66.76 & 0.16 & 78.72 & 0.09 \\
         & llama-embed-nemotron-8b & 67.74 & 0.21 & 71.37 & 0.24 & 77.14 & 0.11 & 84.68 & 0.06 & 77.09 & 0.11 & 77.15 & 0.11 & 72.35 & 0.19 & 77.93 & 0.11 & 73.97 & 0.14 & 78.13 & 0.11 \\
         & Qwen3-Embedding-8B & 64.85 & 0.15 & 74.56 & 0.19 & 72.10 & 0.12 & 82.50 & 0.12 & 72.20 & 0.11 & 73.31 & 0.12 & 67.32 & 0.12 & 74.38 & 0.09 & 70.31 & 0.13 & 75.27 & 0.08 \\
         & KaLM-Embedding-12B & 68.66 & 0.18 & 66.89 & 0.12 & 76.11 & 0.12 & 79.16 & 0.07 & 75.71 & 0.10 & 75.72 & 0.11 & 70.58 & 0.16 & 75.76 & 0.09 & 74.48 & 0.10 & 77.04 & 0.07 \\
        \cmidrule(lr){2-22}
        & \multicolumn{9}{l}{\textit{Cross-lingual Retrieval}} \\
        \cmidrule(lr){2-22}
         & gte-multilingual-base & 41.11 & 0.25 & - & - & 60.45 & 0.14 & - & - & 63.98 & 0.11 & 61.67 & 0.13 & 44.02 & 0.23 & 64.09 & 0.10 & 55.56 & 0.18 & 64.84 & 0.11 \\
 & bge-m3 & 35.34 & 0.37 & - & - & 54.93 & 0.31 & - & - & 55.23 & 0.30 & 53.57 & 0.31 & 40.79 & 0.41 & 54.91 & 0.31 & 44.64 & 0.38 & 54.00 & 0.29 \\
 & Qwen3-Embedding-0.6B & 45.85 & 0.23 & - & - & 65.33 & 0.11 & - & - & 66.56 & 0.10 & 65.90 & 0.13 & 51.56 & 0.19 & 67.32 & 0.11 & 59.87 & 0.16 & 68.43 & 0.09 \\
 & inf-retriever-v1-1.5b & 43.60 & 0.30 & - & - & 64.51 & 0.17 & - & - & 68.19 & 0.13 & 66.04 & 0.16 & 51.07 & 0.26 & 68.50 & 0.14 & 59.11 & 0.18 & 68.84 & 0.12 \\
 & Qwen3-Embedding-4B & 65.11 & 0.14 & - & - & 75.41 & 0.14 & - & - & 74.92 & 0.12 & 75.53 & 0.13 & 69.10 & 0.13 & 76.11 & 0.13 & 72.29 & 0.12 & 76.99 & 0.12 \\
 & inf-retriever-v1 & 57.93 & 0.20 & - & - & 71.70 & 0.13 & - & - & 73.67 & 0.11 & 72.51 & 0.11 & 63.78 & 0.16 & 73.74 & 0.09 & 68.28 & 0.14 & 74.21 & 0.10 \\
 & NV-Embed-v2 & 20.78 & 0.33 & - & - & 64.61 & 0.11 & - & - & 68.83 & 0.13 & 65.27 & 0.12 & 23.73 & 0.26 & 69.30 & 0.10 & 41.04 & 0.16 & 67.86 & 0.09 \\
 & llama-embed-nemotron-8b & 65.40 & 0.15 & - & - & 76.99 & 0.08 & - & - & 77.66 & 0.09 & 77.59 & 0.09 & 70.22 & 0.15 & 78.29 & 0.07 & 73.22 & 0.12 & 78.62 & 0.07 \\
 & Qwen3-Embedding-8B & 68.63 & 0.12 & - & - & 77.43 & 0.13 & - & - & 77.57 & 0.12 & 77.53 & 0.12 & 72.25 & 0.13 & 78.27 & 0.12 & 74.80 & 0.12 & 78.73 & 0.12 \\
 & KaLM-Embedding-12B & 66.68 & 0.15 & - & - & 74.57 & 0.11 & - & - & 75.01 & 0.09 & 74.95 & 0.09 & 68.85 & 0.12 & 75.43 & 0.08 & 72.79 & 0.09 & 75.30 & 0.07 \\
        \midrule
        \multirow{22}{*}{\rotatebox{90}{Q2(1024)}} & \multicolumn{9}{l}{\textit{Multilingual Retrieval}} \\
        \cmidrule(lr){2-22}
        & gte-multilingual-base & 35.75 & 0.58 & 52.22 & 0.28 & 49.87 & 0.48 & 62.56 & 0.24 & 50.17 & 0.46 & 49.25 & 0.47 & 39.15 & 0.49 & 50.84 & 0.47 & 46.55 & 0.51 & 51.54 & 0.46 \\
 & bge-m3 & 34.85 & 0.51 & 46.42 & 0.37 & 47.28 & 0.40 & 51.56 & 0.37 & 44.15 & 0.43 & 43.06 & 0.44 & 37.91 & 0.43 & 44.71 & 0.45 & 41.04 & 0.46 & 43.80 & 0.44 \\
 & Qwen3-Embedding-0.6B & 43.35 & 0.48 & 56.50 & 0.36 & 54.98 & 0.37 & 65.69 & 0.29 & 56.37 & 0.36 & 55.52 & 0.39 & 46.30 & 0.41 & 56.61 & 0.37 & 51.47 & 0.43 & 57.29 & 0.38 \\
 & inf-retriever-v1-1.5b & 46.65 & 0.39 & 62.60 & 0.22 & 58.51 & 0.30 & 69.11 & 0.17 & 60.21 & 0.27 & 58.98 & 0.29 & 50.97 & 0.30 & 60.83 & 0.30 & 55.46 & 0.28 & 60.68 & 0.26 \\
 & Qwen3-Embedding-4B & 55.19 & 0.36 & 59.42 & 0.32 & 64.72 & 0.28 & 70.65 & 0.24 & 63.50 & 0.29 & 64.69 & 0.28 & 57.69 & 0.30 & 65.58 & 0.29 & 61.23 & 0.31 & 66.40 & 0.28 \\
 & inf-retriever-v1 & 55.56 & 0.25 & 67.03 & 0.22 & 65.75 & 0.21 & 72.93 & 0.16 & 66.37 & 0.21 & 66.08 & 0.19 & 58.47 & 0.25 & 67.28 & 0.22 & 63.62 & 0.19 & 67.24 & 0.21 \\
 & NV-Embed-v2 & 11.80 & 0.84 & 41.32 & 0.61 & 54.51 & 0.47 & 72.56 & 0.11 & 57.53 & 0.43 & 55.16 & 0.53 & 24.85 & 0.77 & 56.90 & 0.47 & 37.44 & 0.67 & 59.14 & 0.43 \\
 & llama-embed-nemotron-8b & 55.91 & 0.22 & 59.46 & 0.20 & 66.67 & 0.16 & 73.56 & 0.14 & 65.87 & 0.18 & 65.97 & 0.16 & 58.45 & 0.19 & 66.80 & 0.18 & 62.00 & 0.17 & 66.88 & 0.16 \\
 & Qwen3-Embedding-8B & 58.28 & 0.26 & 59.76 & 0.31 & 65.66 & 0.23 & 71.31 & 0.21 & 65.61 & 0.19 & 65.96 & 0.21 & 59.36 & 0.21 & 66.67 & 0.21 & 63.13 & 0.25 & 67.58 & 0.21 \\
 & KaLM-Embedding-12B & 43.63 & 0.33 & 54.69 & 0.21 & 56.04 & 0.25 & 71.77 & 0.17 & 54.72 & 0.26 & 54.35 & 0.25 & 44.52 & 0.33 & 55.23 & 0.27 & 52.84 & 0.26 & 58.59 & 0.25 \\
        \cmidrule(lr){2-22}
        & \multicolumn{9}{l}{\textit{Cross-lingual Retrieval}} \\
        \cmidrule(lr){2-22}
         & gte-multilingual-base & 31.88 & 0.41 & - & - & 51.87 & 0.29 & - & - & 54.34 & 0.28 & 52.38 & 0.30 & 33.13 & 0.39 & 54.10 & 0.27 & 45.32 & 0.34 & 54.80 & 0.30 \\
 & bge-m3 & 23.61 & 0.56 & - & - & 43.68 & 0.43 & - & - & 41.91 & 0.42 & 40.63 & 0.46 & 27.17 & 0.51 & 41.57 & 0.42 & 31.54 & 0.53 & 40.26 & 0.45 \\
 & Qwen3-Embedding-0.6B & 34.76 & 0.47 & - & - & 54.07 & 0.35 & - & - & 55.18 & 0.35 & 54.22 & 0.33 & 39.07 & 0.41 & 55.64 & 0.34 & 47.54 & 0.40 & 56.29 & 0.33 \\
 & inf-retriever-v1-1.5b & 34.80 & 0.37 & - & - & 57.08 & 0.26 & - & - & 59.37 & 0.23 & 57.53 & 0.24 & 42.48 & 0.29 & 59.63 & 0.23 & 49.28 & 0.27 & 59.79 & 0.22 \\
 & Qwen3-Embedding-4B & 52.24 & 0.34 & - & - & 63.89 & 0.29 & - & - & 62.95 & 0.30 & 63.62 & 0.30 & 55.76 & 0.33 & 64.16 & 0.29 & 59.46 & 0.31 & 65.05 & 0.28 \\
 & inf-retriever-v1 & 52.74 & 0.28 & - & - & 65.53 & 0.19 & - & - & 66.86 & 0.20 & 66.05 & 0.21 & 56.76 & 0.24 & 67.24 & 0.21 & 62.12 & 0.23 & 67.37 & 0.22 \\
 & NV-Embed-v2 & 24.07 & 0.22 & - & - & 60.85 & 0.13 & - & - & 62.60 & 0.11 & 62.05 & 0.13 & 22.51 & 0.15 & 63.08 & 0.11 & 42.00 & 0.16 & 62.89 & 0.12 \\
 & llama-embed-nemotron-8b & 52.25 & 0.26 & - & - & 65.85 & 0.18 & - & - & 66.03 & 0.17 & 65.97 & 0.18 & 55.72 & 0.22 & 66.50 & 0.18 & 59.91 & 0.23 & 66.72 & 0.17 \\
 & Qwen3-Embedding-8B & 54.79 & 0.28 & - & - & 65.10 & 0.25 & - & - & 65.02 & 0.26 & 64.63 & 0.26 & 57.65 & 0.28 & 65.50 & 0.24 & 61.34 & 0.27 & 66.15 & 0.24 \\
 & KaLM-Embedding-12B & 56.99 & 0.27 & - & - & 66.42 & 0.21 & - & - & 66.66 & 0.21 & 66.38 & 0.21 & 59.67 & 0.24 & 67.06 & 0.21 & 63.86 & 0.23 & 66.89 & 0.21 \\
        \midrule
        \multirow{22}{*}{\rotatebox{90}{Q3(1536)}} & \multicolumn{9}{l}{\textit{Multilingual Retrieval}} \\
        \cmidrule(lr){2-22}
        & gte-multilingual-base & 24.71 & 0.74 & 45.26 & 0.29 & 40.32 & 0.58 & 54.46 & 0.32 & 41.32 & 0.58 & 40.12 & 0.62 & 27.32 & 0.70 & 41.18 & 0.59 & 36.08 & 0.62 & 41.30 & 0.60 \\
 & bge-m3 & 27.38 & 0.61 & 39.08 & 0.40 & 38.73 & 0.43 & 42.27 & 0.44 & 35.67 & 0.48 & 34.03 & 0.51 & 30.06 & 0.50 & 36.39 & 0.50 & 33.23 & 0.49 & 34.44 & 0.52 \\
 & Qwen3-Embedding-0.6B & 37.40 & 0.57 & 49.86 & 0.39 & 49.06 & 0.47 & 57.28 & 0.39 & 50.53 & 0.46 & 49.69 & 0.48 & 40.35 & 0.53 & 51.00 & 0.47 & 44.60 & 0.52 & 51.30 & 0.47 \\
 & inf-retriever-v1-1.5b & 41.36 & 0.47 & 59.90 & 0.19 & 54.13 & 0.31 & 63.99 & 0.16 & 55.98 & 0.33 & 54.67 & 0.36 & 46.88 & 0.41 & 56.03 & 0.33 & 50.07 & 0.35 & 56.01 & 0.35 \\
 & Qwen3-Embedding-4B & 48.34 & 0.47 & 51.28 & 0.41 & 59.17 & 0.37 & 62.94 & 0.35 & 58.01 & 0.38 & 59.32 & 0.39 & 51.92 & 0.41 & 60.35 & 0.37 & 55.18 & 0.41 & 60.85 & 0.36 \\
 & inf-retriever-v1 & 48.63 & 0.19 & 64.91 & 0.19 & 60.67 & 0.16 & 67.77 & 0.15 & 61.25 & 0.17 & 60.18 & 0.18 & 52.40 & 0.17 & 62.16 & 0.15 & 57.94 & 0.15 & 62.32 & 0.17 \\
 & NV-Embed-v2 & 1.70 & 0.98 & 22.11 & 0.78 & 30.69 & 0.69 & 61.03 & 0.44 & 33.03 & 0.64 & 29.67 & 0.68 & 5.10 & 0.96 & 31.67 & 0.69 & 14.19 & 0.83 & 34.16 & 0.66 \\
 & llama-embed-nemotron-8b & 47.95 & 0.20 & 54.18 & 0.12 & 60.28 & 0.14 & 66.87 & 0.12 & 59.14 & 0.15 & 59.32 & 0.14 & 51.17 & 0.18 & 60.48 & 0.16 & 55.26 & 0.14 & 60.04 & 0.16 \\
 & Qwen3-Embedding-8B & 53.11 & 0.31 & 50.48 & 0.42 & 61.69 & 0.28 & 62.25 & 0.35 & 61.40 & 0.26 & 61.51 & 0.28 & 55.27 & 0.29 & 62.54 & 0.32 & 58.77 & 0.30 & 62.93 & 0.30 \\
 & KaLM-Embedding-12B & 26.45 & 0.30 & 37.66 & 0.28 & 36.71 & 0.22 & 50.08 & 0.23 & 35.00 & 0.27 & 34.90 & 0.30 & 24.21 & 0.36 & 32.93 & 0.26 & 35.65 & 0.25 & 37.52 & 0.30 \\
        \cmidrule(lr){2-22}
        & \multicolumn{9}{l}{\textit{Cross-lingual Retrieval}} \\
        \cmidrule(lr){2-22}
         & gte-multilingual-base & 24.59 & 0.42 & - & - & 45.21 & 0.34 & - & - & 47.21 & 0.33 & 45.94 & 0.36 & 25.56 & 0.34 & 47.43 & 0.35 & 37.18 & 0.37 & 47.94 & 0.35 \\
 & bge-m3 & 17.04 & 0.64 & - & - & 36.77 & 0.46 & - & - & 34.46 & 0.48 & 32.87 & 0.50 & 20.65 & 0.55 & 33.85 & 0.51 & 24.59 & 0.51 & 32.20 & 0.53 \\
 & Qwen3-Embedding-0.6B & 27.73 & 0.57 & - & - & 47.01 & 0.46 & - & - & 48.26 & 0.49 & 47.34 & 0.47 & 31.08 & 0.50 & 48.47 & 0.45 & 39.70 & 0.46 & 48.85 & 0.46 \\
 & inf-retriever-v1-1.5b & 30.59 & 0.27 & - & - & 53.46 & 0.24 & - & - & 55.78 & 0.22 & 53.70 & 0.23 & 37.07 & 0.29 & 55.22 & 0.25 & 45.03 & 0.20 & 55.60 & 0.24 \\
 & Qwen3-Embedding-4B & 44.18 & 0.47 & - & - & 56.18 & 0.41 & - & - & 55.86 & 0.45 & 56.46 & 0.43 & 47.26 & 0.39 & 57.26 & 0.43 & 51.86 & 0.44 & 57.72 & 0.42 \\
 & inf-retriever-v1 & 49.09 & 0.23 & - & - & 61.91 & 0.18 & - & - & 63.02 & 0.18 & 62.29 & 0.16 & 51.99 & 0.19 & 63.36 & 0.18 & 58.40 & 0.19 & 63.62 & 0.18 \\
 & NV-Embed-v2 & 19.80 & 0.61 & - & - & 51.56 & 0.48 & - & - & 51.97 & 0.47 & 52.47 & 0.47 & 17.34 & 0.64 & 52.57 & 0.46 & 34.65 & 0.54 & 53.02 & 0.49 \\
 & llama-embed-nemotron-8b & 44.91 & 0.19 & - & - & 59.04 & 0.17 & - & - & 59.26 & 0.13 & 59.46 & 0.14 & 47.49 & 0.14 & 59.94 & 0.18 & 52.96 & 0.17 & 59.91 & 0.15 \\
 & Qwen3-Embedding-8B & 43.82 & 0.47 & - & - & 56.09 & 0.37 & - & - & 56.22 & 0.39 & 55.92 & 0.39 & 47.01 & 0.43 & 56.41 & 0.42 & 51.55 & 0.40 & 57.03 & 0.39 \\
 & KaLM-Embedding-12B & 28.01 & 0.33 & - & - & 40.94 & 0.26 & - & - & 41.78 & 0.27 & 41.24 & 0.27 & 32.68 & 0.27 & 41.07 & 0.28 & 37.02 & 0.31 & 42.28 & 0.29 \\
        \bottomrule
        \end{tabular}
    }

\end{table}

\begin{table}[t]
    \centering
        \caption{Detailed evaluation results of 10 multilingual retrieval models on \posir (Part 2 of 2). In both multilingual retrieval and cross-lingual retrieval (translated queries retrieving English documents) settings, the results for each language are weighted-averaged across 31 domains. Q1--Q4 represent query buckets partitioned by the token length of positive documents (512-token intervals). ``KaLM-Embedding-12B'' denotes the ``KaLM-Embedding-Gemma3-12B-2511'' model.
    }
    \label{tab:detailed_experiment_results_part2}
    \setlength{\extrarowheight}{2pt}
    \resizebox{\linewidth}{!}{
        \begin{tabular}{c l | c c |c c|c c|c c|c c|c c|c c|c c|c c|c c}
        \toprule
        & \multirow{2}{*}{\textbf{Model}} & \multicolumn{2}{c}{\textbf{Arabic}}  & \multicolumn{2}{c}{\textbf{Chinese}}  & \multicolumn{2}{c}{\textbf{German}}  & \multicolumn{2}{c}{\textbf{English}} & \multicolumn{2}{c}{\textbf{French}} & \multicolumn{2}{c}{\textbf{Italian}} & \multicolumn{2}{c}{\textbf{Korean}} & \multicolumn{2}{c}{\textbf{Portuguese}} & \multicolumn{2}{c}{\textbf{Russian}} & \multicolumn{2}{c}{\textbf{Spanish}} \\
        & & nDCG@10 & PSI & nDCG@10 & PSI & nDCG@10 & PSI & nDCG@10 & PSI & nDCG@10 & PSI & nDCG@10 & PSI & nDCG@10 & PSI & nDCG@10 & PSI & nDCG@10 & PSI & nDCG@10 & PSI \\
        \midrule
        \multirow{22}{*}{\rotatebox{90}{Q4(2048)}} & \multicolumn{9}{l}{\textit{Multilingual Retrieval}} \\
        \cmidrule(lr){2-22}
& gte-multilingual-base & 18.41 & 0.78 & 39.21 & 0.40 & 32.10 & 0.67 & 48.46 & 0.38 & 35.11 & 0.62 & 31.76 & 0.67 & 20.19 & 0.74 & 32.57 & 0.64 & 28.78 & 0.69 & 33.50 & 0.63 \\
 & bge-m3 & 22.82 & 0.48 & 34.22 & 0.42 & 32.62 & 0.42 & 34.30 & 0.40 & 30.98 & 0.40 & 28.76 & 0.46 & 25.69 & 0.44 & 30.57 & 0.47 & 28.17 & 0.42 & 29.08 & 0.46 \\
 & Qwen3-Embedding-0.6B & 34.18 & 0.57 & 43.84 & 0.44 & 44.13 & 0.51 & 50.39 & 0.39 & 45.78 & 0.47 & 45.08 & 0.49 & 37.03 & 0.52 & 45.55 & 0.48 & 41.43 & 0.53 & 45.79 & 0.49 \\
 & inf-retriever-v1-1.5b & 39.18 & 0.36 & 57.73 & 0.14 & 51.06 & 0.30 & 61.13 & 0.16 & 53.37 & 0.26 & 51.65 & 0.30 & 44.03 & 0.32 & 52.83 & 0.28 & 47.65 & 0.26 & 53.40 & 0.31 \\
 & Qwen3-Embedding-4B & 43.76 & 0.48 & 43.94 & 0.49 & 53.04 & 0.41 & 54.94 & 0.41 & 52.09 & 0.43 & 54.05 & 0.42 & 47.61 & 0.47 & 54.40 & 0.40 & 50.46 & 0.45 & 55.29 & 0.39 \\
 & inf-retriever-v1 & 43.55 & 0.20 & 63.03 & 0.17 & 55.31 & 0.16 & 63.61 & 0.15 & 56.13 & 0.20 & 54.21 & 0.16 & 46.81 & 0.20 & 55.77 & 0.17 & 52.39 & 0.20 & 55.99 & 0.16 \\
 & NV-Embed-v2 & 0.21 & 1.00 & 9.78 & 0.92 & 20.25 & 0.74 & 47.37 & 0.57 & 18.90 & 0.74 & 16.97 & 0.74 & 0.66 & 1.00 & 19.00 & 0.75 & 9.18 & 0.85 & 20.42 & 0.75 \\
 & llama-embed-nemotron-8b & 43.13 & 0.27 & 49.47 & 0.28 & 54.75 & 0.21 & 61.77 & 0.15 & 54.05 & 0.20 & 54.11 & 0.21 & 45.84 & 0.24 & 55.01 & 0.19 & 49.57 & 0.25 & 55.13 & 0.20 \\
 & Qwen3-Embedding-8B & 48.73 & 0.33 & 43.69 & 0.48 & 56.83 & 0.30 & 54.68 & 0.40 & 56.97 & 0.27 & 56.64 & 0.30 & 51.44 & 0.30 & 57.14 & 0.32 & 54.86 & 0.32 & 57.76 & 0.33 \\
 & KaLM-Embedding-12B & 20.88 & 0.40 & 30.88 & 0.29 & 27.99 & 0.29 & 41.33 & 0.17 & 27.51 & 0.30 & 26.83 & 0.36 & 18.55 & 0.37 & 24.85 & 0.35 & 28.67 & 0.34 & 29.00 & 0.30 \\
        \cmidrule(lr){2-22}
        & \multicolumn{9}{l}{\textit{Cross-lingual Retrieval}} \\
        \cmidrule(lr){2-22}
 & gte-multilingual-base & 21.80 & 0.49 & - & - & 39.50 & 0.41 & - & - & 41.98 & 0.41 & 40.45 & 0.43 & 22.27 & 0.47 & 41.37 & 0.40 & 32.99 & 0.43 & 42.19 & 0.40 \\
 & bge-m3 & 14.47 & 0.53 & - & - & 31.06 & 0.44 & - & - & 28.02 & 0.42 & 26.87 & 0.46 & 17.50 & 0.53 & 26.98 & 0.45 & 19.85 & 0.50 & 25.76 & 0.45 \\
 & Qwen3-Embedding-0.6B & 23.99 & 0.61 & - & - & 41.08 & 0.48 & - & - & 42.09 & 0.45 & 41.08 & 0.47 & 27.83 & 0.59 & 41.76 & 0.48 & 34.97 & 0.48 & 42.65 & 0.50 \\
 & inf-retriever-v1-1.5b & 30.48 & 0.24 & - & - & 50.96 & 0.18 & - & - & 53.05 & 0.14 & 51.61 & 0.13 & 36.60 & 0.27 & 52.93 & 0.17 & 43.84 & 0.20 & 53.06 & 0.19 \\
 & Qwen3-Embedding-4B & 37.33 & 0.50 & - & - & 48.60 & 0.44 & - & - & 48.43 & 0.45 & 48.86 & 0.44 & 40.74 & 0.52 & 49.31 & 0.46 & 44.62 & 0.48 & 49.84 & 0.44 \\
 & inf-retriever-v1 & 46.25 & 0.18 & - & - & 58.48 & 0.16 & - & - & 59.01 & 0.16 & 58.27 & 0.16 & 49.89 & 0.19 & 58.80 & 0.13 & 54.42 & 0.16 & 59.36 & 0.14 \\
 & NV-Embed-v2 & 16.81 & 0.66 & - & - & 41.08 & 0.59 & - & - & 40.98 & 0.60 & 40.78 & 0.59 & 14.19 & 0.70 & 40.48 & 0.59 & 27.57 & 0.64 & 41.57 & 0.59 \\
 & llama-embed-nemotron-8b & 41.60 & 0.19 & - & - & 54.61 & 0.18 & - & - & 54.99 & 0.17 & 55.18 & 0.17 & 44.35 & 0.20 & 55.45 & 0.17 & 48.75 & 0.19 & 55.58 & 0.20 \\
 & Qwen3-Embedding-8B & 37.03 & 0.47 & - & - & 48.14 & 0.44 & - & - & 48.58 & 0.42 & 48.04 & 0.45 & 40.78 & 0.48 & 48.71 & 0.45 & 44.48 & 0.45 & 49.36 & 0.45 \\
 & KaLM-Embedding-12B & 20.69 & 0.26 & - & - & 32.23 & 0.22 & - & - & 32.94 & 0.21 & 32.26 & 0.25 & 25.30 & 0.26 & 32.42 & 0.23 & 28.67 & 0.24 & 33.43 & 0.22 \\
        \bottomrule
        \end{tabular}
    }

\end{table}

\begin{table}[!h]
    \centering
    \caption{Detailed information on all of the models appearing in our paper.}
    \label{tab:model_information}
    \small
    \resizebox{\textwidth}{!}{
        \begin{tabular}{l|c|c|c}
        \toprule
        \textbf{Model} & \textbf{Size} & \textbf{Open-Sourced} & \textbf{Model Link} \\
        \midrule
        \multicolumn{4}{l}{\textit{Tokenizer}} \\
        \midrule
        Qwen3 tokenizer~\cite{yang2025qwen3technicalreport} & - & \Checkmark & \url{https://huggingface.co/Qwen/Qwen3-30B-A3B-Instruct-2507/blob/main/tokenizer.json} \\
        \midrule
        \multicolumn{4}{l}{\textit{Embedding Model}} \\
        \midrule
        gte-multilingual-base~\cite{zhang-etal-2024-mgte} & 305M & \Checkmark & \url{https://huggingface.co/Alibaba-NLP/gte-multilingual-base} \\
        bge-m3~\cite{chen-etal-2024-m3} & 568M & \Checkmark & \url{https://huggingface.co/BAAI/bge-m3} \\
        Qwen3-Embedding-0.6B~\cite{zhang2025qwen3embeddingadvancingtext} & 595M & \Checkmark & \url{https://huggingface.co/Qwen/Qwen3-Embedding-0.6B} \\
        inf-retriever-v1-1.5b & 2B & \Checkmark & \url{https://huggingface.co/infly/inf-retriever-v1-1.5b} \\
        Qwen3-Embedding-4B~\cite{zhang2025qwen3embeddingadvancingtext} & 4B & \Checkmark & \url{https://huggingface.co/Qwen/Qwen3-Embedding-4B} \\
        inf-retriever-v1 & 7B & \Checkmark & \url{https://huggingface.co/infly/inf-retriever-v1} \\
        QZhou-Embedding~\cite{yu2025qzhouembeddingtechnicalreport} & 7B & \Checkmark & \url{https://huggingface.co/Kingsoft-LLM/QZhou-Embedding} \\
        NV-Embed-v2~\cite{lee2025nvembed} & 8B  & \Checkmark & \url{https://huggingface.co/nvidia/NV-Embed-v2} \\
        llama-embed-nemotron-8b~\cite{babakhin2025llamaembednemotron8buniversaltextembedding} & 8B & \Checkmark & \url{https://huggingface.co/nvidia/llama-embed-nemotron-8b} \\
         Qwen3-Embedding-8B~\cite{zhang2025qwen3embeddingadvancingtext} & 8B & \Checkmark & \url{https://huggingface.co/Qwen/Qwen3-Embedding-8B} \\
         KaLM-Embedding-Gemma3-12B-2511~\cite{zhao2025kalmembeddingv2superiortrainingtechniques} & 12B & \Checkmark & \url{https://huggingface.co/tencent/KaLM-Embedding-Gemma3-12B-2511} \\
        \midrule
        \multicolumn{4}{l}{\textit{Re-ranking Model}} \\
        \midrule
        bge-reranker-v2-m3~\cite{chen-etal-2024-m3} & 0.6B & \Checkmark & \url{https://huggingface.co/BAAI/bge-reranker-v2-m3} \\
        Qwen3-Reranker-0.6B~\cite{zhang2025qwen3embeddingadvancingtext} & 0.6B  & \Checkmark & \url{https://huggingface.co/Qwen/Qwen3-Reranker-0.6B} \\
        bge-reranker-v2-minicpm-layerwise~\cite{chen-etal-2024-m3} & 3B  & \Checkmark & \url{https://huggingface.co/BAAI/bge-reranker-v2-minicpm-layerwise} \\
        Qwen3-Reranker-4B~\cite{zhang2025qwen3embeddingadvancingtext} & 4B  & \Checkmark & \url{https://huggingface.co/Qwen/Qwen3-Reranker-4B} \\
        \midrule
        \multicolumn{4}{l}{\textit{Large Language Model}} \\
        \midrule
        Hunyuan-MT-7B~\cite{zheng2025hunyuanmttechnicalreport} & 8B  & \Checkmark & \url{https://huggingface.co/tencent/Hunyuan-MT-7B} \\
        Qwen3-30B-A3B-Instruct-2507~\cite{yang2025qwen3technicalreport} & 31B  & \Checkmark & \url{https://huggingface.co/Qwen/Qwen3-30B-A3B-Instruct-2507} \\
        DeepSeek-V3.1~\cite{deepseekai2025deepseekv3technicalreport} & 685B & \Checkmark & \url{https://huggingface.co/deepseek-ai/DeepSeek-V3.1} \\
        GPT-4o~\cite{openai2024gpt4technicalreport} & - & \XSolidBrush & \url{https://platform.openai.com/docs/models/gpt-4o} \\
        \midrule
        \multicolumn{4}{l}{\textit{Online Service}} \\
        \midrule
        Google Translate & - & \XSolidBrush & \url{https://translate.google.com} \\
        \bottomrule
        \end{tabular}
    }

\end{table}

\begin{table}[htbp]
\centering
\caption{Statistics for domain: Accommodation Catering Hotel (Part 1 of 31).}
\label{tab:dataset_stats_accommodationcateringhotel}
\small
\resizebox{\textwidth}{!}{
\begin{tabular}{l|c|c|c|c|c|c}
\toprule
\textbf{Language} & \textbf{\#Corpus} & \textbf{\begin{tabular}[c]{@{}c@{}}Avg. Tokens\\per Document\end{tabular}} & \textbf{\begin{tabular}[c]{@{}c@{}}Ratio\\(Rel. English)\end{tabular}} & \textbf{\#Queries} & \textbf{\begin{tabular}[c]{@{}c@{}}Avg. Tokens\\per Query\end{tabular}} & \textbf{\begin{tabular}[c]{@{}c@{}}Ratio\\(Rel. English)\end{tabular}} \\
\midrule
Arabic & 42,596 & 1511.2 & 166.0\% & 1458 & 24.4 & 193.6\% \\
Chinese & 55,969 & 1009.2 & - & 1488 & 11.8 & --- \\
German & 42,690 & 1339.6 & 147.1\% & 1459 & 20.7 & 164.2\% \\
English & 42,708 & 910.6 & - & 1459 & 12.6 & --- \\
French & 42,699 & 1342.6 & 147.4\% & 1459 & 22.0 & 174.6\% \\
Italian & 42,690 & 1360.8 & 149.4\% & 1459 & 21.5 & 170.0\% \\
Korean & 42,602 & 1543.8 & 169.5\% & 1456 & 26.2 & 207.4\% \\
Portuguese & 42,696 & 1264.2 & 138.8\% & 1459 & 19.8 & 156.5\% \\
Russian & 42,671 & 1503.1 & 165.1\% & 1456 & 24.4 & 193.1\% \\
Spanish & 42,672 & 1250.5 & 137.3\% & 1459 & 20.7 & 163.8\% \\
\bottomrule
\end{tabular}
}

\end{table}


\begin{table}[htbp]
\centering
\caption{Statistics for domain: Aerospace (Part 2 of 31).}
\small
\resizebox{\textwidth}{!}{
\begin{tabular}{l|c|c|c|c|c|c}
\toprule
\textbf{Language} & \textbf{\#Corpus} & \textbf{\begin{tabular}[c]{@{}c@{}}Avg. Tokens\\per Document\end{tabular}} & \textbf{\begin{tabular}[c]{@{}c@{}}Ratio\\(Rel. English)\end{tabular}} & \textbf{\#Queries} & \textbf{\begin{tabular}[c]{@{}c@{}}Avg. Tokens\\per Query\end{tabular}} & \textbf{\begin{tabular}[c]{@{}c@{}}Ratio\\(Rel. English)\end{tabular}} \\
\midrule
Arabic & 57,393 & 1520.6 & 165.7\% & 1339 & 26.0 & 193.7\% \\
Chinese & 63,398 & 933.5 & - & 1492 & 12.8 & --- \\
German & 57,655 & 1394.7 & 152.0\% & 1340 & 22.9 & 170.7\% \\
English & 57,671 & 917.4 & - & 1340 & 13.4 & --- \\
French & 57,661 & 1387.6 & 151.2\% & 1340 & 24.2 & 180.7\% \\
Italian & 57,658 & 1421.0 & 154.9\% & 1340 & 23.7 & 176.7\% \\
Korean & 57,545 & 1516.9 & 165.3\% & 1335 & 27.0 & 201.5\% \\
Portuguese & 57,649 & 1319.8 & 143.9\% & 1340 & 21.7 & 161.8\% \\
Russian & 57,643 & 1577.4 & 171.9\% & 1340 & 26.5 & 197.6\% \\
Spanish & 57,660 & 1312.5 & 143.1\% & 1340 & 22.7 & 169.4\% \\
\bottomrule
\end{tabular}
}
\label{tab:dataset_stats_aerospace}
\end{table}


\begin{table}[htbp]
\centering
\caption{Statistics for domain: Agriculture Forestry Animal Husbandry Fishery (Part 3 of 31).}
\label{tab:dataset_stats_agricultureforestryanimalhusbandryfishery}
\small
\resizebox{\textwidth}{!}{
\begin{tabular}{l|c|c|c|c|c|c}
\toprule
\textbf{Language} & \textbf{\#Corpus} & \textbf{\begin{tabular}[c]{@{}c@{}}Avg. Tokens\\per Document\end{tabular}} & \textbf{\begin{tabular}[c]{@{}c@{}}Ratio\\(Rel. English)\end{tabular}} & \textbf{\#Queries} & \textbf{\begin{tabular}[c]{@{}c@{}}Avg. Tokens\\per Query\end{tabular}} & \textbf{\begin{tabular}[c]{@{}c@{}}Ratio\\(Rel. English)\end{tabular}} \\
\midrule
Arabic & 57,327 & 1550.9 & 166.8\% & 1613 & 26.0 & 195.8\% \\
Chinese & 64,363 & 930.0 & - & 1649 & 11.8 & --- \\
German & 57,466 & 1423.2 & 153.1\% & 1615 & 23.1 & 173.9\% \\
English & 57,490 & 929.6 & - & 1615 & 13.3 & --- \\
French & 57,469 & 1426.3 & 153.4\% & 1615 & 24.6 & 185.4\% \\
Italian & 57,477 & 1445.9 & 155.5\% & 1615 & 23.8 & 179.0\% \\
Korean & 57,362 & 1539.8 & 165.6\% & 1615 & 26.9 & 202.5\% \\
Portuguese & 57,474 & 1338.9 & 144.0\% & 1615 & 21.9 & 164.6\% \\
Russian & 57,438 & 1597.3 & 171.8\% & 1615 & 26.6 & 200.6\% \\
Spanish & 57,463 & 1332.1 & 143.3\% & 1608 & 23.0 & 172.9\% \\
\bottomrule
\end{tabular}}

\end{table}


\begin{table}[htbp]
\centering
\caption{Statistics for domain: Artificial Intelligence Machine Learning (Part 4 of 31).}
\label{tab:dataset_stats_artificialintelligencemachinelearning}
\small
\resizebox{\textwidth}{!}{
\begin{tabular}{l|c|c|c|c|c|c}
\toprule
\textbf{Language} & \textbf{\#Corpus} & \textbf{\begin{tabular}[c]{@{}c@{}}Avg. Tokens\\per Document\end{tabular}} & \textbf{\begin{tabular}[c]{@{}c@{}}Ratio\\(Rel. English)\end{tabular}} & \textbf{\#Queries} & \textbf{\begin{tabular}[c]{@{}c@{}}Avg. Tokens\\per Query\end{tabular}} & \textbf{\begin{tabular}[c]{@{}c@{}}Ratio\\(Rel. English)\end{tabular}} \\
\midrule
Arabic & 54,352 & 1460.8 & 160.0\% & 1079 & 26.2 & 207.6\% \\
Chinese & 57,698 & 1008.1 & - & 1419 & 12.2 & --- \\
German & 54,918 & 1358.2 & 148.7\% & 1081 & 23.1 & 183.2\% \\
English & 54,969 & 913.1 & - & 1083 & 12.6 & --- \\
French & 54,921 & 1331.2 & 145.8\% & 1083 & 24.2 & 191.9\% \\
Italian & 54,931 & 1356.8 & 148.6\% & 1081 & 23.6 & 187.0\% \\
Korean & 54,786 & 1423.8 & 155.9\% & 1081 & 25.8 & 204.9\% \\
Portuguese & 54,927 & 1255.0 & 137.4\% & 1083 & 21.0 & 166.8\% \\
Russian & 54,904 & 1435.1 & 157.2\% & 1083 & 25.4 & 201.6\% \\
Spanish & 54,912 & 1241.2 & 135.9\% & 1081 & 22.0 & 174.5\% \\
\bottomrule
\end{tabular}}

\end{table}


\begin{table}[htbp]
\centering
\caption{Statistics for domain: Automobile (Part 5 of 31).}
\label{tab:dataset_stats_automobile}
\small
\resizebox{\textwidth}{!}{
\begin{tabular}{l|c|c|c|c|c|c}
\toprule
\textbf{Language} & \textbf{\#Corpus} & \textbf{\begin{tabular}[c]{@{}c@{}}Avg. Tokens\\per Document\end{tabular}} & \textbf{\begin{tabular}[c]{@{}c@{}}Ratio\\(Rel. English)\end{tabular}} & \textbf{\#Queries} & \textbf{\begin{tabular}[c]{@{}c@{}}Avg. Tokens\\per Query\end{tabular}} & \textbf{\begin{tabular}[c]{@{}c@{}}Ratio\\(Rel. English)\end{tabular}} \\
\midrule
Arabic & 55,866 & 1471.7 & 165.7\% & 1300 & 26.0 & 196.2\% \\
Chinese & 62,074 & 943.6 & - & 1329 & 12.3 & --- \\
German & 55,929 & 1341.8 & 151.1\% & 1299 & 22.6 & 170.7\% \\
English & 55,948 & 888.1 & - & 1303 & 13.3 & --- \\
French & 55,933 & 1312.9 & 147.8\% & 1303 & 24.0 & 180.6\% \\
Italian & 55,931 & 1367.2 & 154.0\% & 1303 & 23.6 & 178.0\% \\
Korean & 55,846 & 1452.9 & 163.6\% & 1301 & 27.3 & 205.7\% \\
Portuguese & 55,936 & 1257.3 & 141.6\% & 1303 & 21.5 & 162.2\% \\
Russian & 55,922 & 1462.4 & 164.7\% & 1303 & 25.6 & 192.6\% \\
Spanish & 55,932 & 1251.5 & 140.9\% & 1303 & 22.7 & 170.9\% \\
\bottomrule
\end{tabular}}

\end{table}


\begin{table}[htbp]
\centering
\caption{Statistics for domain: Biomedicine (Part 6 of 31).}
\label{tab:dataset_stats_biomedicine}
\small
\resizebox{\textwidth}{!}{
\begin{tabular}{l|c|c|c|c|c|c}
\toprule
\textbf{Language} & \textbf{\#Corpus} & \textbf{\begin{tabular}[c]{@{}c@{}}Avg. Tokens\\per Document\end{tabular}} & \textbf{\begin{tabular}[c]{@{}c@{}}Ratio\\(Rel. English)\end{tabular}} & \textbf{\#Queries} & \textbf{\begin{tabular}[c]{@{}c@{}}Avg. Tokens\\per Query\end{tabular}} & \textbf{\begin{tabular}[c]{@{}c@{}}Ratio\\(Rel. English)\end{tabular}} \\
\midrule
Arabic & 61,392 & 1546.3 & 167.9\% & 1406 & 27.7 & 200.9\% \\
Chinese & 63,133 & 930.7 & - & 1254 & 12.2 & --- \\
German & 61,546 & 1405.8 & 152.7\% & 1411 & 23.8 & 172.9\% \\
English & 61,582 & 920.8 & - & 1411 & 13.8 & --- \\
French & 61,562 & 1394.0 & 151.4\% & 1411 & 25.2 & 183.2\% \\
Italian & 61,562 & 1410.9 & 153.2\% & 1411 & 24.0 & 174.5\% \\
Korean & 61,465 & 1477.3 & 160.4\% & 1411 & 27.4 & 199.1\% \\
Portuguese & 61,562 & 1318.6 & 143.2\% & 1411 & 22.3 & 162.2\% \\
Russian & 61,545 & 1571.4 & 170.7\% & 1411 & 27.0 & 195.8\% \\
Spanish & 61,561 & 1308.2 & 142.1\% & 1411 & 23.3 & 169.1\% \\
\bottomrule
\end{tabular}}

\end{table}


\begin{table}[htbp]
\centering
\caption{Statistics for domain: Computer Communication (Part 7 of 31).}
\label{tab:dataset_stats_computercommunication}
\small
\resizebox{\textwidth}{!}{
\begin{tabular}{l|c|c|c|c|c|c}
\toprule
\textbf{Language} & \textbf{\#Corpus} & \textbf{\begin{tabular}[c]{@{}c@{}}Avg. Tokens\\per Document\end{tabular}} & \textbf{\begin{tabular}[c]{@{}c@{}}Ratio\\(Rel. English)\end{tabular}} & \textbf{\#Queries} & \textbf{\begin{tabular}[c]{@{}c@{}}Avg. Tokens\\per Query\end{tabular}} & \textbf{\begin{tabular}[c]{@{}c@{}}Ratio\\(Rel. English)\end{tabular}} \\
\midrule
Arabic & 61,435 & 1512.8 & 164.0\% & 1384 & 25.7 & 200.5\% \\
Chinese & 64,004 & 931.5 & - & 1657 & 11.6 & --- \\
German & 61,852 & 1383.2 & 149.9\% & 1388 & 22.7 & 176.8\% \\
English & 61,885 & 922.6 & - & 1389 & 12.8 & --- \\
French & 61,861 & 1352.0 & 146.5\% & 1387 & 23.5 & 183.5\% \\
Italian & 61,856 & 1385.5 & 150.2\% & 1388 & 22.8 & 177.6\% \\
Korean & 61,800 & 1500.8 & 162.7\% & 1387 & 26.6 & 207.0\% \\
Portuguese & 61,860 & 1285.5 & 139.3\% & 1386 & 20.9 & 162.6\% \\
Russian & 61,854 & 1480.0 & 160.4\% & 1389 & 24.6 & 191.8\% \\
Spanish & 61,847 & 1265.8 & 137.2\% & 1389 & 21.8 & 169.6\% \\
\bottomrule
\end{tabular}}

\end{table}


\begin{table}[htbp]
\centering
\caption{Statistics for domain: Computer Programming Code (Part 8 of 31).}
\label{tab:dataset_stats_computerprogrammingcode}
\small
\resizebox{\textwidth}{!}{
\begin{tabular}{l|c|c|c|c|c|c}
\toprule
\textbf{Language} & \textbf{\#Corpus} & \textbf{\begin{tabular}[c]{@{}c@{}}Avg. Tokens\\per Document\end{tabular}} & \textbf{\begin{tabular}[c]{@{}c@{}}Ratio\\(Rel. English)\end{tabular}} & \textbf{\#Queries} & \textbf{\begin{tabular}[c]{@{}c@{}}Avg. Tokens\\per Query\end{tabular}} & \textbf{\begin{tabular}[c]{@{}c@{}}Ratio\\(Rel. English)\end{tabular}} \\
\midrule
Arabic & 50,902 & 1515.4 & 154.0\% & 615 & 24.5 & 198.5\% \\
Chinese & 56,802 & 995.7 & - & 945 & 11.7 & --- \\
German & 51,499 & 1394.2 & 141.7\% & 617 & 21.6 & 175.3\% \\
English & 51,612 & 983.9 & - & 624 & 12.3 & --- \\
French & 51,511 & 1365.0 & 138.7\% & 624 & 22.0 & 178.6\% \\
Italian & 51,544 & 1383.9 & 140.7\% & 618 & 21.6 & 175.0\% \\
Korean & 51,385 & 1485.0 & 150.9\% & 617 & 24.5 & 198.8\% \\
Portuguese & 51,540 & 1284.6 & 130.6\% & 624 & 19.5 & 158.3\% \\
Russian & 51,524 & 1424.9 & 144.8\% & 615 & 22.8 & 184.8\% \\
Spanish & 51,496 & 1272.4 & 129.3\% & 623 & 20.5 & 166.4\% \\
\bottomrule
\end{tabular}}

\end{table}


\begin{table}[htbp]
\centering
\caption{Statistics for domain: Current Affairs Government Administration (Part 9 of 31).}
\label{tab:dataset_stats_currentaffairsgovernmentadministration}
\small
\resizebox{\textwidth}{!}{
\begin{tabular}{l|c|c|c|c|c|c}
\toprule
\textbf{Language} & \textbf{\#Corpus} & \textbf{\begin{tabular}[c]{@{}c@{}}Avg. Tokens\\per Document\end{tabular}} & \textbf{\begin{tabular}[c]{@{}c@{}}Ratio\\(Rel. English)\end{tabular}} & \textbf{\#Queries} & \textbf{\begin{tabular}[c]{@{}c@{}}Avg. Tokens\\per Query\end{tabular}} & \textbf{\begin{tabular}[c]{@{}c@{}}Ratio\\(Rel. English)\end{tabular}} \\
\midrule
Arabic & 60,224 & 1485.1 & 159.9\% & 1270 & 25.5 & 185.5\% \\
Chinese & 63,941 & 927.4 & - & 1280 & 12.3 & --- \\
German & 60,374 & 1442.3 & 155.3\% & 1284 & 24.3 & 176.5\% \\
English & 60,390 & 928.9 & - & 1286 & 13.7 & --- \\
French & 60,375 & 1423.4 & 153.2\% & 1286 & 25.3 & 184.0\% \\
Italian & 60,382 & 1445.6 & 155.6\% & 1286 & 24.5 & 178.3\% \\
Korean & 60,216 & 1536.2 & 165.4\% & 1282 & 27.8 & 202.0\% \\
Portuguese & 60,375 & 1329.8 & 143.2\% & 1286 & 22.3 & 162.6\% \\
Russian & 60,356 & 1604.8 & 172.8\% & 1286 & 27.9 & 203.3\% \\
Spanish & 60,365 & 1319.8 & 142.1\% & 1286 & 23.2 & 168.7\% \\
\bottomrule
\end{tabular}}

\end{table}


\begin{table}[htbp]
\centering
\caption{Statistics for domain: Electric Power Energy (Part 10 of 31).}
\label{tab:dataset_stats_electricpowerenergy}
\small
\resizebox{\textwidth}{!}{
\begin{tabular}{l|c|c|c|c|c|c}
\toprule
\textbf{Language} & \textbf{\#Corpus} & \textbf{\begin{tabular}[c]{@{}c@{}}Avg. Tokens\\per Document\end{tabular}} & \textbf{\begin{tabular}[c]{@{}c@{}}Ratio\\(Rel. English)\end{tabular}} & \textbf{\#Queries} & \textbf{\begin{tabular}[c]{@{}c@{}}Avg. Tokens\\per Query\end{tabular}} & \textbf{\begin{tabular}[c]{@{}c@{}}Ratio\\(Rel. English)\end{tabular}} \\
\midrule
Arabic & 59,471 & 1540.7 & 167.8\% & 1283 & 26.9 & 199.3\% \\
Chinese & 63,487 & 932.5 & - & 1535 & 12.5 & --- \\
German & 59,557 & 1433.1 & 156.0\% & 1285 & 24.0 & 177.8\% \\
English & 59,580 & 918.4 & - & 1288 & 13.5 & --- \\
French & 59,562 & 1436.6 & 156.4\% & 1288 & 25.6 & 189.3\% \\
Italian & 59,561 & 1469.3 & 160.0\% & 1288 & 25.0 & 184.8\% \\
Korean & 59,486 & 1526.2 & 166.2\% & 1288 & 27.5 & 204.0\% \\
Portuguese & 59,568 & 1346.8 & 146.7\% & 1288 & 22.6 & 167.4\% \\
Russian & 59,551 & 1612.0 & 175.5\% & 1288 & 27.9 & 206.6\% \\
Spanish & 59,566 & 1333.6 & 145.2\% & 1288 & 23.5 & 174.2\% \\
\bottomrule
\end{tabular}}

\end{table}


\begin{table}[htbp]
\centering
\caption{Statistics for domain: Film Entertainment (Part 11 of 31).}
\label{tab:dataset_stats_filmentertainment}
\small
\resizebox{\textwidth}{!}{
\begin{tabular}{l|c|c|c|c|c|c}
\toprule
\textbf{Language} & \textbf{\#Corpus} & \textbf{\begin{tabular}[c]{@{}c@{}}Avg. Tokens\\per Document\end{tabular}} & \textbf{\begin{tabular}[c]{@{}c@{}}Ratio\\(Rel. English)\end{tabular}} & \textbf{\#Queries} & \textbf{\begin{tabular}[c]{@{}c@{}}Avg. Tokens\\per Query\end{tabular}} & \textbf{\begin{tabular}[c]{@{}c@{}}Ratio\\(Rel. English)\end{tabular}} \\
\midrule
Arabic & 58,987 & 1530.5 & 164.5\% & 1679 & 30.6 & 227.9\% \\
Chinese & 63,899 & 932.2 & - & 1739 & 12.6 & --- \\
German & 59,235 & 1330.3 & 143.0\% & 1687 & 21.4 & 159.4\% \\
English & 59,250 & 930.3 & - & 1687 & 13.4 & --- \\
French & 59,236 & 1320.7 & 142.0\% & 1687 & 22.6 & 167.9\% \\
Italian & 59,236 & 1330.7 & 143.0\% & 1687 & 21.8 & 162.0\% \\
Korean & 59,079 & 1579.6 & 169.8\% & 1687 & 27.9 & 208.0\% \\
Portuguese & 59,235 & 1246.7 & 134.0\% & 1687 & 20.0 & 148.7\% \\
Russian & 59,205 & 1538.4 & 165.4\% & 1684 & 26.3 & 196.1\% \\
Spanish & 59,213 & 1241.6 & 133.5\% & 1687 & 21.1 & 157.3\% \\
\bottomrule
\end{tabular}}

\end{table}


\begin{table}[htbp]
\centering
\caption{Statistics for domain: Finance Economics (Part 12 of 31).}
\label{tab:dataset_stats_financeeconomics}
\small
\resizebox{\textwidth}{!}{
\begin{tabular}{l|c|c|c|c|c|c}
\toprule
\textbf{Language} & \textbf{\#Corpus} & \textbf{\begin{tabular}[c]{@{}c@{}}Avg. Tokens\\per Document\end{tabular}} & \textbf{\begin{tabular}[c]{@{}c@{}}Ratio\\(Rel. English)\end{tabular}} & \textbf{\#Queries} & \textbf{\begin{tabular}[c]{@{}c@{}}Avg. Tokens\\per Query\end{tabular}} & \textbf{\begin{tabular}[c]{@{}c@{}}Ratio\\(Rel. English)\end{tabular}} \\
\midrule
Arabic & 60,258 & 1518.8 & 163.5\% & 1315 & 26.6 & 193.5\% \\
Chinese & 64,184 & 931.8 & - & 1118 & 12.1 & --- \\
German & 60,373 & 1443.0 & 155.4\% & 1315 & 24.3 & 176.4\% \\
English & 60,395 & 928.7 & - & 1315 & 13.8 & --- \\
French & 60,375 & 1423.6 & 153.3\% & 1315 & 25.5 & 185.2\% \\
Italian & 60,383 & 1453.1 & 156.5\% & 1315 & 24.7 & 179.1\% \\
Korean & 60,287 & 1535.0 & 165.3\% & 1312 & 27.5 & 199.9\% \\
Portuguese & 60,382 & 1338.0 & 144.1\% & 1315 & 22.6 & 164.0\% \\
Russian & 60,358 & 1569.8 & 169.0\% & 1315 & 27.4 & 199.1\% \\
Spanish & 60,369 & 1320.5 & 142.2\% & 1313 & 23.3 & 169.1\% \\
\bottomrule
\end{tabular}}

\end{table}


\begin{table}[htbp]
\centering
\caption{Statistics for domain: Fineweb (Part 13 of 31).}
\label{tab:dataset_stats_fineweb}
\small
\resizebox{\textwidth}{!}{
\begin{tabular}{l|c|c|c|c|c|c}
\toprule
\textbf{Language} & \textbf{\#Corpus} & \textbf{\begin{tabular}[c]{@{}c@{}}Avg. Tokens\\per Document\end{tabular}} & \textbf{\begin{tabular}[c]{@{}c@{}}Ratio\\(Rel. English)\end{tabular}} & \textbf{\#Queries} & \textbf{\begin{tabular}[c]{@{}c@{}}Avg. Tokens\\per Query\end{tabular}} & \textbf{\begin{tabular}[c]{@{}c@{}}Ratio\\(Rel. English)\end{tabular}} \\
\midrule
Arabic & 69,570 & 1512.8 & 163.5\% & 1717 & 24.2 & 192.1\% \\
Chinese & 69,998 & 931.4 & - & 1218 & 12.3 & --- \\
German & 69,918 & 1361.3 & 147.1\% & 1728 & 20.8 & 164.9\% \\
English & 69,999 & 925.1 & - & 1728 & 12.6 & --- \\
French & 69,933 & 1348.3 & 145.7\% & 1728 & 22.1 & 175.4\% \\
Italian & 69,912 & 1368.2 & 147.9\% & 1728 & 21.5 & 170.7\% \\
Korean & 69,643 & 1537.2 & 166.2\% & 1717 & 25.8 & 204.9\% \\
Portuguese & 69,939 & 1271.1 & 137.4\% & 1728 & 19.8 & 156.8\% \\
Russian & 69,838 & 1506.6 & 162.9\% & 1728 & 24.3 & 192.5\% \\
Spanish & 69,883 & 1258.1 & 136.0\% & 1728 & 20.7 & 164.2\% \\
\bottomrule
\end{tabular}}

\end{table}


\begin{table}[htbp]
\centering
\caption{Statistics for domain: Fire Safety Food Safety (Part 14 of 31).}
\label{tab:dataset_stats_firesafetyfoodsafety}
\small
\resizebox{\textwidth}{!}{
\begin{tabular}{l|c|c|c|c|c|c}
\toprule
\textbf{Language} & \textbf{\#Corpus} & \textbf{\begin{tabular}[c]{@{}c@{}}Avg. Tokens\\per Document\end{tabular}} & \textbf{\begin{tabular}[c]{@{}c@{}}Ratio\\(Rel. English)\end{tabular}} & \textbf{\#Queries} & \textbf{\begin{tabular}[c]{@{}c@{}}Avg. Tokens\\per Query\end{tabular}} & \textbf{\begin{tabular}[c]{@{}c@{}}Ratio\\(Rel. English)\end{tabular}} \\
\midrule
Arabic & 29,634 & 1573.6 & 166.6\% & 1294 & 25.7 & 196.6\% \\
Chinese & 46,710 & 966.3 & - & 1195 & 12.0 & --- \\
German & 29,687 & 1455.5 & 154.1\% & 1295 & 23.4 & 178.7\% \\
English & 29,703 & 944.6 & - & 1296 & 13.1 & --- \\
French & 29,700 & 1459.0 & 154.5\% & 1294 & 25.3 & 193.7\% \\
Italian & 29,696 & 1503.4 & 159.2\% & 1296 & 24.9 & 190.5\% \\
Korean & 29,644 & 1559.7 & 165.1\% & 1294 & 26.7 & 204.6\% \\
Portuguese & 29,696 & 1385.8 & 146.7\% & 1296 & 22.9 & 175.0\% \\
Russian & 29,684 & 1630.1 & 172.6\% & 1296 & 27.5 & 210.8\% \\
Spanish & 29,692 & 1360.0 & 144.0\% & 1296 & 23.3 & 178.2\% \\
\bottomrule
\end{tabular}}

\end{table}


\begin{table}[htbp]
\centering
\caption{Statistics for domain: Game (Part 15 of 31).}
\label{tab:dataset_stats_game}
\small
\resizebox{\textwidth}{!}{
\begin{tabular}{l|c|c|c|c|c|c}
\toprule
\textbf{Language} & \textbf{\#Corpus} & \textbf{\begin{tabular}[c]{@{}c@{}}Avg. Tokens\\per Document\end{tabular}} & \textbf{\begin{tabular}[c]{@{}c@{}}Ratio\\(Rel. English)\end{tabular}} & \textbf{\#Queries} & \textbf{\begin{tabular}[c]{@{}c@{}}Avg. Tokens\\per Query\end{tabular}} & \textbf{\begin{tabular}[c]{@{}c@{}}Ratio\\(Rel. English)\end{tabular}} \\
\midrule
Arabic & 44,213 & 1549.6 & 163.5\% & 1276 & 24.7 & 197.1\% \\
Chinese & 44,119 & 1100.7 & - & 1499 & 12.1 & --- \\
German & 44,341 & 1386.4 & 146.3\% & 1281 & 20.4 & 162.8\% \\
English & 44,373 & 947.7 & - & 1281 & 12.5 & --- \\
French & 44,353 & 1361.2 & 143.6\% & 1281 & 21.5 & 171.4\% \\
Italian & 44,358 & 1397.4 & 147.4\% & 1281 & 20.8 & 166.0\% \\
Korean & 44,272 & 1591.5 & 167.9\% & 1278 & 26.3 & 210.1\% \\
Portuguese & 44,353 & 1292.4 & 136.4\% & 1281 & 19.3 & 153.6\% \\
Russian & 44,299 & 1490.8 & 157.3\% & 1275 & 23.0 & 183.6\% \\
Spanish & 44,316 & 1277.2 & 134.8\% & 1281 & 20.1 & 160.8\% \\
\bottomrule
\end{tabular}}

\end{table}


\begin{table}[htbp]
\centering
\caption{Statistics for domain: Law Judiciary (Part 16 of 31).}
\label{tab:dataset_stats_lawjudiciary}
\small
\resizebox{\textwidth}{!}{
\begin{tabular}{l|c|c|c|c|c|c}
\toprule
\textbf{Language} & \textbf{\#Corpus} & \textbf{\begin{tabular}[c]{@{}c@{}}Avg. Tokens\\per Document\end{tabular}} & \textbf{\begin{tabular}[c]{@{}c@{}}Ratio\\(Rel. English)\end{tabular}} & \textbf{\#Queries} & \textbf{\begin{tabular}[c]{@{}c@{}}Avg. Tokens\\per Query\end{tabular}} & \textbf{\begin{tabular}[c]{@{}c@{}}Ratio\\(Rel. English)\end{tabular}} \\
\midrule
Arabic & 59,765 & 1514.9 & 163.7\% & 1287 & 26.6 & 195.7\% \\
Chinese & 64,276 & 931.6 & - & 1354 & 12.1 & --- \\
German & 59,954 & 1453.8 & 157.1\% & 1283 & 24.6 & 180.4\% \\
English & 59,987 & 925.5 & - & 1293 & 13.6 & --- \\
French & 59,962 & 1410.8 & 152.4\% & 1286 & 25.6 & 188.1\% \\
Italian & 59,965 & 1450.2 & 156.7\% & 1293 & 24.9 & 183.1\% \\
Korean & 59,773 & 1504.3 & 162.5\% & 1293 & 27.8 & 204.0\% \\
Portuguese & 59,959 & 1321.4 & 142.8\% & 1286 & 22.2 & 163.2\% \\
Russian & 59,950 & 1592.2 & 172.0\% & 1286 & 28.4 & 208.8\% \\
Spanish & 59,953 & 1312.7 & 141.8\% & 1286 & 23.2 & 170.4\% \\
\bottomrule
\end{tabular}}

\end{table}


\begin{table}[htbp]
\centering
\caption{Statistics for domain: Literature Emotion (Part 17 of 31).}
\label{tab:dataset_stats_literatureemotion}
\small
\resizebox{\textwidth}{!}{
\begin{tabular}{l|c|c|c|c|c|c}
\toprule
\textbf{Language} & \textbf{\#Corpus} & \textbf{\begin{tabular}[c]{@{}c@{}}Avg. Tokens\\per Document\end{tabular}} & \textbf{\begin{tabular}[c]{@{}c@{}}Ratio\\(Rel. English)\end{tabular}} & \textbf{\#Queries} & \textbf{\begin{tabular}[c]{@{}c@{}}Avg. Tokens\\per Query\end{tabular}} & \textbf{\begin{tabular}[c]{@{}c@{}}Ratio\\(Rel. English)\end{tabular}} \\
\midrule
Arabic & 59,078 & 1499.8 & 160.2\% & 1579 & 24.3 & 180.3\% \\
Chinese & 64,410 & 935.4 & - & 1455 & 12.4 & --- \\
German & 59,407 & 1352.2 & 144.5\% & 1586 & 20.8 & 154.4\% \\
English & 59,439 & 936.0 & - & 1586 & 13.5 & --- \\
French & 59,415 & 1343.0 & 143.5\% & 1586 & 22.4 & 165.7\% \\
Italian & 59,410 & 1348.0 & 144.0\% & 1586 & 21.3 & 157.7\% \\
Korean & 59,189 & 1557.0 & 166.3\% & 1586 & 26.9 & 199.1\% \\
Portuguese & 59,408 & 1265.4 & 135.2\% & 1586 & 19.5 & 144.8\% \\
Russian & 59,380 & 1534.6 & 164.0\% & 1581 & 25.6 & 189.6\% \\
Spanish & 59,402 & 1257.8 & 134.4\% & 1586 & 20.7 & 153.5\% \\
\bottomrule
\end{tabular}}

\end{table}


\begin{table}[htbp]
\centering
\caption{Statistics for domain: Mathematics Statistics (Part 18 of 31).}
\label{tab:dataset_stats_mathematicsstatistics}
\small
\resizebox{\textwidth}{!}{
\begin{tabular}{l|c|c|c|c|c|c}
\toprule
\textbf{Language} & \textbf{\#Corpus} & \textbf{\begin{tabular}[c]{@{}c@{}}Avg. Tokens\\per Document\end{tabular}} & \textbf{\begin{tabular}[c]{@{}c@{}}Ratio\\(Rel. English)\end{tabular}} & \textbf{\#Queries} & \textbf{\begin{tabular}[c]{@{}c@{}}Avg. Tokens\\per Query\end{tabular}} & \textbf{\begin{tabular}[c]{@{}c@{}}Ratio\\(Rel. English)\end{tabular}} \\
\midrule
Arabic & 62,943 & 1398.9 & 149.4\% & 1032 & 26.1 & 186.5\% \\
Chinese & 50,306 & 1014.6 & - & 977 & 12.8 & --- \\
German & 63,178 & 1307.3 & 139.6\% & 1041 & 24.0 & 171.6\% \\
English & 63,420 & 936.4 & - & 1041 & 14.0 & --- \\
French & 63,214 & 1278.4 & 136.5\% & 1039 & 24.6 & 176.2\% \\
Italian & 63,196 & 1290.9 & 137.9\% & 1041 & 24.0 & 172.0\% \\
Korean & 63,049 & 1354.1 & 144.6\% & 1036 & 26.5 & 189.3\% \\
Portuguese & 63,201 & 1221.4 & 130.4\% & 1041 & 22.1 & 158.3\% \\
Russian & 63,188 & 1380.2 & 147.4\% & 1035 & 26.8 & 191.9\% \\
Spanish & 63,169 & 1216.4 & 129.9\% & 1041 & 23.1 & 165.4\% \\
\bottomrule
\end{tabular}}

\end{table}


\begin{table}[htbp]
\centering
\caption{Statistics for domain: Medicine Health Psychology Traditional Chinese Medicine (Part 19 of 31).}
\label{tab:dataset_stats_medicinehealthpsychologytraditionalchinesemedicine}
\small
\resizebox{\textwidth}{!}{
\begin{tabular}{l|c|c|c|c|c|c}
\toprule
\textbf{Language} & \textbf{\#Corpus} & \textbf{\begin{tabular}[c]{@{}c@{}}Avg. Tokens\\per Document\end{tabular}} & \textbf{\begin{tabular}[c]{@{}c@{}}Ratio\\(Rel. English)\end{tabular}} & \textbf{\#Queries} & \textbf{\begin{tabular}[c]{@{}c@{}}Avg. Tokens\\per Query\end{tabular}} & \textbf{\begin{tabular}[c]{@{}c@{}}Ratio\\(Rel. English)\end{tabular}} \\
\midrule
Arabic & 63,143 & 1554.0 & 167.8\% & 1385 & 27.5 & 204.1\% \\
Chinese & 64,519 & 929.2 & - & 1469 & 11.6 & --- \\
German & 63,285 & 1426.6 & 154.0\% & 1385 & 24.4 & 180.7\% \\
English & 63,314 & 926.3 & - & 1385 & 13.5 & --- \\
French & 63,297 & 1420.3 & 153.3\% & 1385 & 25.6 & 189.9\% \\
Italian & 63,292 & 1433.9 & 154.8\% & 1385 & 24.7 & 183.2\% \\
Korean & 63,197 & 1505.5 & 162.5\% & 1378 & 27.4 & 203.7\% \\
Portuguese & 63,297 & 1323.2 & 142.8\% & 1383 & 22.5 & 166.9\% \\
Russian & 63,271 & 1598.4 & 172.5\% & 1385 & 27.8 & 206.5\% \\
Spanish & 63,288 & 1320.7 & 142.6\% & 1385 & 23.7 & 175.7\% \\
\bottomrule
\end{tabular}}

\end{table}


\begin{table}[htbp]
\centering
\caption{Statistics for domain: Mining (Part 20 of 31).}
\label{tab:dataset_stats_mining}
\small
\resizebox{\textwidth}{!}{
\begin{tabular}{l|c|c|c|c|c|c}
\toprule
\textbf{Language} & \textbf{\#Corpus} & \textbf{\begin{tabular}[c]{@{}c@{}}Avg. Tokens\\per Document\end{tabular}} & \textbf{\begin{tabular}[c]{@{}c@{}}Ratio\\(Rel. English)\end{tabular}} & \textbf{\#Queries} & \textbf{\begin{tabular}[c]{@{}c@{}}Avg. Tokens\\per Query\end{tabular}} & \textbf{\begin{tabular}[c]{@{}c@{}}Ratio\\(Rel. English)\end{tabular}} \\
\midrule
Arabic & 47,403 & 1435.2 & 165.1\% & 1319 & 26.3 & 195.2\% \\
Chinese & 58,159 & 975.6 & - & 1474 & 12.6 & --- \\
German & 47,500 & 1329.7 & 152.9\% & 1319 & 23.7 & 176.1\% \\
English & 47,521 & 869.5 & - & 1319 & 13.5 & --- \\
French & 47,500 & 1325.8 & 152.5\% & 1319 & 24.8 & 183.9\% \\
Italian & 47,503 & 1351.8 & 155.5\% & 1319 & 24.0 & 178.4\% \\
Korean & 47,438 & 1448.9 & 166.6\% & 1319 & 33.9 & 251.5\% \\
Portuguese & 47,510 & 1254.8 & 144.3\% & 1319 & 22.2 & 164.6\% \\
Russian & 47,497 & 1495.8 & 172.0\% & 1319 & 27.5 & 204.3\% \\
Spanish & 47,497 & 1240.2 & 142.6\% & 1319 & 23.0 & 170.3\% \\
\bottomrule
\end{tabular}}

\end{table}


\begin{table}[htbp]
\centering
\caption{Statistics for domain: News Media (Part 21 of 31).}
\label{tab:dataset_stats_newsmedia}
\small
\resizebox{\textwidth}{!}{
\begin{tabular}{l|c|c|c|c|c|c}
\toprule
\textbf{Language} & \textbf{\#Corpus} & \textbf{\begin{tabular}[c]{@{}c@{}}Avg. Tokens\\per Document\end{tabular}} & \textbf{\begin{tabular}[c]{@{}c@{}}Ratio\\(Rel. English)\end{tabular}} & \textbf{\#Queries} & \textbf{\begin{tabular}[c]{@{}c@{}}Avg. Tokens\\per Query\end{tabular}} & \textbf{\begin{tabular}[c]{@{}c@{}}Ratio\\(Rel. English)\end{tabular}} \\
\midrule
Arabic & 53,784 & 1497.8 & 162.1\% & 1264 & 25.6 & 190.8\% \\
Chinese & 55,021 & 1024.2 & - & 1453 & 12.1 & --- \\
German & 53,905 & 1376.1 & 148.9\% & 1270 & 22.6 & 168.3\% \\
English & 53,931 & 924.1 & - & 1271 & 13.4 & --- \\
French & 53,918 & 1376.2 & 148.9\% & 1271 & 24.0 & 178.9\% \\
Italian & 53,922 & 1391.0 & 150.5\% & 1271 & 23.1 & 172.0\% \\
Korean & 53,785 & 1549.1 & 167.6\% & 1266 & 28.2 & 210.5\% \\
Portuguese & 53,918 & 1292.1 & 139.8\% & 1271 & 21.2 & 157.8\% \\
Russian & 53,897 & 1561.0 & 168.9\% & 1271 & 27.0 & 201.5\% \\
Spanish & 53,915 & 1278.5 & 138.3\% & 1271 & 22.0 & 163.8\% \\
\bottomrule
\end{tabular}}

\end{table}


\begin{table}[htbp]
\centering
\caption{Statistics for domain: Other Information Services Information Security (Part 22 of 31).}
\label{tab:dataset_stats_otherinformationservicesinformationsecurity}
\small
\resizebox{\textwidth}{!}{
\begin{tabular}{l|c|c|c|c|c|c}
\toprule
\textbf{Language} & \textbf{\#Corpus} & \textbf{\begin{tabular}[c]{@{}c@{}}Avg. Tokens\\per Document\end{tabular}} & \textbf{\begin{tabular}[c]{@{}c@{}}Ratio\\(Rel. English)\end{tabular}} & \textbf{\#Queries} & \textbf{\begin{tabular}[c]{@{}c@{}}Avg. Tokens\\per Query\end{tabular}} & \textbf{\begin{tabular}[c]{@{}c@{}}Ratio\\(Rel. English)\end{tabular}} \\
\midrule
Arabic & 31,329 & 1571.8 & 165.4\% & 839 & 25.2 & 201.3\% \\
Chinese & 38,947 & 989.5 & - & 1044 & 11.8 & --- \\
German & 31,360 & 1494.3 & 157.2\% & 839 & 22.8 & 182.5\% \\
English & 31,375 & 950.5 & - & 840 & 12.5 & --- \\
French & 31,369 & 1478.1 & 155.5\% & 839 & 24.2 & 193.1\% \\
Italian & 31,366 & 1515.0 & 159.4\% & 839 & 23.6 & 188.7\% \\
Korean & 31,341 & 1566.9 & 164.8\% & 833 & 25.9 & 207.2\% \\
Portuguese & 31,367 & 1376.3 & 144.8\% & 839 & 21.1 & 168.4\% \\
Russian & 31,365 & 1635.4 & 172.1\% & 839 & 26.1 & 208.7\% \\
Spanish & 31,368 & 1359.9 & 143.1\% & 839 & 21.9 & 174.8\% \\
\bottomrule
\end{tabular}}

\end{table}


\begin{table}[htbp]
\centering
\caption{Statistics for domain: Other Manufacturing (Part 23 of 31).}
\label{tab:dataset_stats_othermanufacturing}
\small
\resizebox{\textwidth}{!}{
\begin{tabular}{l|c|c|c|c|c|c}
\toprule
\textbf{Language} & \textbf{\#Corpus} & \textbf{\begin{tabular}[c]{@{}c@{}}Avg. Tokens\\per Document\end{tabular}} & \textbf{\begin{tabular}[c]{@{}c@{}}Ratio\\(Rel. English)\end{tabular}} & \textbf{\#Queries} & \textbf{\begin{tabular}[c]{@{}c@{}}Avg. Tokens\\per Query\end{tabular}} & \textbf{\begin{tabular}[c]{@{}c@{}}Ratio\\(Rel. English)\end{tabular}} \\
\midrule
Arabic & 59,176 & 1472.4 & 163.3\% & 1609 & 25.8 & 196.9\% \\
Chinese & 63,160 & 928.7 & - & 1496 & 12.1 & --- \\
German & 59,231 & 1370.3 & 151.9\% & 1615 & 22.9 & 174.5\% \\
English & 59,269 & 901.9 & - & 1615 & 13.1 & --- \\
French & 59,248 & 1338.9 & 148.5\% & 1615 & 24.1 & 183.5\% \\
Italian & 59,250 & 1383.4 & 153.4\% & 1615 & 23.7 & 180.5\% \\
Korean & 59,170 & 1433.7 & 159.0\% & 1615 & 26.4 & 201.4\% \\
Portuguese & 59,253 & 1282.2 & 142.2\% & 1615 & 21.7 & 165.3\% \\
Russian & 59,235 & 1478.8 & 164.0\% & 1615 & 25.6 & 195.3\% \\
Spanish & 59,252 & 1268.2 & 140.6\% & 1615 & 22.7 & 172.7\% \\
\bottomrule
\end{tabular}}

\end{table}


\begin{table}[htbp]
\centering
\caption{Statistics for domain: Petrochemical (Part 24 of 31).}
\label{tab:dataset_stats_petrochemical}
\small
\resizebox{\textwidth}{!}{
\begin{tabular}{l|c|c|c|c|c|c}
\toprule
\textbf{Language} & \textbf{\#Corpus} & \textbf{\begin{tabular}[c]{@{}c@{}}Avg. Tokens\\per Document\end{tabular}} & \textbf{\begin{tabular}[c]{@{}c@{}}Ratio\\(Rel. English)\end{tabular}} & \textbf{\#Queries} & \textbf{\begin{tabular}[c]{@{}c@{}}Avg. Tokens\\per Query\end{tabular}} & \textbf{\begin{tabular}[c]{@{}c@{}}Ratio\\(Rel. English)\end{tabular}} \\
\midrule
Arabic & 57,032 & 1499.3 & 164.9\% & 1245 & 27.0 & 196.8\% \\
Chinese & 63,395 & 930.0 & - & 1605 & 12.7 & --- \\
German & 57,116 & 1389.7 & 152.8\% & 1245 & 24.1 & 175.9\% \\
English & 57,150 & 909.4 & - & 1245 & 13.7 & --- \\
French & 57,131 & 1393.0 & 153.2\% & 1245 & 25.7 & 187.0\% \\
Italian & 57,133 & 1421.4 & 156.3\% & 1245 & 24.9 & 181.3\% \\
Korean & 57,034 & 1487.5 & 163.6\% & 1240 & 27.7 & 201.9\% \\
Portuguese & 57,127 & 1317.5 & 144.9\% & 1245 & 23.3 & 169.5\% \\
Russian & 57,114 & 1559.4 & 171.5\% & 1243 & 27.8 & 202.9\% \\
Spanish & 57,116 & 1302.2 & 143.2\% & 1245 & 23.9 & 174.2\% \\
\bottomrule
\end{tabular}}

\end{table}


\begin{table}[htbp]
\centering
\caption{Statistics for domain: Real Estate Construction (Part 25 of 31).}
\label{tab:dataset_stats_realestateconstruction}
\small
\resizebox{\textwidth}{!}{
\begin{tabular}{l|c|c|c|c|c|c}
\toprule
\textbf{Language} & \textbf{\#Corpus} & \textbf{\begin{tabular}[c]{@{}c@{}}Avg. Tokens\\per Document\end{tabular}} & \textbf{\begin{tabular}[c]{@{}c@{}}Ratio\\(Rel. English)\end{tabular}} & \textbf{\#Queries} & \textbf{\begin{tabular}[c]{@{}c@{}}Avg. Tokens\\per Query\end{tabular}} & \textbf{\begin{tabular}[c]{@{}c@{}}Ratio\\(Rel. English)\end{tabular}} \\
\midrule
Arabic & 55,904 & 1529.9 & 166.1\% & 1618 & 25.2 & 191.3\% \\
Chinese & 63,618 & 931.3 & - & 1698 & 12.0 & --- \\
German & 56,015 & 1403.1 & 152.3\% & 1618 & 22.7 & 172.1\% \\
English & 56,034 & 921.1 & - & 1618 & 13.2 & --- \\
French & 56,024 & 1385.8 & 150.4\% & 1618 & 23.7 & 180.0\% \\
Italian & 56,018 & 1429.6 & 155.2\% & 1618 & 23.1 & 175.7\% \\
Korean & 55,928 & 1544.4 & 167.7\% & 1618 & 26.9 & 204.1\% \\
Portuguese & 56,020 & 1314.9 & 142.8\% & 1618 & 21.3 & 161.3\% \\
Russian & 55,999 & 1572.9 & 170.8\% & 1617 & 26.2 & 198.8\% \\
Spanish & 56,015 & 1307.3 & 141.9\% & 1618 & 22.2 & 168.7\% \\
\bottomrule
\end{tabular}}

\end{table}


\begin{table}[htbp]
\centering
\caption{Statistics for domain: Sports (Part 26 of 31).}
\label{tab:dataset_stats_sports}
\small
\resizebox{\textwidth}{!}{
\begin{tabular}{l|c|c|c|c|c|c}
\toprule
\textbf{Language} & \textbf{\#Corpus} & \textbf{\begin{tabular}[c]{@{}c@{}}Avg. Tokens\\per Document\end{tabular}} & \textbf{\begin{tabular}[c]{@{}c@{}}Ratio\\(Rel. English)\end{tabular}} & \textbf{\#Queries} & \textbf{\begin{tabular}[c]{@{}c@{}}Avg. Tokens\\per Query\end{tabular}} & \textbf{\begin{tabular}[c]{@{}c@{}}Ratio\\(Rel. English)\end{tabular}} \\
\midrule
Arabic & 55,871 & 1499.9 & 165.0\% & 1372 & 32.5 & 238.5\% \\
Chinese & 62,963 & 931.1 & - & 1429 & 12.3 & --- \\
German & 56,023 & 1314.6 & 144.6\% & 1374 & 22.1 & 162.1\% \\
English & 56,048 & 908.8 & - & 1374 & 13.6 & --- \\
French & 56,035 & 1323.3 & 145.6\% & 1374 & 23.8 & 174.6\% \\
Italian & 56,030 & 1336.0 & 147.0\% & 1374 & 22.9 & 167.9\% \\
Korean & 55,898 & 1542.0 & 169.7\% & 1368 & 34.4 & 252.4\% \\
Portuguese & 56,029 & 1246.1 & 137.1\% & 1374 & 21.2 & 155.8\% \\
Russian & 55,991 & 1519.3 & 167.2\% & 1374 & 33.4 & 244.9\% \\
Spanish & 56,024 & 1243.1 & 136.8\% & 1374 & 22.4 & 164.2\% \\
\bottomrule
\end{tabular}}

\end{table}


\begin{table}[htbp]
\centering
\caption{Statistics for domain: Subject Education Education (Part 27 of 31).}
\label{tab:dataset_stats_subjecteducationeducation}
\small
\resizebox{\textwidth}{!}{
\begin{tabular}{l|c|c|c|c|c|c}
\toprule
\textbf{Language} & \textbf{\#Corpus} & \textbf{\begin{tabular}[c]{@{}c@{}}Avg. Tokens\\per Document\end{tabular}} & \textbf{\begin{tabular}[c]{@{}c@{}}Ratio\\(Rel. English)\end{tabular}} & \textbf{\#Queries} & \textbf{\begin{tabular}[c]{@{}c@{}}Avg. Tokens\\per Query\end{tabular}} & \textbf{\begin{tabular}[c]{@{}c@{}}Ratio\\(Rel. English)\end{tabular}} \\
\midrule
Arabic & 61,343 & 1531.2 & 164.5\% & 1669 & 25.5 & 190.1\% \\
Chinese & 64,739 & 932.0 & - & 1751 & 11.4 & --- \\
German & 61,544 & 1410.7 & 151.5\% & 1672 & 22.7 & 169.2\% \\
English & 61,577 & 931.0 & - & 1672 & 13.4 & --- \\
French & 61,554 & 1408.9 & 151.3\% & 1672 & 24.3 & 180.9\% \\
Italian & 61,549 & 1432.4 & 153.9\% & 1672 & 23.2 & 172.7\% \\
Korean & 61,466 & 1549.0 & 166.4\% & 1672 & 27.1 & 202.3\% \\
Portuguese & 61,561 & 1318.0 & 141.6\% & 1672 & 21.1 & 157.2\% \\
Russian & 61,525 & 1590.1 & 170.8\% & 1672 & 26.4 & 197.0\% \\
Spanish & 61,551 & 1308.9 & 140.6\% & 1672 & 22.1 & 164.4\% \\
\bottomrule
\end{tabular}}

\end{table}


\begin{table}[htbp]
\centering
\caption{Statistics for domain: Technology Scientific Research (Part 28 of 31).}
\label{tab:dataset_stats_technologyscientificresearch}
\small
\resizebox{\textwidth}{!}{
\begin{tabular}{l|c|c|c|c|c|c}
\toprule
\textbf{Language} & \textbf{\#Corpus} & \textbf{\begin{tabular}[c]{@{}c@{}}Avg. Tokens\\per Document\end{tabular}} & \textbf{\begin{tabular}[c]{@{}c@{}}Ratio\\(Rel. English)\end{tabular}} & \textbf{\#Queries} & \textbf{\begin{tabular}[c]{@{}c@{}}Avg. Tokens\\per Query\end{tabular}} & \textbf{\begin{tabular}[c]{@{}c@{}}Ratio\\(Rel. English)\end{tabular}} \\
\midrule
Arabic & 61,031 & 1510.4 & 162.8\% & 1398 & 26.8 & 196.6\% \\
Chinese & 63,950 & 930.5 & - & 1594 & 12.0 & --- \\
German & 61,220 & 1396.1 & 150.5\% & 1401 & 23.7 & 173.7\% \\
English & 61,254 & 927.7 & - & 1401 & 13.6 & --- \\
French & 61,229 & 1381.2 & 148.9\% & 1401 & 25.0 & 183.3\% \\
Italian & 61,233 & 1405.4 & 151.5\% & 1401 & 24.1 & 176.9\% \\
Korean & 61,141 & 1490.1 & 160.6\% & 1401 & 27.4 & 201.0\% \\
Portuguese & 61,220 & 1311.7 & 141.4\% & 1401 & 22.2 & 163.0\% \\
Russian & 61,201 & 1550.6 & 167.1\% & 1399 & 27.2 & 199.7\% \\
Spanish & 61,218 & 1299.2 & 140.0\% & 1401 & 23.3 & 170.6\% \\
\bottomrule
\end{tabular}
}
\end{table}


\begin{table}[htbp]
\centering
\caption{Statistics for domain: Tourism Geography (Part 29 of 31).}
\label{tab:dataset_stats_tourismgeography}
\small
\resizebox{\textwidth}{!}{
\begin{tabular}{l|c|c|c|c|c|c}
\toprule
\textbf{Language} & \textbf{\#Corpus} & \textbf{\begin{tabular}[c]{@{}c@{}}Avg. Tokens\\per Document\end{tabular}} & \textbf{\begin{tabular}[c]{@{}c@{}}Ratio\\(Rel. English)\end{tabular}} & \textbf{\#Queries} & \textbf{\begin{tabular}[c]{@{}c@{}}Avg. Tokens\\per Query\end{tabular}} & \textbf{\begin{tabular}[c]{@{}c@{}}Ratio\\(Rel. English)\end{tabular}} \\
\midrule
Arabic & 53,946 & 1499.7 & 163.8\% & 1504 & 23.6 & 182.2\% \\
Chinese & 62,844 & 935.1 & - & 1456 & 11.7 & --- \\
German & 54,127 & 1338.0 & 146.2\% & 1504 & 20.5 & 157.9\% \\
English & 54,144 & 915.4 & - & 1504 & 13.0 & --- \\
French & 54,127 & 1346.8 & 147.1\% & 1504 & 22.0 & 169.7\% \\
Italian & 54,124 & 1363.1 & 148.9\% & 1504 & 21.1 & 163.2\% \\
Korean & 53,981 & 1524.5 & 166.5\% & 1487 & 25.2 & 194.7\% \\
Portuguese & 54,124 & 1279.5 & 139.8\% & 1504 & 19.7 & 152.2\% \\
Russian & 54,037 & 1538.5 & 168.1\% & 1503 & 24.5 & 188.9\% \\
Spanish & 54,099 & 1277.9 & 139.6\% & 1504 & 20.8 & 160.2\% \\
\bottomrule
\end{tabular}}

\end{table}


\begin{table}[htbp]
\centering
\caption{Statistics for domain: Transportation (Part 30 of 31).}
\label{tab:dataset_stats_transportation}
\small
\resizebox{\textwidth}{!}{
\begin{tabular}{l|c|c|c|c|c|c}
\toprule
\textbf{Language} & \textbf{\#Corpus} & \textbf{\begin{tabular}[c]{@{}c@{}}Avg. Tokens\\per Document\end{tabular}} & \textbf{\begin{tabular}[c]{@{}c@{}}Ratio\\(Rel. English)\end{tabular}} & \textbf{\#Queries} & \textbf{\begin{tabular}[c]{@{}c@{}}Avg. Tokens\\per Query\end{tabular}} & \textbf{\begin{tabular}[c]{@{}c@{}}Ratio\\(Rel. English)\end{tabular}} \\
\midrule
Arabic & 53,730 & 1516.6 & 166.7\% & 1418 & 26.1 & 195.6\% \\
Chinese & 63,115 & 931.9 & - & 1370 & 12.2 & --- \\
German & 53,812 & 1401.2 & 154.0\% & 1418 & 23.4 & 174.8\% \\
English & 53,829 & 909.9 & - & 1418 & 13.4 & --- \\
French & 53,820 & 1383.5 & 152.1\% & 1418 & 24.4 & 182.9\% \\
Italian & 53,821 & 1435.5 & 157.8\% & 1418 & 24.4 & 182.4\% \\
Korean & 53,745 & 1513.9 & 166.4\% & 1415 & 27.3 & 204.5\% \\
Portuguese & 53,821 & 1317.5 & 144.8\% & 1418 & 22.0 & 164.8\% \\
Russian & 53,801 & 1580.0 & 173.6\% & 1418 & 27.3 & 204.2\% \\
Spanish & 53,818 & 1311.4 & 144.1\% & 1418 & 23.1 & 172.7\% \\
\bottomrule
\end{tabular}}

\end{table}


\begin{table}[htbp]
\centering
\caption{Statistics for domain: Water Resources Ocean (Part 31 of 31).}
\label{tab:dataset_stats_waterresourcesocean}
\small
\resizebox{\textwidth}{!}{
\begin{tabular}{l|c|c|c|c|c|c}
\toprule
\textbf{Language} & \textbf{\#Corpus} & \textbf{\begin{tabular}[c]{@{}c@{}}Avg. Tokens\\per Document\end{tabular}} & \textbf{\begin{tabular}[c]{@{}c@{}}Ratio\\(Rel. English)\end{tabular}} & \textbf{\#Queries} & \textbf{\begin{tabular}[c]{@{}c@{}}Avg. Tokens\\per Query\end{tabular}} & \textbf{\begin{tabular}[c]{@{}c@{}}Ratio\\(Rel. English)\end{tabular}} \\
\midrule
Arabic & 53,749 & 1514.7 & 166.9\% & 1328 & 26.3 & 195.4\% \\
Chinese & 63,027 & 929.0 & - & 1489 & 12.2 & --- \\
German & 53,839 & 1403.8 & 154.7\% & 1328 & 23.8 & 176.6\% \\
English & 53,854 & 907.5 & - & 1328 & 13.5 & --- \\
French & 53,838 & 1415.5 & 156.0\% & 1328 & 25.4 & 188.7\% \\
Italian & 53,842 & 1439.7 & 158.6\% & 1328 & 24.8 & 184.4\% \\
Korean & 53,746 & 1487.7 & 163.9\% & 1325 & 26.7 & 198.5\% \\
Portuguese & 53,841 & 1325.7 & 146.1\% & 1328 & 22.5 & 167.0\% \\
Russian & 53,813 & 1579.1 & 174.0\% & 1328 & 27.4 & 203.2\% \\
Spanish & 53,843 & 1323.7 & 145.9\% & 1328 & 23.5 & 174.7\% \\
\bottomrule
\end{tabular}}

\end{table}

\end{document}